\documentclass[PAPER, cernpreprint, texlive=2023, coverpage=false, UKenglish, orcidlogo, texmf, orcidlogo]{atlasdoc}

\usepackage[subcaption,siunitx=false]{atlaspackage}
\usepackage[backref=false,block=none]{atlasbiblatex}
 
\usepackage{atlasphysics}

\graphicspath{{logos/}{figures/}}
 
\usepackage{ANA-SIMU-2020-04-PAPER-defs}

\usepackage[binary-units]{siunitx}

\usepackage{placeins}
 
\usepackage{tikz}
\usetikzlibrary{3d,calc,decorations.text,shapes,positioning,shadows,decorations.pathreplacing}
\usetikzlibrary{shapes.geometric,arrows.meta,decorations.markings,arrows,patterns}
\usetikzlibrary{external}
\usetikzlibrary{spy}
\usepackage{tikzscale}
\usepackage{tikzpagenodes}
 
\tikzstyle{only in spy node}=[
transform canvas={
shift={(tikzspyinnode)},
scale=\magn,
}
]
 
\tikzset{external/system call={pdflatex -shell-escape
\tikzexternalcheckshellescape -halt-on-error
-interaction=batchmode -jobname "\image" "\texsource"}}
\tikzset{
double arrow/.style args={#1 colored by #2 and #3}{
-stealth,line width=#1,#2, 
postaction={draw,-stealth,#3,line width=(#1)/3,
shorten <=(#1)/3,shorten >=2*(#1)/3}, 
}
}
 
\tikzset{
double -latex/.style args={#1 colored by #2 and #3}{
-latex,line width=#1,#2,
postaction={draw,-latex,#3,line width=(#1)/3,shorten <=(#1)/4,shorten >=4.5*(#1)/3},
}
}
 
\tikzset{
dim above/.style={to path={\pgfextra{
\pgfinterruptpath
\draw[>=latex,|<->|] let
\p1=($(\tikztostart)!2mm!90:(\tikztotarget)$),
\p2=($(\tikztotarget)!2mm!-90:(\tikztostart)$)
in(\p1) -- (\p2) node[pos=.5,sloped,above]{#1};
\endpgfinterruptpath
}(\tikztostart) -- (\tikztotarget) \tikztonodes
}
},
dim below/.style={to path={\pgfextra{
\pgfinterruptpath
\draw[>=latex,|<->|] let
\p1=($(\tikztostart)!2mm!90:(\tikztotarget)$),
\p2=($(\tikztotarget)!2mm!-90:(\tikztostart)$)
in (\p1) -- (\p2) node[pos=.5,sloped,below]{#1};
\endpgfinterruptpath
}(\tikztostart) -- (\tikztotarget) \tikztonodes
}
},
}
 
\tikzset{
hatch distance/.store in=\hatchdistance,
hatch distance=10pt,
hatch thickness/.store in=\hatchthickness,
hatch thickness=2pt
}
 
\makeatletter
\pgfdeclarepatternformonly[\hatchdistance,\hatchthickness]{flexible hatch}{\pgfqpoint{0pt}{0pt}}{\pgfqpoint{\hatchdistance}{\hatchdistance}}{\pgfpoint{\hatchdistance-1pt}{\hatchdistance-1pt}}{
\pgfsetcolor{\tikz@pattern@color}
\pgfsetlinewidth{\hatchthickness}
\pgfpathmoveto{\pgfqpoint{0pt}{0pt}}
\pgfpathlineto{\pgfqpoint{\hatchdistance}{\hatchdistance}}
\pgfusepath{stroke}
}
 
\usepackage{bm}
\usepackage{multirow}
 
\usepackage{rotating}
\usepackage{wrapfig}
\usepackage{floatrow}
\newfloatcommand{capbtabbox}{table}[][\FBwidth]
\usepackage{changepage}
\usepackage{afterpage}
\usepackage{textcomp}

\AtlasTitle{Deep generative models for fast photon shower simulation in ATLAS}
 
\AtlasAbstract{
The need for large-scale production of highly accurate simulated event samples for the extensive physics programme of the ATLAS experiment at the Large Hadron Collider motivates the development of new simulation techniques. Building on the recent success of deep learning algorithms, variational autoencoders and generative adversarial networks are investigated for modelling the response of the central region of the ATLAS electromagnetic calorimeter to photons of various energies. The properties of synthesised showers are compared with showers from a full detector simulation using \textsc{Geant4}.
Both variational autoencoders and generative adversarial networks are capable of quickly simulating electromagnetic showers with correct total energies and stochasticity, though the modelling of some shower shape distributions requires more refinement.
This feasibility study demonstrates the potential of using such algorithms for ATLAS fast calorimeter simulation in the future and shows a possible way to complement current simulation techniques.
}
 
\author{The ATLAS Collaboration}
 
\AtlasRefCode{SIMU-2020-04}
 
\AtlasDate{\today}

\AtlasJournal{Computing and Software for Big Science}
\AtlasJournalRef{Comput Softw Big Sci 8, 7 (2024)}
\AtlasDOI{10.1007/s41781-023-00106-9}
 
\PreprintIdNumber{CERN-EP-2022-185}

\hypersetup{pdftitle={Deep generative models for fast photon shower simulation in ATLAS},pdfauthor={The ATLAS Collaboration}}
 
\begin{document}
 
\maketitle
 
\tableofcontents
 
\section{Introduction}
\label{sec:intro}

 
The extensive physics programme of the ATLAS experiment~\cite{PERF-2007-01} at the Large Hadron Collider~\cite{Evans:2008zzb} relies on highly accurate Monte Carlo (MC) simulation as a basis for hypothesis tests that compare physics models to collision data.
Part of the simulation chain includes the interaction of incoming particles with the ATLAS detector.
Calorimeters are a key detector technology used for measuring the energy and position of charged and neutral particles; however, it is computationally intensive to simulate their response to incoming particles.
ATLAS employs sampling calorimeters where particles lose their energy in a cascade (called a shower) of electromagnetic and hadronic interactions with a dense absorber material. The number of particles produced in these interactions is subsequently measured in thin sampling layers of an active medium interleaved with the absorber material.
 
The deposition of energy from a developing shower in the calorimeter is a stochastic process that cannot be calculated from first principles but rather is modelled with a precise simulation. It requires the calculation of interactions of particles with matter at the microscopic level as implemented using the \GEANT toolkit~\cite{Agostinelli:2002hh}. This simulation process is inherently slow and thus becomes a bottleneck in the ATLAS simulation pipeline~\cite{SOFT-2010-01,Calafiura:2729668}.
The large number of particles produced in showers within a calorimeter, as well as the complexity of the geometry of a calorimeter, lead to the large computational complexity in simulations.
To meet the growing demands of physics analyses on MC simulation, ATLAS already relies strongly on fast calorimeter simulation techniques based on thousands of individual parameterisations of the calorimeter response in the directions longitudinal and transverse to the axis of the shower, given the pseudorapidity and initial energy of a single particle~\cite{atlfast3}.

In recent years, it has been demonstrated that deep neural networks can accurately model the underlying distributions of rich, structured data for a wide range of applications, notably in the areas of computer vision, natural language processing and signal processing. The ability to embed complex distributions in a low-dimensional manifold has been leveraged to generate samples of higher dimensionality and approximate the underlying probability densities. Among the most promising approaches to generative models are variational autoencoders (VAEs)~\cite{2013arXiv1312.6114K,2014arXiv1401.4082J} and generative adversarial networks (GANs)~\cite{2014arXiv1406.2661G}, which have been used to simulate the response of idealised and future high-granularity calorimeters~\cite{deOliveira:2017pjk,Paganini:2017hrr,Paganini:2017dwg,Buhmann:2020pmy,Erdmann:2018jxd}.
 
This paper presents novel applications of generative models as alternatives to techniques currently used~\cite{atlfast3} in the fast simulation of the ATLAS calorimeter response.
Notably, the two models presented in this paper address the challenge of reproducing the true distributions of shower properties when using simple generative models.
Alongside the model presented in Ref.~\cite{atlfast3}, this is one of the first applications of generative models using the real ATLAS detector geometry instead of a simplified model and is the first to use the readout of energy deposits recorded by the cells of the ATLAS calorimeter.
 
The simulation approach used in ATLAS focuses on single-particle simulations resulting in additive energy deposits in the calorimeter. Fast-simulation approaches replace \GEANT for the simulation of individual particles showering in the calorimeter.
The ATLAS calorimeter is particularly challenging due to its complex structure, which comprises several layers of cells with varying geometry and depth, following an accordion shape in the radial direction in several layers.
This paper focuses on generating showers for single photons covering a range of energies from 1~\GeV\ to 262~\GeV\ in a narrow slice of the central pseudorapidity region of the electromagnetic calorimeter.\footnote{\AtlasCoordFootnote} The calorimeter geometry in the central region is more uniform than in other regions, and the electromagnetic showers of photons are intrinsically less complex than showers initiated by hadrons.
By focusing on a narrow region of the calorimeter it is possible to neglect the dependence of the calorimeter response on the pseudorapidity of the incident particle.
These approaches demonstrate the feasibility of using such models to generate particle showers that reproduce numerous details of the \GEANT simulation, and this work shows possible ways to complement current techniques.
 
The paper is organised as follows. Section~\ref{sec:calorimeter} describes the design of the ATLAS detector, and the geometry of the calorimeter system in particular. The Monte Carlo simulation samples are described in Section~\ref{sec:samples}. Section~\ref{sec:algorithms} presents a comprehensive summary of the deep generative models exploited in this work, as well as their optimisation. The results obtained by these algorithms are reviewed in Section~\ref{sec:results}, with the challenges faced discussed in Section~\ref{sec:challenges}. Section~\ref{sec:conclusion} summarises the achievements and challenges presented in the paper.


\section{ATLAS calorimeter}
\label{sec:calorimeter}

The ATLAS experiment at the LHC is a multipurpose particle detector with a forward--backward symmetric cylindrical geometry, covering the full $4\pi$ solid angle almost hermetically by combining several subdetector systems installed in layers around the interaction point. 
 
Starting at the beamline and moving radially outwards, the ATLAS detector comprises an inner tracker covering the central $|\eta|$ region, a thin superconducting solenoid, a calorimeter system, and a muon spectrometer based on a system of air-core toroidal superconducting magnets. The calorimeter system consists of separate electromagnetic~(EM) and hadronic calorimeters, which are designed to capture and record the energy from different types of incident particles, and a forward calorimeter covering a pseudorapidity range of $3.2< |\eta| <4.9$. The EM and hadronic calorimeters covering the lower $|\eta|$ range are divided into the barrel and endcap regions.

\begin{figure}[tb]
\centering
\includegraphics[width=0.67\textwidth]{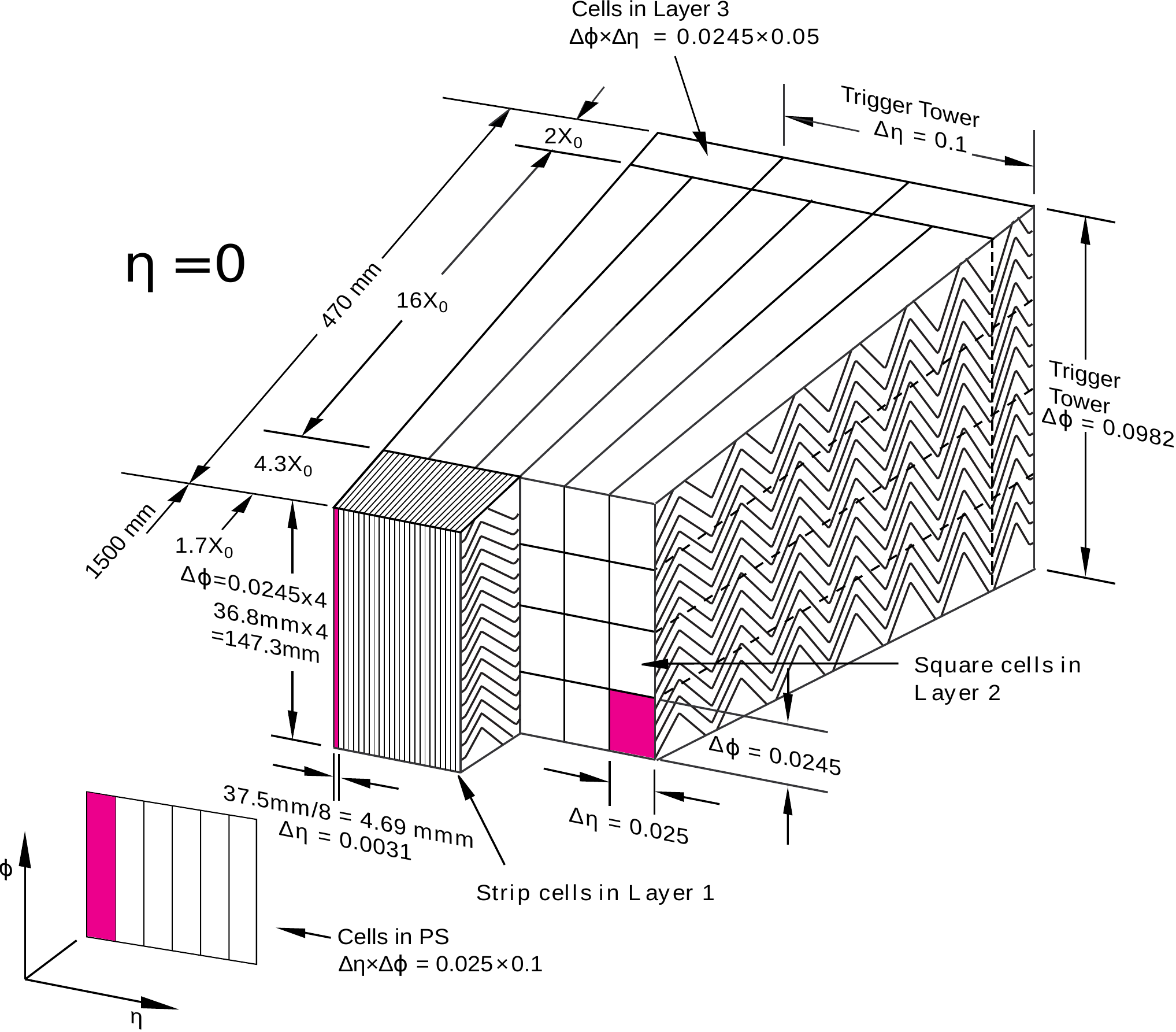}
\caption{Schematic of a slice of the ATLAS electromagnetic calorimeter at $\eta=0$~\cite{Collaboration:2285582}. The cell geometry and relative depths of the individual layers are labelled for the front (Layer 1), middle (Layer 2) and back (Layer 3). The accordion structure can be seen along the depths of all sampling layers. The size and geometry of the cells in the presampler (PS) are shown to scale.}
\label{fig:atlas:emcal}
\end{figure}

The EM calorimeter in the barrel region, which covers $|\eta| < 1.475$, is a lead/liquid-argon~(LAr) sampling calorimeter with an accordion geometry in the $\phi$ direction. It is constructed from accordion-shaped kapton electrodes and lead absorber plates with varying size in different regions of the detector, as shown in~Figure~\ref{fig:atlas:emcal}. This geometry allows modules to overlap each other and eliminates intermodule gaps. The design provides EM energy measurements with high granularity and longitudinal segmentation into multiple layers, capturing the shower development in depth. The three layers in the central region of the EM barrel are labelled as front, middle and back. In the region $|\eta| < 1.8$, the ATLAS experiment is equipped with a LAr presampler detector to correct for the energy lost by electrons and photons upstream of the calorimeter and thus to improve the energy measurements in this region. The EM calorimeter is surrounded by a hadronic sampling calorimeter consisting of steel absorbers and active scintillator tiles, covering the central pseudorapidity range ($|\eta| < 1.7$).

The ATLAS EM calorimeter is segmented into three-dimensional cells, which are rectangular in the $\eta$--$\phi$ plane, follow the accordion shape in $r$, and vary in shape and size in the ($r/z$, $\eta$, $\phi$) space. The EM calorimeter is over 22 electromagnetic radiation lengths in depth at $|\eta| \approx 0$, ensuring that there is little leakage of EM showers into the hadronic calorimeter. 
Throughout the paper, the raw $\eta$ and $\phi$ values of the calorimeter cells are used. These do not include corrections for imperfections in the detector, such as deformations or misalignment.

Among the three EM barrel layers, the middle layer is the thickest and receives the largest energy deposit from EM showers. The front layer is thinner and exhibits a fine granularity in $|\eta| < 1.475$ (eight times finer than in the middle layer), but is four times less granular in $\phi$. In the back layer, less energy is deposited than in the front and middle layers.

\section{Monte Carlo samples and preprocessing}
\label{sec:samples}

The ATLAS simulation infrastructure, consisting of event generation, detector simulation and digitisation, was used to produce and validate the samples used in the studies presented in this paper.
Dedicated samples of single photons were simulated in the ATLAS calorimeter using \GEANT with the standard ATLAS Run~2 geometry, allowing the models to focus on simulating the showers initiated by photons which did not convert in the ATLAS inner tracking detector (ID).
This simulation infrastructure is a component of an extensive software suite~\cite{ATL-SOFT-PUB-2021-001} which is also used in the reconstruction and analysis of real and simulated data, in detector operations, and in the trigger and data acquisition systems of the experiment.
 
Training samples of individual photon showers were generated for nine discrete particle energies logarithmically spaced between 1024 and 262\!\;144~\MeV, with the energy doubling in each step. These energy points are labelled by the nearest integer value in \GeV.
The showers are uniformly distributed in $0.20 < |\eta| < 0.25$, which corresponds to the width in $\eta$ of one cell in the back layer, and uniformly spread across the whole range of azimuthal angle $\phi$.
Each simulated sample contains approximately 10\,000 generated events, for a total of nearly 90\,000 events.
The initial `MC-truth' particles were generated on the front surface of the calorimeter, representing the handover boundary between the ID and calorimeter in the Integrated Simulation Framework \cite{SOFT-2010-01}. Thus, the showers are not subject to energy losses in the ID or the LAr cryostat.
The generated samples do not include displacements corresponding to the expected beam spread in the ATLAS interaction region, and their direction of propagation is consistent with production at the interaction point.
In the digitisation of the \GEANT hit output for the training samples, some detector effects like electronic noise, crosstalk between neighbouring cells and dead cells are turned off.
The true energy deposits considered in each cell only those in the sensitive material, which for the electromagnetic calorimeter is liquid argon, with a timing cut of 300~ns applied to the readout, and with Birks' law~\cite{birks1951scintillations,BURDIN2012111,AMORUSO2004275} applied to the energy response. The sampling fraction of the calorimeter layer is subsequently applied to scale up the energy deposits in order to account for the energy deposited in the inactive material.

Since \GEANT is used for the simulation of the ID in the default ATLAS simulation chain, an additional set of testing samples, each with 5000 photon showers per energy point, was produced. These samples are used to assess the performance of the models when incorporated into the full simulation chain to replace only the simulation of the calorimeter. These samples contain photons generated on the front surface of the calorimeter across a range to emulate the typical spread of collision vertex positions along the $z$-axis.
The photons were generated in the simulation as originating directly from the surface of the calorimeter instead of from the interaction point in order to explicitly exclude energy losses in the ID. This makes possible a modular combined simulation of the ID and the calorimeter in the ATLAS simulation framework, each with their own simulation code, e.g.\ full detailed simulation in the ID with \GEANT and fast simulation approaches in the calorimeter.
The lack of ID simulation does not impact the assessment of the performance, as in all cases the models will produce showers for the energies of photons as they reach the calorimeter and are independent of the previous stages.
All detector and digitisation effects are included, in contrast to the samples produced for training. Finally, photon reconstruction~\cite{Aad:2019tso} is applied to the simulated showers, and all shower shape variables used to assess the performance of the simulation approaches are calculated as part of the photon reconstruction algorithm.

The showers originating from photons deposit the majority of their energy in the EM calorimeter, especially those from low-energy photons. Therefore, to simplify the task, only layers of the EM calorimeter are considered. As a further simplification in the samples used to train the generative models, the calorimeter cells are approximated by cuboids and do not follow the accordion shape, and only the energy deposits within a rectangular selection are considered for each layer. The dimensions of the rectangular selection in the middle layer are chosen to be $7 \times 7$ cells in $\eta \times \phi$, containing more than 99\% of the total energy deposited by a typical electromagnetic shower in this layer. This selection is consistent with other fast calorimeter simulations used in ATLAS.
 
The dimensions of the rectangular selections in the remaining three layers are chosen such that they cover the spread in $\eta$ and $\phi$ of the cells in the middle layer.
As the cells in each layer do not have the same size in $\eta$ and $\phi$, it is not possible to select a single window with the same dimensions in $\Delta\eta$ and $\Delta\phi$ for all layers which lines up with cell edges in a 7 $\times$ 7 selection in the middle layer.
To select the cell windows in each layer, two slightly different approaches are employed. In both approaches the cells are selected with respect to the impact cell, defined as the cell closest to the extrapolated trajectory of the photon. In both approaches a fixed window size is used for each layer. The two approaches, described in the following paragraphs, differ only in how to select the cell window in the front and back layers, which has only a small impact on the overall performance. The two models aim to reproduce the energies deposited in each of the cells within the windows across all layers, after accounting for the sampling fraction in each layer.

\begin{figure}[tb]
\centering
\includegraphics[width=0.6\textwidth]{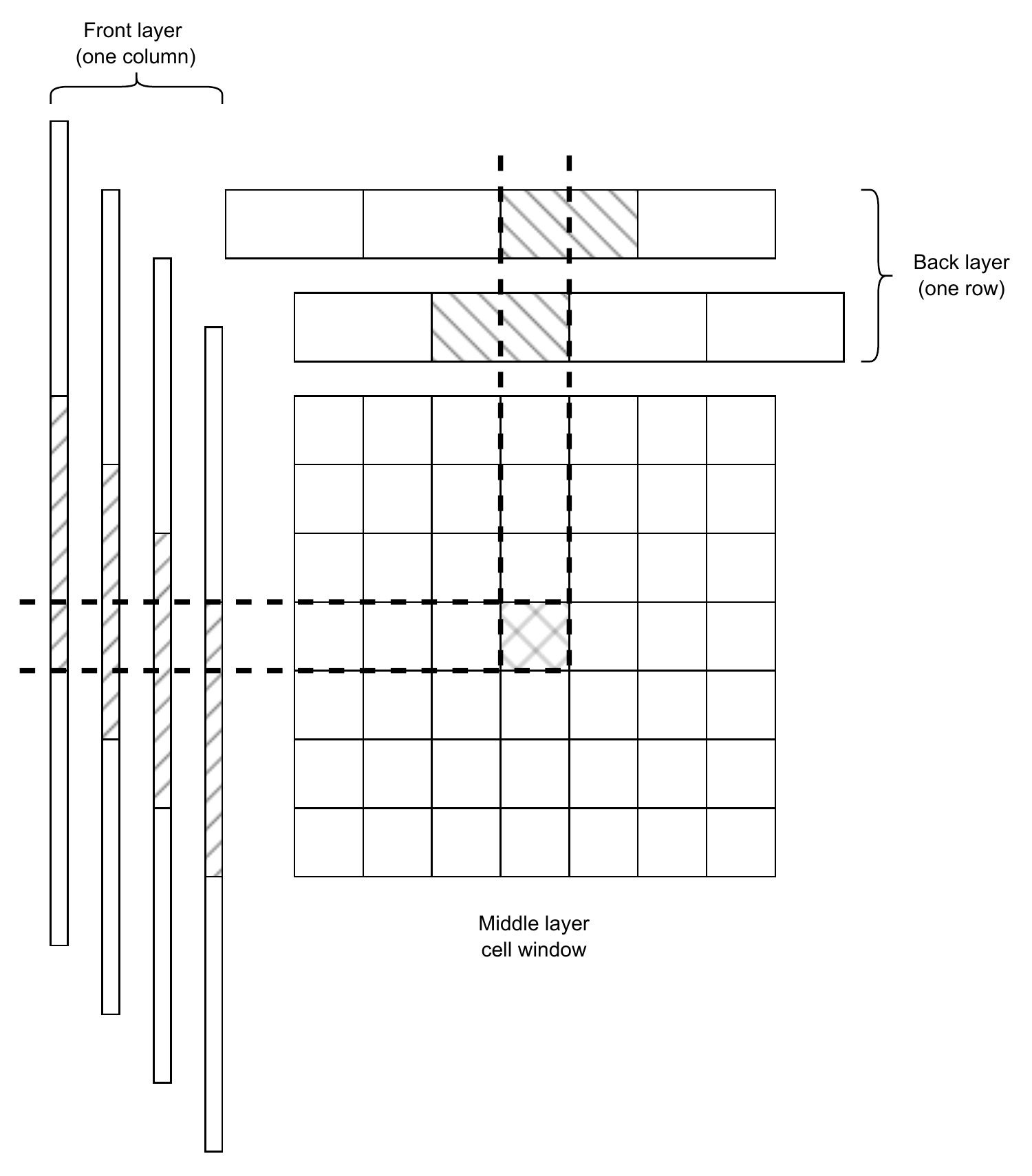}
\caption{The possible alignments of the cell windows in the front and back layers with respect to the middle layer.
A single column (row) of cells in $\phi$ ($\eta$) for the front (back) layer are shown, with the centre cells in each layer shaded, illustrating the four (two) possible alignments and their resulting overhang.}
\label{fig:atlas:align}
\end{figure}

In the first approach, used by the GAN, the cells in the presampler, front and back layers are chosen such that they cover the area of all cells in the 7 $\times$ 7 window in the middle layer; this results in cell windows which are larger in $\phi$ ($\eta$) in the front (back) layer.
The window size in number of cells for the presampler, front and back layer are $7 \times 3$, $56 \times 3$ and $4 \times 7$, respectively. In total, the energy deposits in 266 cells are considered.
Due to the different sizes of the cells in the front and back layers, and because the cell edges are designed to align with those in other layers, the cell windows in the front and back layers are not centred on the window in the middle layer but instead `overhang' in the $\phi$ or $\eta$ direction respectively. 
This can be seen in Fig.~\ref{fig:atlas:emcal} where the middle layer has $4 \times 4$ cells which corresponds to exactly   $32 \times 1$ cells in the front layer and $3 \times 4$ cells in the $\eta$ direction for the back layer. For a $7 \times 7$ window in the middle layer, there are therefore four (two) possible alignments of the cell window in the front (back) layer relative to the centre cell in the middle layer, as shown in Fig.~\ref{fig:atlas:align}.

The second approach, which is used by the VAE, makes use of the impact position in each layer to select the centre cell for the cell window in that layer. Since an odd number of cells in both $\eta$ and $\phi$ per layer are required in order to have only one centre cell per layer, this results in an additional row of cells in $\eta$ in each of the front and back layers compared to the first approach, with window sizes of $57 \times 3$ and $5 \times 7$, respectively.
This results in an additional 10 cells, and removes the need to consider the alignment of the cell windows between layers.
In total, the energy deposits in 276 cells are considered.
In both cell-window approaches, when simulating new showers the energies of cells outside of the selected cell windows are set to zero.

In this paper the detailed simulation samples are produced with \GEANT without any simplifications applied; training samples are produced with \GEANT with simplifications applied when selecting the cell windows; and generated data are showers produced by the two models, the VAE and the GAN. All three samples are simulated data.


\section{Models}
\label{sec:algorithms}

The architecture of the studied neural networks, the objective functions used in the training, the tuning of the hyperparameters and their impact on the shower simulation are discussed in this section. A general introduction to machine learning is given, for example, in Refs.~\cite{Bishop:2006:PRM:1162264,Hastie}. Two fundamentally different architectures were studied, a VAE and a GAN. It should be emphasised that in both cases the training data are given as flattened one-dimensional vector of calorimeter cell energies, along with a few additional condition parameters. When the trained generative networks are integrated into the ATLAS simulation infrastructure using the Lightweight Trained Neural Network package~\cite{daniel_hay_guest_2019_3366463}, the flattening of the cells is performed in the same way as for training of the networks. The correlations between deposits in neighbouring cells, within the same layer or in depth, are learned directly from the training data without being reflected in the architecture.
Although convolutional layers were studied to exploit information about the local structure, they did not bring any improvement over dense layers.
Furthermore, determining which trained model produces the most realistic showers across all ranges of energies and distributions is a fundamental challenge in the optimisation of generative models for use in calorimeter simulation and requires visual inspection of a range of key distributions. It is not trivial to quantify the performance trade off in different observables and across the different energy values.

\subsection{Variational autoencoders}
\label{sec:vae}

\afterpage{
 
\begin{figure}[tb]
\centering
\includegraphics[width=0.9\textwidth]{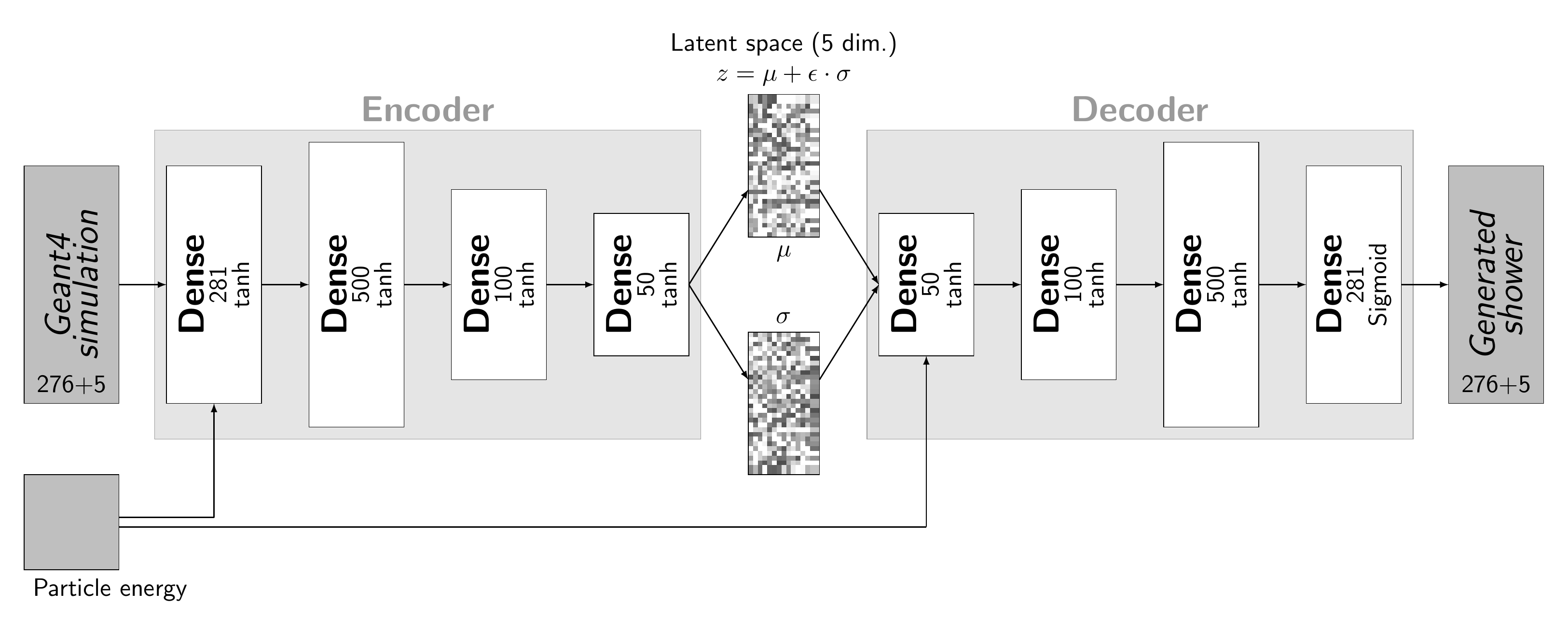}
\caption{Schematic representation of the architecture of the VAE presented in this paper. It is composed of two stacked neural networks, each consisting of four hidden layers with 281, 500, 100, and 50 nodes, arranged so that the networks acts as an encoder and decoder (with the layer structure reversed).
The encoder compresses the input, comprising 276 cell energies and the five energy fractions, into a five-dimensional latent space, with encoded events constrained to follow a multidimensional Gaussian distribution. The latent variable $z$, whose prior distribution is chosen to be a multivariate normal distribution, is formed by combining an encoded mean $\mu$ and standard deviation $\sigma$ for each encoded event with a factor $\epsilon$ sampled from the unit normal distribution.
The decoder takes the latent space as an input and tries to reconstruct the original inputs.
The model uses a tanh activation function for the hidden layers and a sigmoid function for the output layer. The implemented algorithm is conditioned on the energy of the incident particle to generate showers corresponding to a specific energy. The condition is passed to both the encoder and decoder networks alongside their inputs.}
\label{fig:vaeArch}
\end{figure}
}

VAEs~\cite{2013arXiv1312.6114K,2014arXiv1401.4082J} are a class of unsupervised learning models combining deep learning with variational inference. They are composed of two stacked neural networks, acting as an encoder and a decoder respectively, and can be used as generative models. VAEs are latent-variable models that introduce a set of random variables that are not directly observed but are used to reveal underlying structures in data. As such, the encoder $q_\theta(z|x)$ compresses the input data $x$, in this case the energy deposits of a calorimeter shower, into a lower-dimensional latent variable $z$. The learnable parameters of the encoder are denoted by $\theta$. It should be explicitly noted that this latent representation is stochastic, i.e.\ $q_\theta(z|x)$ maps $x$ to a full distribution rather than being a function $x \mapsto z$. The decoder $p_\phi(x|z)$ learns the inverse mapping, thus reconstructing the original input from this latent representation. The learnable parameters of the decoder are denoted by $\phi$. Once the model is trained, the decoder can be used independently of the encoder to generate data $\tilde x$; new calorimeter showers are synthesised by sampling $z$ according to the prior probability density function $p(z)$, which they are assumed to follow, and then sampling $\tilde{x}$ from $p_\phi(x|z)$. The prior distribution of the latent variable $p(z)$ is chosen to be a multivariate normal distribution with a covariance equal to the identity matrix.
 
The model explored in this paper is composed of two stacked neural networks each comprising four hidden layers.
The input to the encoder is the energy of each of the cells in the window selected for each layer, as introduced in Section~\ref{sec:samples}. As a result, the output of the decoder consists of the energies of 276 cells for each shower. The architecture is illustrated in Figure~\ref{fig:vaeArch}. The implemented model is conditioned~\cite{NIPS2015_5775,2016arXiv160607873W} on the energy of the incident particle to generate showers corresponding to a specific energy.
The encoder and decoder networks are connected and trained together, but only the decoder is used to generate new showers. The training maximises the variational lower bound on the marginal log-likelihood for the data, approximated with the reconstruction loss, and the Kullback--Leibler divergence~\cite{kullback1951} ($D_\text{KL}$) of the distribution of the encoded latent space with respect to $p(z)$. The total loss is given by
 
\begin{equation}
\label{eq:lower bound}
\log p(x) \geqslant \underbrace{E_{z \sim q_\theta (z|x)} \left[\log p_\phi(x|z)\right]}_{\text{Variational lower bound}} - \underbrace{D_\text{KL}(q_\theta(z|x)||p(z))}_{\text{Kullback--Leibler divergence}}.
\end{equation}
 
The variational lower bound $E_{z \sim q_\theta(z|x)} [\log p_\phi(x|z)]$ measures the expectation value of the log probability of inputs reconstructed by the decoder $p_\phi(x|z)$, with $z$ sampled from the encoded latent space distribution $q_\theta(z|x)$ for the given input data $x$.
This penalises the VAE for generating output distributions different from the input training data $x$ and thus characterises the capacity of the decoder to recover data from the latent representation.
Assuming that the reconstructed input data $\tilde{x}$ differ from the true input samples $x$ in a normally distributed way, the mean squared error $||\tilde{x}-x||^2_2$ is used as the reconstruction loss.
The second term, the negative Kullback--Leibler divergence, measures the divergence of the encoded latent space distribution $q_\theta(z|x)$ from the chosen prior probability density function $p(z)$. It is defined as
 
\begin{equation*}
-D_\text{KL}(q_\theta(z|x)||p(z)) = \frac{1}{2}\sum_{i=1}^{n} \left(1+\log\left(\sigma_i^2(x)\right) - \mu_i^2(x) - \sigma_i^2(x)\right),
\end{equation*}
 
where $\mu_i$ and $\sigma_i$ enter $q_\theta(z|x)$ as the normal distribution $\mathcal{N}(z|{\mu},{\sigma})$ and $n$ is the number of latent-space dimensions.
 
For fast calorimeter simulation, the objective function is scaled by a weight for each term, controlling the relative importance of the contributions during the optimisation of the model, as proposed by the $\beta$-VAE model~\cite{Higgins2017betaVAELB}.
The $D_\text{KL}$ term acts as a regularisation, and changing its weight directly affects the generated distributions. In contrast to the values proposed in Ref.~\cite{Higgins2017betaVAELB}, $D_\text{KL}$ is multiplied by a weight $w_\text{KL}$ in the interval $(0,1]$ to prioritise reconstruction performance over accurately matching the latent space distribution to the chosen prior $p(z)$. At the maximum value for the weight, the loss becomes equivalent to the true variational lower bound, as seen in~Eq.~(\ref{eq:lower bound}).
 
The VAE in this paper is implemented in Keras~2.3.1~\cite{chollet2015keras}, using TensorFlow~1.15.0~\cite{tensorflow2015-whitepaper} as the backend with mini-batch gradient descent via the Adam optimiser~\cite{kingma2014adam}. It is trained on an NVIDIA\textsuperscript{\textregistered} Titan X Pascal graphics card with a processing power of $3584$ cores, each clocked at $\SI{1417}{\mega\hertz}$, for a total of 1000 epochs, which results in a maximum training time of around 4~hours for the networks. A hyperparameter scan is performed to find the best network architecture and hyperparameter settings.
The final network is chosen by minimising the total loss after training for a set number of epochs and inspecting the physics observables of new showers generated by sampling from the latent space according to the prior distribution $p(z)$.

\subsubsection*{Data reparametrisation}
 
When training the VAE on the absolute energies in all of the cells, even with the condition of the photon energy, it was found that the network could accurately model either the total energy of the shower or the distribution of cell energies within each layer, but not both with the same model. In order to address this issue the representation of the inputs entering the network was reparametrised to simplify the learning process.
Per shower, the energies of the cells within each layer are normalised to the total energy recorded in that layer. For a cell $j$ in layer $i$, its new value is given by the ratio
 
\begin{equation*}
R_{\text{Cell}(i,j)} = \frac{E_{\text{Cell}(i,j)} }{ E_{\text{Layer}(i)} } =
\frac{E_{\text{Cell}(i,j)} }{\sum_{j=1}^{N_i} E_{\text{Cell}(i,j)} },
\label{equation:ratio}
\end{equation*}
 
where $N_i$ is the number of cells in the selected window for each of the four layers.

In order to preserve the correlations of the energies across layers, and to retrieve the absolute energy of each cell from the output of the decoder, the VAE is tasked with learning the energy deposited in each layer and the total energy of the shower. This is encoded as five additional features provided as input to the model.
For each event the total energy of the shower divided by the MC-truth energy of the photon, 
and the energy in each layer divided by the total energy of the shower, 
are provided as additional inputs which are to be reconstructed.
Consequently, the input and output of the model each consist of 281 features, alongside the MC-truth energy of the input particle used to condition the model.
 
Furthermore, to improve the performance in reconstructing the shape of the showers within each layer a modification is made to the reconstruction loss term in the loss function. To capture the importance of accurately reconstructing the energies of cells closer to the shower centre while accommodating the variability of cell energies at the edges of the shower, a weight term for each of the $m$ input features is incorporated into the reconstruction loss:
\begin{equation*}
||\tilde{x} - x||^2_{w} = \frac{1}{m}\sum_{i=1}^{m=281} w_i (x_i - \tilde{x}_i)^2.
\end{equation*}
 
The weights $w_{i}$ are derived from the width of the input distribution per feature $i$ over all training events. To assign a higher importance to input features with a narrower spread of values, the weights $w_{i}$ are calculated per feature from the inverse of the standard deviation calculated over all showers. These weights are derived in the same way for all 281 features. The full loss function used in the training is therefore
 
\begin{equation*}
L_\text{VAE}(x, \tilde{x}) =  ||\tilde{x} - x||^2_{w} + w_\text{KL} \frac{1}{2} \left(1+\log\left(\sigma^2\left(x\right)\right) - \mu^2(x) - \sigma^2(x)\right).
\end{equation*}


\subsection{Generative adversarial networks}
\label{sec:gan}

\afterpage{
 
\begin{figure}
\centering
\includegraphics{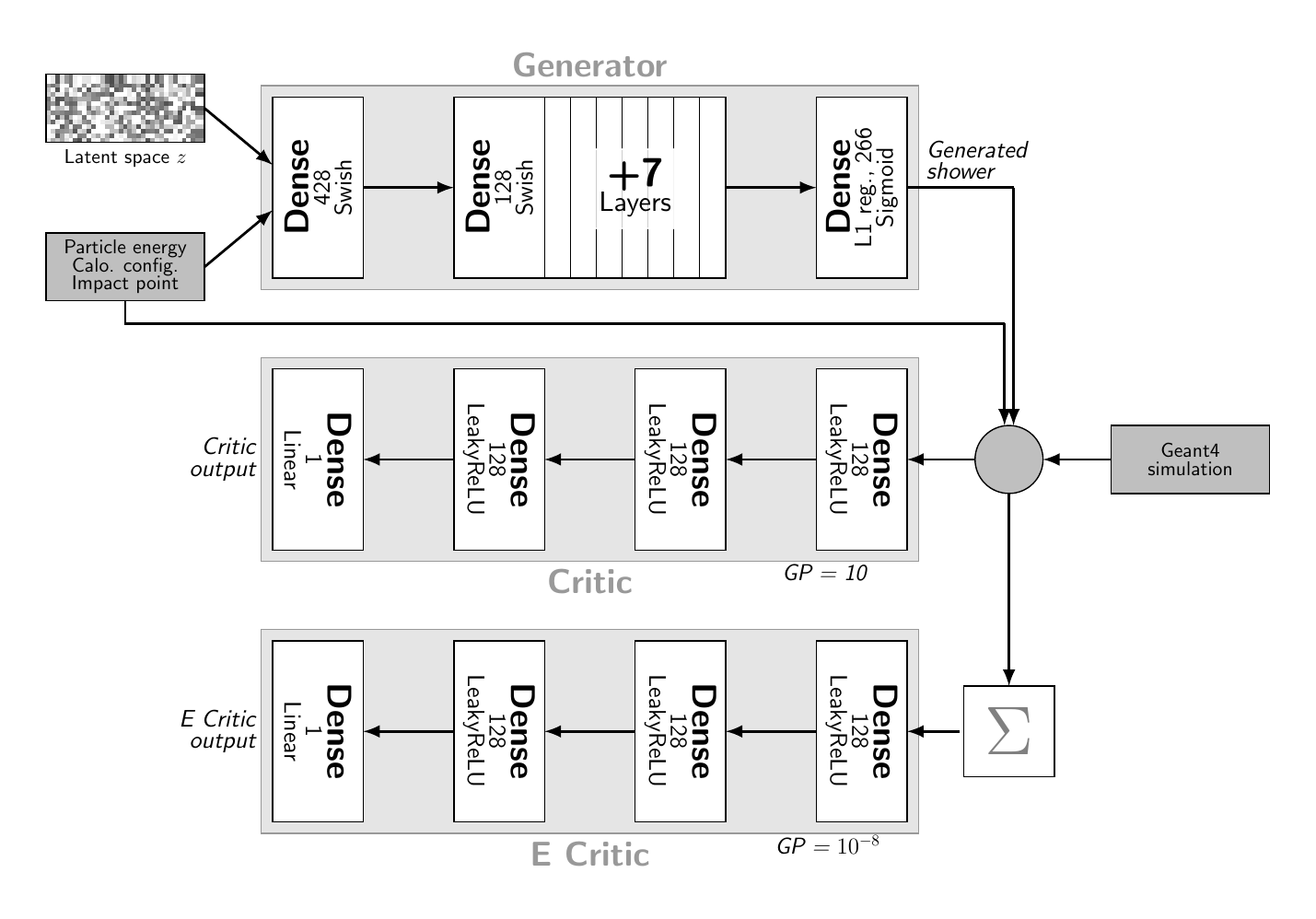}
\caption{Schematic representation of the architecture of the GAN used in this paper. It is composed of three neural networks. The generator takes as input 300 random numbers drawn from the multivariate normal distribution which forms the latent space and is conditioned on the energy of the input particle, extrapolated position within the impact cell, and information about the alignment of the calorimeter cell windows. The critic compares synthesised showers from the generator with showers generated by \GEANT while the energy critic compares only the total energy of the showers. The model uses trainable Swish activation functions~\cite{ramach2017searching} for the hidden layers of the generator and leaky rectified linear units (LeakyReLUs)~\cite{Maas13rectifiernonlinearities} as activation functions for the hidden layers of the critics. The output layers of the generator and two critics have sigmoid and linear activations, respectively.
At the grey circle, the training procedure switches between batches drawn from the generator, the simulation and the mixed sample used to evaluate the WGAN-GP losses.}
\label{fig:GAN_Architecture}
\end{figure}

}

GANs are unsupervised learning algorithms implemented as a deep generative neural network taking feedback from an additional discriminative neural network. Originally developed as a minimax game~\cite{Neumann2} for generating realistic-looking natural images~\cite{2014arXiv1406.2661G}, GANs have a wide range of applications, including calorimeter simulation~\cite{deOliveira:2017pjk,Paganini:2017hrr,Paganini:2017dwg}. The gradient-penalty-based Wasserstein GAN (WGAN-GP) is a slightly modified algorithm that is widely used because it is more stable during training~\cite{2017arXiv170107875A,DBLP:journals/corr/GulrajaniAADC17}. This is the flavour of GAN used in this paper and from now on is referred to as the GAN.
The GAN is composed of three neural networks: a generator, a standard critic and an additional critic for the total energy. The model is conditioned on the energy of the incident particle, its position within the impact cell and the alignments of the cell windows across the layers in $\eta$ and $\phi$, discussed in Section~\ref{sec:samples}. The detailed architecture of the three networks is illustrated in Figure~\ref{fig:GAN_Architecture}.
 
A series of preprocessing steps are applied to each of the neural network inputs to make the learning easier. Firstly, the cell energies for each shower are normalised by the true energy of the incident particle. Secondly, the particle position information is computed with respect to the centre of the impact cell, and then standard-normalised, whereby the mean and standard deviation of the distribution across all training samples are adjusted to be equal to zero and one, respectively. Finally, a log-transform followed by standard-normalisation is applied to the true energy of the particle.
The cell window alignment is passed to the network using two one-hot encoding vectors, with the encoded bits indicating the alignments of cell edges in the front and back layers with respect to the cell window in the middle layer.

The loss function for each of the critics is
 
\begin{equation*}
L_\text{Critic} = E_{\tilde{x} \sim p_\text{gen}} \left[D({\tilde{x}})\right] - E_{x \sim p_\text{\GEANT}} \left[D({x})\right] + \beta E_{\hat{x} \sim p_{\hat{x}}} \left[(||\nabla_{\hat{x}}D({\hat{x}})||_{2} - 1)^{2}\right].
\label{eqn:WGAN-GP}
\end{equation*}

Here, $p_\text{\GEANT}$ is the \GEANT probability distribution, $p_\text{gen}$ is the probability distribution of outputs from the generator network (which takes a standard-normal-distributed latent space as input), $D({x})$ is the output of the critic network for real showers from \GEANT and $D(\tilde{x})$ is the output for showers generated by the generator. During training, the critic $D$ tries to maximise the distance between the predictions for the showers drawn from \GEANT and those from the generator.
The term $E_{\tilde{x} \sim p_\text{gen}} [D({\tilde{x}})]$ represents the critic's ability to correctly identify synthesised showers, while the term $E_{x \sim p_\text{\GEANT}} [D({x})]$ represents the ability of the critic to correctly identify showers from \GEANT.
 
The last term in the loss function, $\beta E_{\hat{x} \sim p_{\hat{x}}} [(||\nabla_{\hat{x}}D({\hat{x}})||_{2} - 1)^{2}]$, is the two-sided gradient penalty (which leads to the `GP' in WGAN-GP). Here $D(\hat{x})$ is the output of the critic for a mixture of real and generated showers, where $\hat{x}$ can be interpreted as a random point along the straight line connecting a point from the real distribution $p_\text{\GEANT}$ and generator distribution $p_\text{gen}$. The value of the hyperparameter $\beta$ indicates the relative importance of the final term in the loss function and is known as the `gradient penalty weight' (GPW).
The gradient penalty softly enforces the Lipschitz constraint on the critic, which allows the critic to estimate the Wasserstein-1 distance between real and generated images.
The algorithm is further extended to estimate conditional probabilities, leaving the evaluation of the gradient penalty over the showers unchanged.
 
For the energy critic, the input image of the shower is replaced by the total energy of the three-dimensional image. The loss function remains the same, although ${x}$, $\tilde{x}$ and ${\hat{x}}$ are now taken as the sums over all respective values.

A regulariser on the sum of magnitudes of the outputs is applied to the final layer of the generator to encourage the generation of sparse energy deposits.
The GAN is trained for 25\,000 epochs, but the best epoch is chosen manually due to epoch-to-epoch fluctuations in performance. After a primary, loose shortlisting based on the RMS difference between the covariance matrices of the GAN and \GEANT distributions, the best model is chosen through visual inspection of the energy resolution and the ability of the GAN to condition on the calorimeter geometry, manifested by its modelling of the mean $\eta$ in the back layer and width in $\phi$ for the front layer. A similar strategy is also employed in hyperparameter searches. Although it is susceptible to confirmation biases, the strategy was found to be better than using standard statistical distance measures such as Kolmogorov--Smirnov or Anderson--Darling tests, due to the difficulty in quantitatively assessing performance across multiple distributions and a wide range of energies.
 
The training for 25\,000 epochs  was completed in 79~hours using approximately $50\%$ of the available training data on an NVIDIA\textsuperscript{\textregistered}~Kepler\texttrademark~GK210 GPU with a processing power of $2496$ cores, each clocked at $\SI{562}{\mega\hertz}$. The card has a video RAM size of $\SI{12}{\giga\byte}$ with a clock speed of $\SI{5}{\giga\hertz}$. The training data are read from memory. Although trained on different hardware to the VAE, this has a negligible impact on comparisons of model performance.
The model is implemented in Keras~2.0.8~\cite{chollet2015keras}, using TensorFlow~1.3.0~\cite{tensorflow2015-whitepaper} as the backend, and the networks are trained with mini-batch gradient descent using the RMSProp optimiser~\cite{rmsprop}.
 
\subsubsection{Energy critic}
When training a WGAN-GP with a single critic, it was found that the model failed to simultaneously reproduce the energy resolution of the detector and the lateral shower shapes. This is because in gradient-penalty-based Wasserstein GANs, the gradient penalty prevents the critic from making sharp decisions based on every cell of the input. For example, a GPW of $10^{-13}$ allowed better modelling of the energy resolution of the detector at the cost of poorer lateral shape distributions and training stability. The final choice was optimised for the overall network performance, not just the total energy.
 
Disentangling the two tasks into separate critic networks allows the injection of domain priorities into the training algorithm. In addition to the original critic which focuses on the shower shapes, an additional energy critic with a GPW of $10^{-8}$ is trained to be blind to all aspects of the calorimeter image except the total energy. The generator network is trained against the critic and energy critic simultaneously with a loss ratio of $1:10^{-6}$, forcing it to learn the shower shapes and the total energy distributions simultaneously. The combined loss for the generator is
\begin{equation*}
L_\text{Generator} = E_{\tilde{x} \sim p_\text{gen}} \left[D({\tilde{x}})\right] + 10^{-6}\cdot E_{\tilde{x} \sim p_\text{gen}} \left[D_{E}\left({\underset{\text{cells}}{\Sigma}\tilde{x}}\right)\right],
\label{eqn:GenLoss}
\end{equation*}
where $D({\tilde{x}})$ and $D_{E}({\underset{\text{cells}}{\Sigma}\tilde{x}})$ are the outputs of the first critic and the energy critic for a generated image $\tilde{x}$.
This concept could be extended further with the insertion of additional physics observables as inputs to the second critic, or additional critics, if necessary, in order to explicitly indicate the importance of modelling particular physics-motivated features to the generator within the training algorithm.


\subsection{Comparison with prior work}
\label{sec:priorwork}

 
In contrast to other approaches in previous work, the models in this paper are trained and evaluated using samples produced with the detailed simulation model used by the ATLAS Collaboration.
 
This is one of the first applications of generative models using the real ATLAS detector geometry instead of a `toy' detector model,
and it is the first to use the readout of the energy deposits recorded by the cells of the ATLAS calorimeter.
Although the study is performed in a small region in $\eta$, the full $\phi$ range is simulated.
This introduces various complications, such as the changing geometry of the calorimeter.
However, it also presents many benefits, such as the ability to perform a direct comparison with the full simulation of interactions in the calorimeter from \GEANT. These comparisons are performed in terms of memory footprint and speed.
Future comparisons with current state-of-the-art fast simulation approaches are also simplified.
Comparisons with reconstructed physics objects can also be performed after implementation in the full simulation chain used by the ATLAS experiment. The effect of calibration, which is tuned using only \GEANT simulations, can also be studied.
Overall, this makes it possible to understand the differences at the level of real physics analyses and not just in the study of a shower simulation.
 
As mentioned previously, the geometry of the ATLAS calorimeter itself presents its own challenges.
The ATLAS calorimeter is constructed from cells arranged in several layers radially around the interaction point, comprising cells of varying size and depth. This creates a varying arrangement of calorimeter cells depending on where in the calorimeter the cell selection is performed. The detector geometries presented in Refs.~\cite{deOliveira:2017pjk,Paganini:2017hrr,Paganini:2017dwg,Buhmann:2020pmy,Erdmann:2018jxd}, in contrast, model a cuboidal block of detector volume which does not require any further window selection. Furthermore, in the models presented in this paper, only the four EM calorimeter layers are considered and information about the depth of a shower is not represented by coordinates in the $r/z$ direction, but rather by a label for each layer.
This has the effect of simplifying the geometry in the radial direction, but it also adds a complication when modelling the propagation of the shower through the detector.
Since the cells in the different layers have different sizes, this adds complexity especially when selecting and ordering the relevant cells.
To address this challenge, the input is simplified by representing the cells as a single concatenated vector with no spatial information or relationship preserved, instead of using three-dimensional convolutional layers in the networks. Any spatial relationship between cells in generated showers is therefore learned by the two models.
 
Similarly to previous approaches which have focused on a block of calorimeter layers, the models in this paper are trained to model showers only in a narrow $\eta$ window in the ATLAS detector. This removes the challenge of modelling the changing combination of layers and varying granularity of cells across the different layers of the ATLAS detector.
 
In comparison to the models presented in Refs.~\cite{Erdmann:2018jxd,Buhmann:2020pmy}, which synthesise particle showers in high-granularity calorimeters, the architectures of the two models presented in this paper are much less complex. Instead, the challenges faced in accurately simulating particle showers, such that they are able to reproduce global distributions constructed from the generated outputs, are addressed by using innovative adaptations. The most notable of these are the additional critic in the WGAN to improve the modelling of the overall calorimeter energy response, and the reparametrisation of the inputs in the VAE whereby the network is trained on relative energies instead of absolute values.
Problems with WGAN-GPs, similar to those addressed by the network presented in this paper, have been encountered in other applications in high-energy physics, and were addressed using a Maximum Mean Discrepancy loss term \cite{Butter:2019cae} and postprocessing networks \cite{Buhmann:2020pmy}.

Furthermore, in contrast to other approaches, only fully connected layers are employed in the networks, and the calorimeter design and structure is
not reflected in the design of the models.



\section{Results}
\label{sec:results}

 
To assess the quality of the generative models described in this paper, generated electromagnetic showers for single photons with various energies are compared with the training samples and the detailed simulation.
Comparisons with the training samples are used to assess whether the energy deposits in the cells are well modelled, before the addition of digitisation noise which would hide the true modelling. The detailed simulation is used to assess the real performance of these models in the setting where they would be employed.
Due to the stochastic nature of shower development in the calorimeter, it is not possible to determine the quality of the shower generation by looking at individual showers.
Instead, relevant distributions used during event reconstruction and particle identification, such as the total energy, the energy deposited in each calorimeter layer and the relative distribution of energies in the calorimeter cells, are compared.
This tests the quality of the shower generation and also whether probabilities and correlations in the shower development are reproduced correctly.
 
All observables studied test the ability of the models not only to generate realistic showers, but also to capture complex underlying correlations between the deposits in different cells in all layers.
This presents a challenge to all generative models, which are normally designed and employed to generate realistic and accurate-looking samples in isolation, rather than capturing and reproducing complex underlying correlations and distributions from the data.
 
Furthermore, all distributions used to assess the performance are unseen during the training, with no optimisation of the models via comparisons with the detailed simulation or for the observables describing the shower substructure. The models were optimised to maximise their ability to reconstruct the total shower energy across all energy points equally well, with performance assessed on only a few simple shower-shape observables constructed from the training samples in addition.
Comparisons with the detailed simulation also present additional challenges, as several effects from the detailed simulation, such as underlying correlations arising from the $z$-spread of the photons (emulating the range of collision vertex positions), are not shown to the models during training and thus they will not learn this behaviour.

Three energy points from the nine used in training the models, 4~\GeV, 65~\GeV and 262~\GeV, are compared. The 262~\GeV\ and 65~\GeV\ energy points are the highest- and third-highest energy points used in training respectively. High-energy showers are known to be the hardest to model due to the higher degree of complexity and substructure which is present in showers from more energetic incident particles.
Showers generated by the two models are also compared for two energy points not seen during training, 25~\GeV and 524~\GeV, showing the ability of the models to interpolate and extrapolate to other values.
Due to the difficulty of the task and the various challenges presented in defining single metrics, as discussed in Section 6, the relative performance of these models is therefore assessed qualitatively by observing the modelling across multiple distributions. 
 
\subsection{Shower generation}
 
\begin{figure}[tb!p]
\centering
\begin{subfigure}[b]{0.49\textwidth}
\centering
\includegraphics[width=1.0\textwidth]{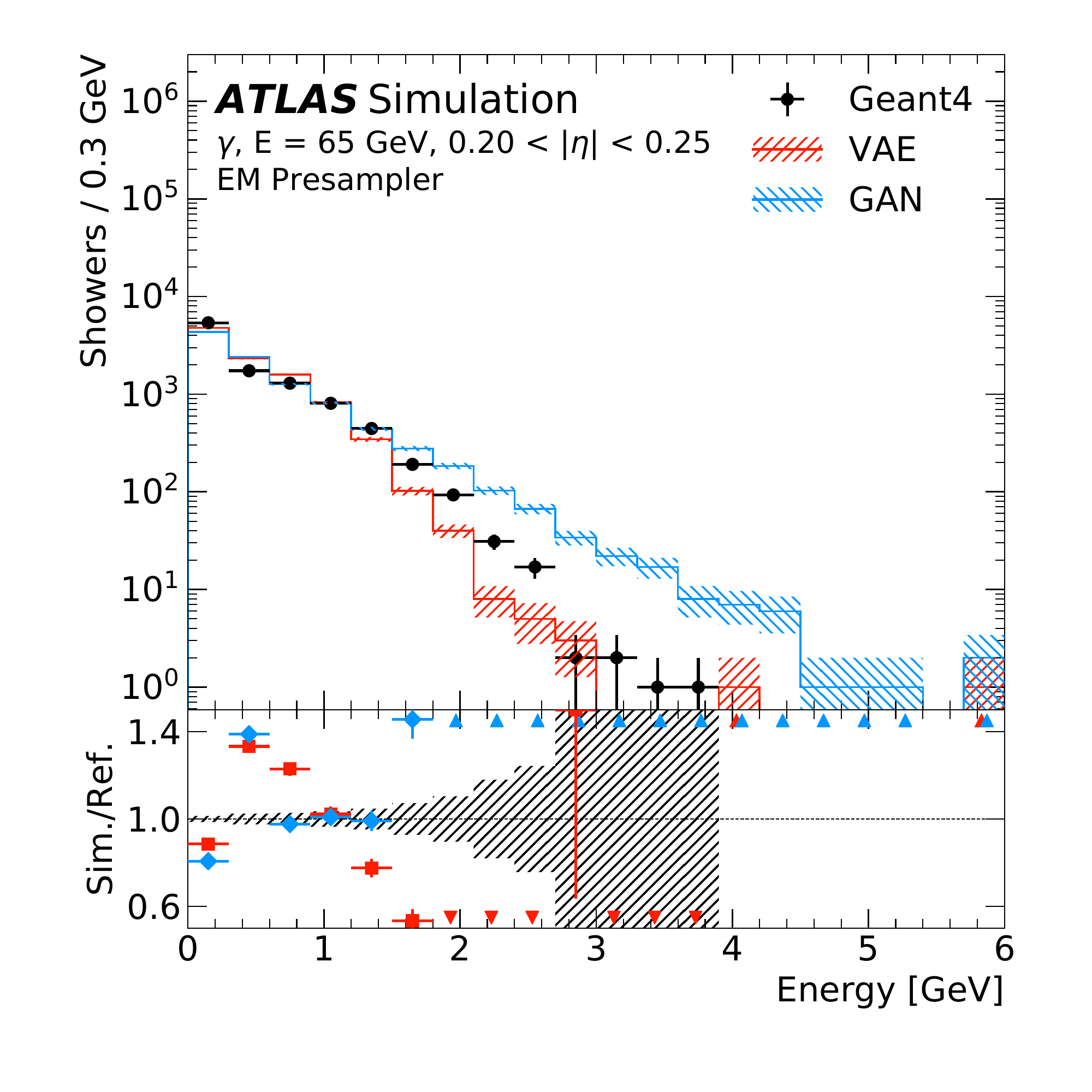}
\caption{}
\label{fig:results:elayers:presampler}
\end{subfigure}
\begin{subfigure}[b]{0.49\textwidth}
\centering
\includegraphics[width=1.0\textwidth]{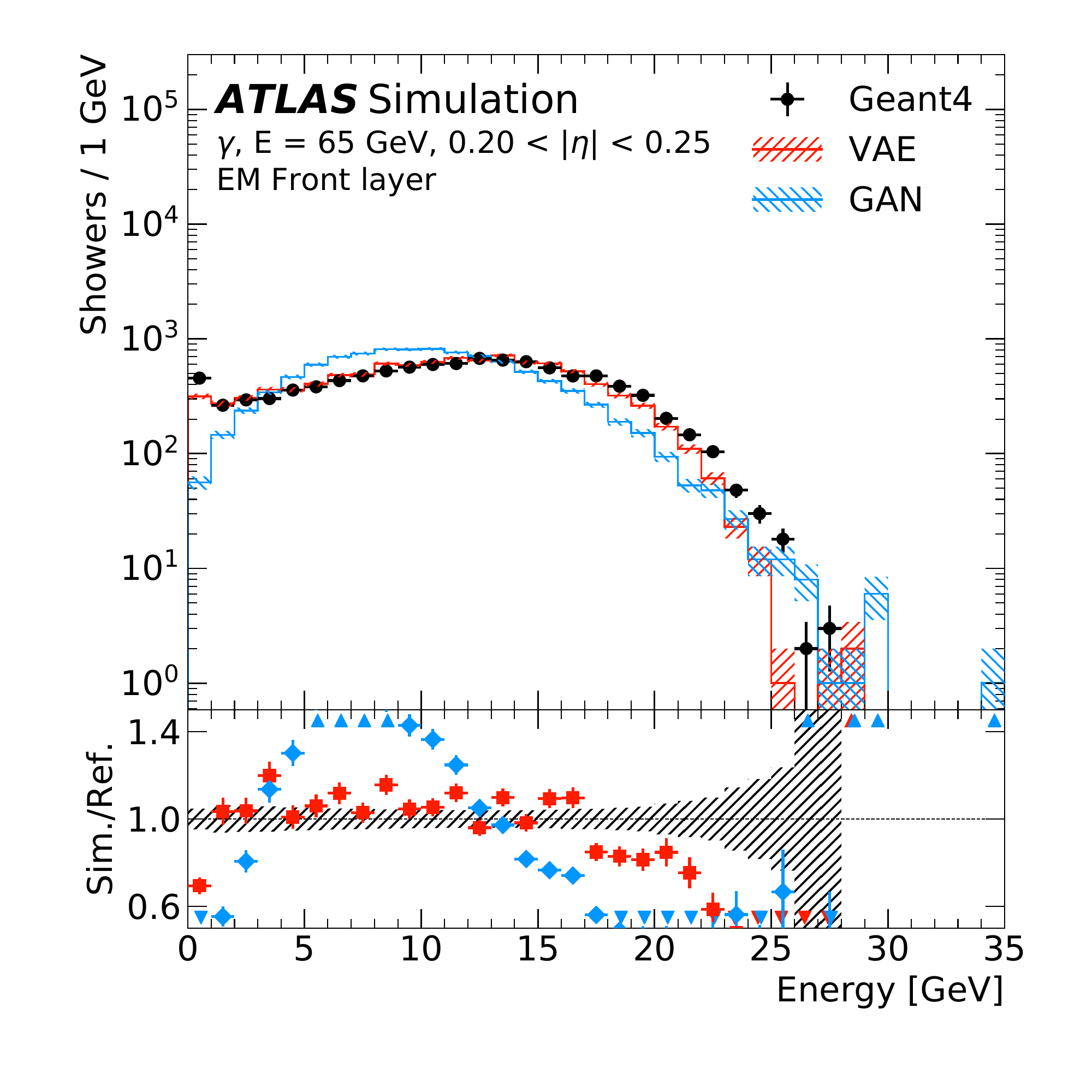}
\caption{}
\label{fig:results:elayers:front}
\end{subfigure}
\begin{subfigure}[b]{0.49\textwidth}
\centering
\includegraphics[width=1.0\textwidth]{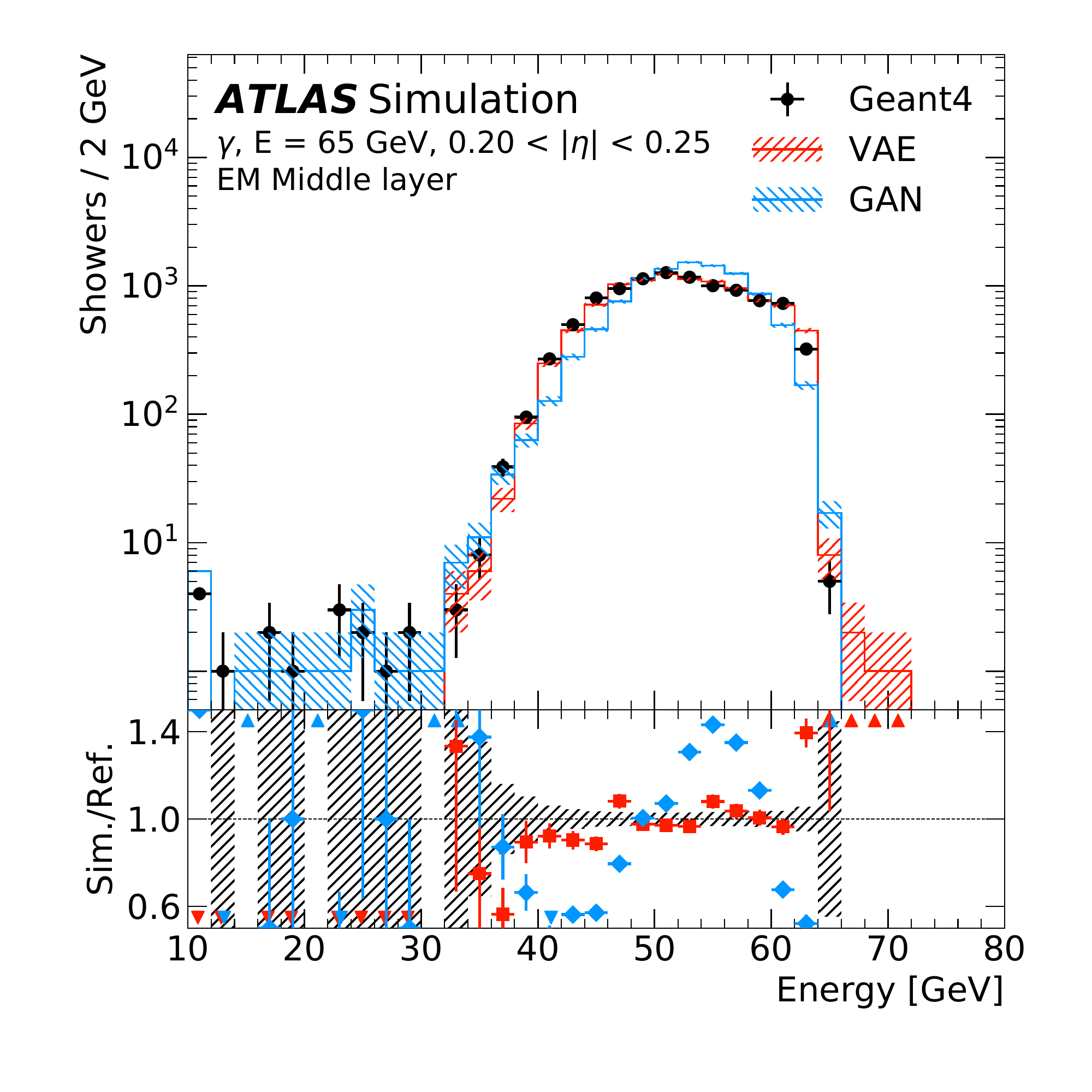}
\caption{}
\label{fig:results:elayers:middle}
\end{subfigure}
\begin{subfigure}[b]{0.49\textwidth}
\centering
\includegraphics[width=1.0\textwidth]{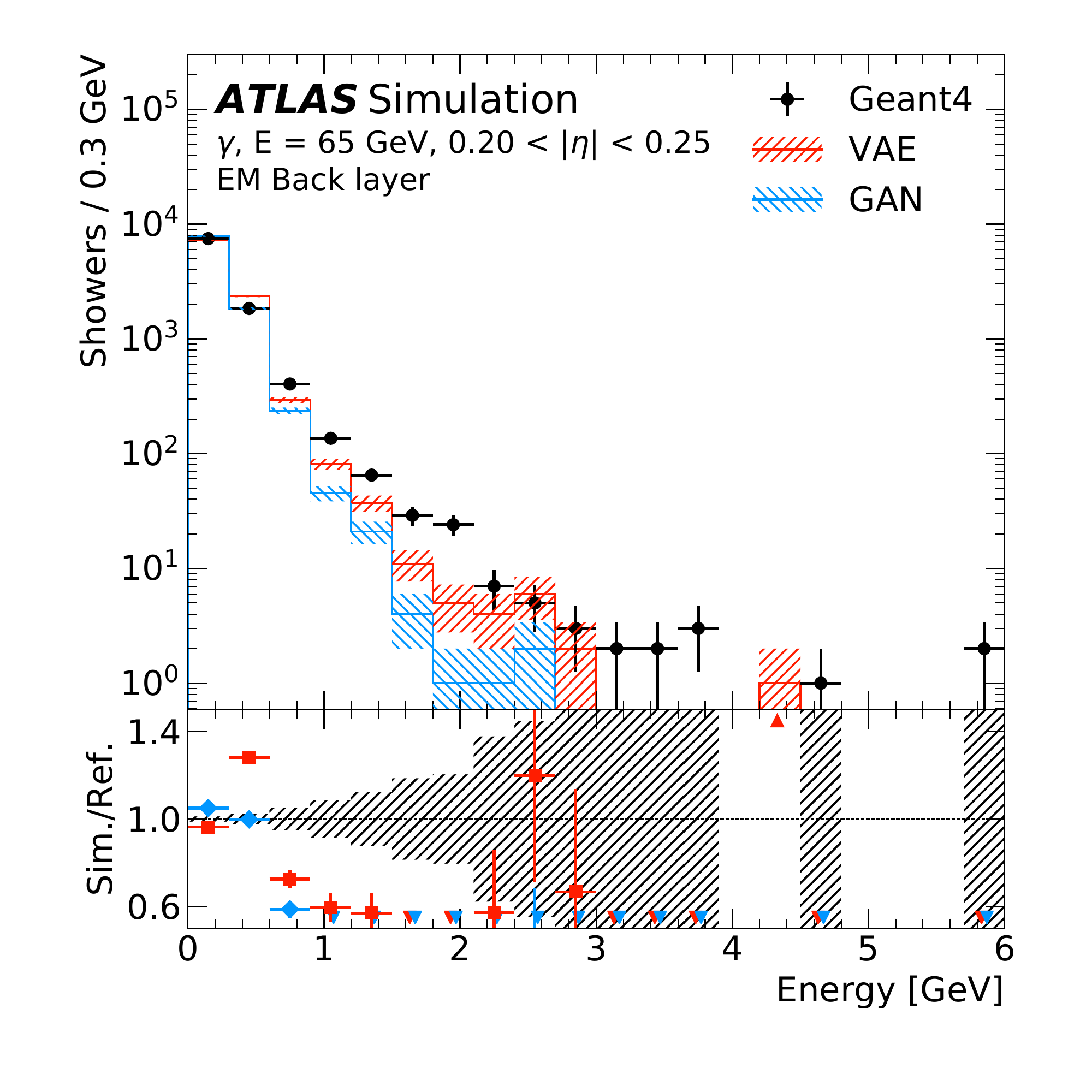}
\caption{}
\label{fig:results:elayers:back}
\end{subfigure}
\caption{Energy deposited in the individual calorimeter layers,  \subref{fig:results:elayers:presampler} the presampler, \subref{fig:results:elayers:front} the front layer, \subref{fig:results:elayers:middle} the middle layer, and \subref{fig:results:elayers:back} the back layer, for photons with an energy of $\SI{65}{\giga\electronvolt}$. The energy deposits from the \GEANT training data (black markers) are shown as reference points and compared with those from a VAE (solid red line) and a GAN (solid blue line). The error bars and the hatched bands indicate the statistical uncertainty of the training data and the synthesised samples, respectively. The underflow and overflow are included in the first and last bin of each distribution, respectively.
Arrows at the top and bottom of the ratio indicate when the values for either model in a bin fall outside of the displayed axis range.
All showers are simulated consistent with the training samples, considering only the energy deposits in the cells of the four electromagnetic calorimeter layers without digitisation and crosstalk noise.}
\label{fig:results:elayers}
\end{figure}

The models presented in this paper are designed to produce the same calorimeter response to a specific photon energy as the full detector simulation.
The energy deposited in the individual calorimeter layers is shown in Figure~\ref{fig:results:elayers} for photons with an energy of $\SI{65}{\giga\electronvolt}$.
Both the VAE and GAN reproduce the range of energy values in the training data. They also produce distributions similar to those in the training data, with the VAE reproducing the distribution of the deposits in individual layers more accurately in the front and middle layer.
Poorer agreement is observed in the presampler and in the back layer, where both models simulate showers with lower energies than in training data.
It should be noted that in comparison with the GAN, the VAE is additionally tasked with reconstructing the fraction of the total shower energy that is recorded in each layer, which has a large impact on the ability to reconstruct these distributions. 
 
\begin{figure}[tb!p]
\centering
\begin{subfigure}[b]{0.49\textwidth}
\centering
\includegraphics[width=1.0\textwidth]{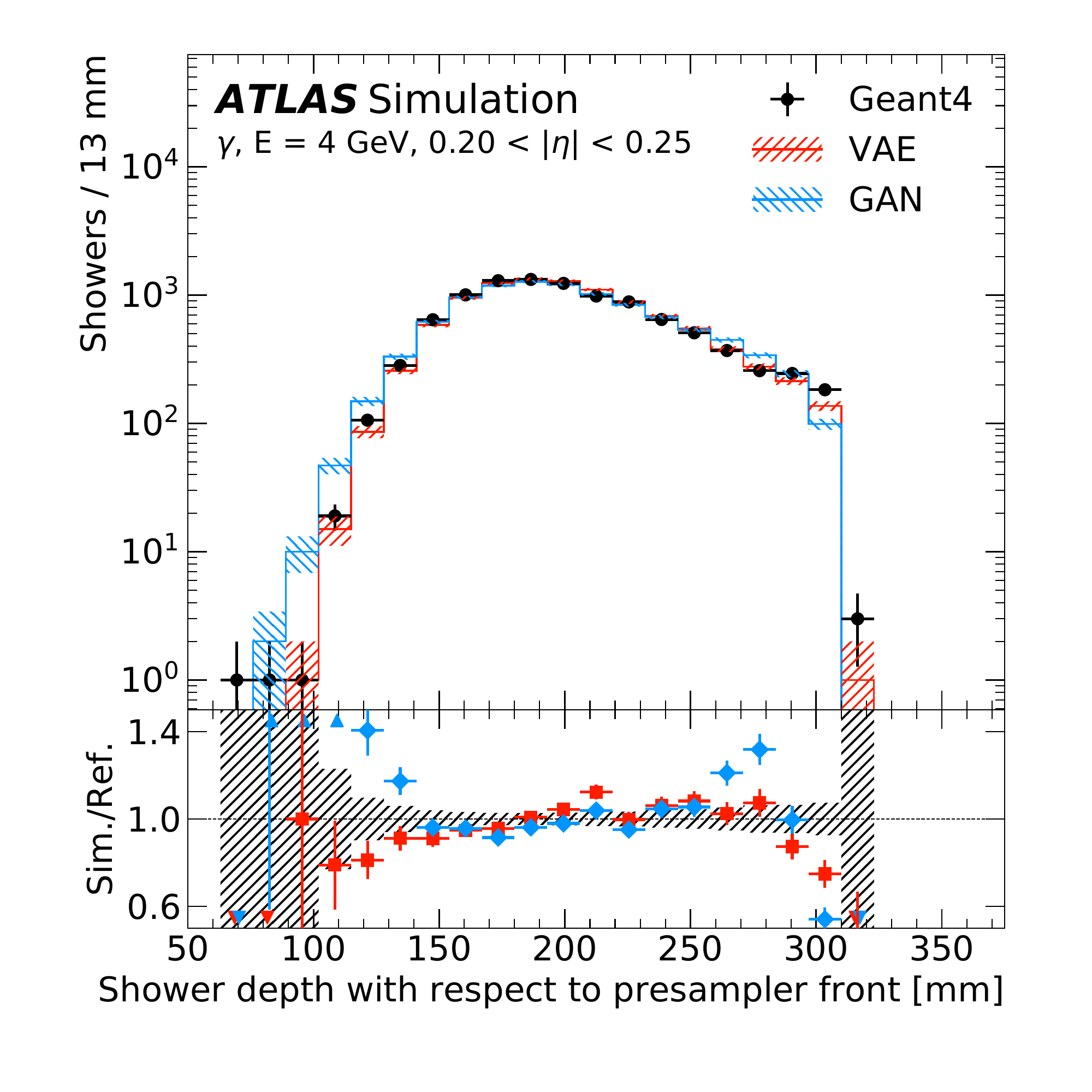}
\caption{}
\label{fig:results:depth:4}
\end{subfigure}
\begin{subfigure}[b]{0.49\textwidth}
\centering
\includegraphics[width=1.0\textwidth]{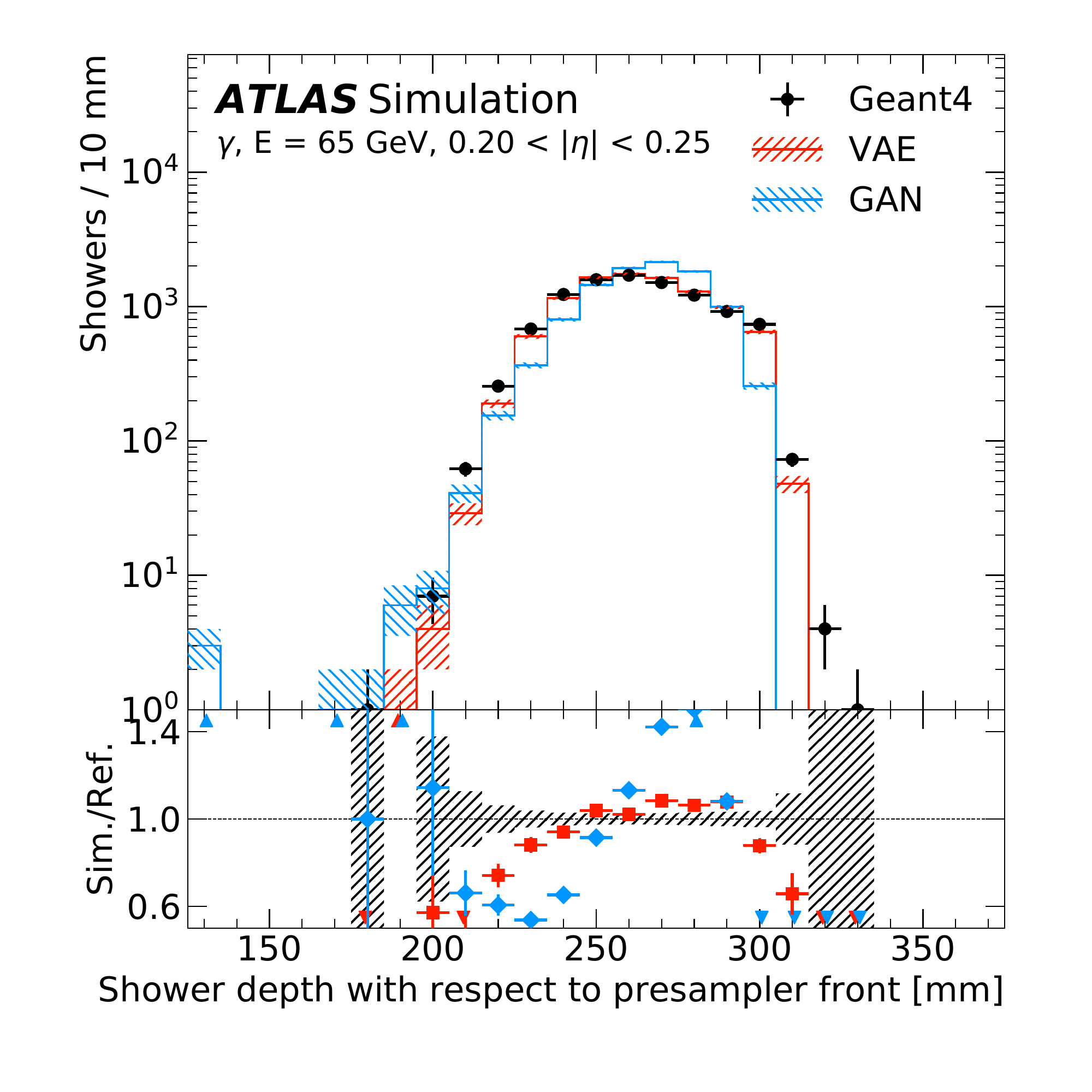}
\caption{}
\label{fig:results:depth:65}
\end{subfigure}
\begin{subfigure}[b]{0.49\textwidth}
\centering
\includegraphics[width=1.0\textwidth]{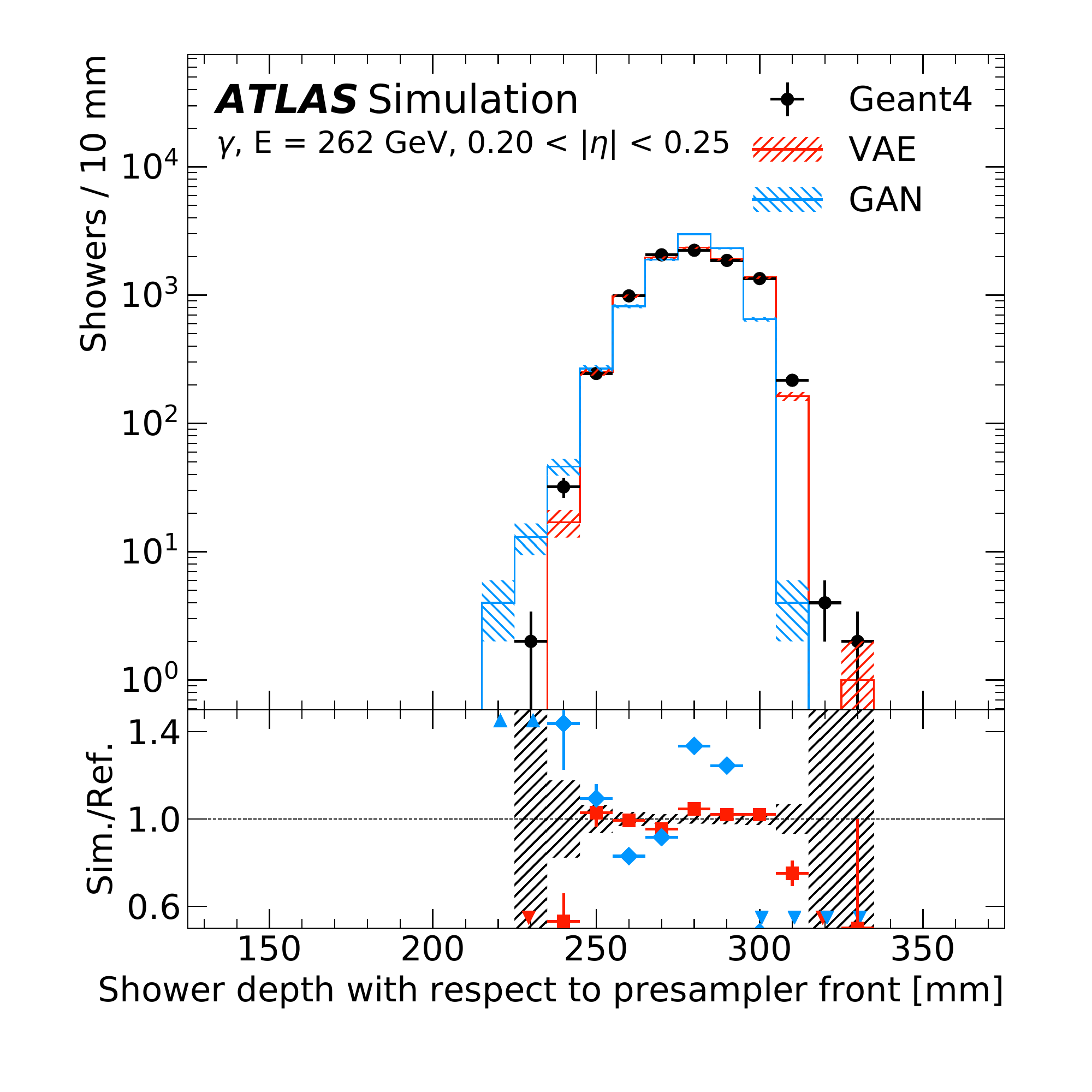}
\caption{}
\label{fig:results:depth:260}
\end{subfigure}
\caption{Reconstructed longitudinal shower centre (shower depth) for photons with an energy of \subref{fig:results:depth:4} $\SI{4}{\giga\electronvolt}$, \subref{fig:results:depth:65} $\SI{65}{\giga\electronvolt}$ and \subref{fig:results:depth:260} $\SI{262}{\giga\electronvolt}$. The shower depth for the \GEANT training data (black markers) are shown as reference points and compared with those from a VAE (solid red line) and a GAN (solid blue line). The error bars and the hatched bands indicate the statistical uncertainty of the training data and the synthesised samples, respectively. The underflow and overflow are included in the first and last bin of each distribution, respectively.
Arrows at the top and bottom of the ratio indicate when the values for either model in a bin fall outside of the displayed axis range.
All showers are simulated consistent with the training samples, considering only the energy deposits in the cells of the four electromagnetic calorimeter layers without digitisation and crosstalk noise.}
\label{fig:results:depth}
\end{figure}

\begin{figure}[tb!p]
\centering
\begin{subfigure}[b]{0.49\textwidth}
\centering
\includegraphics[width=1.0\textwidth]{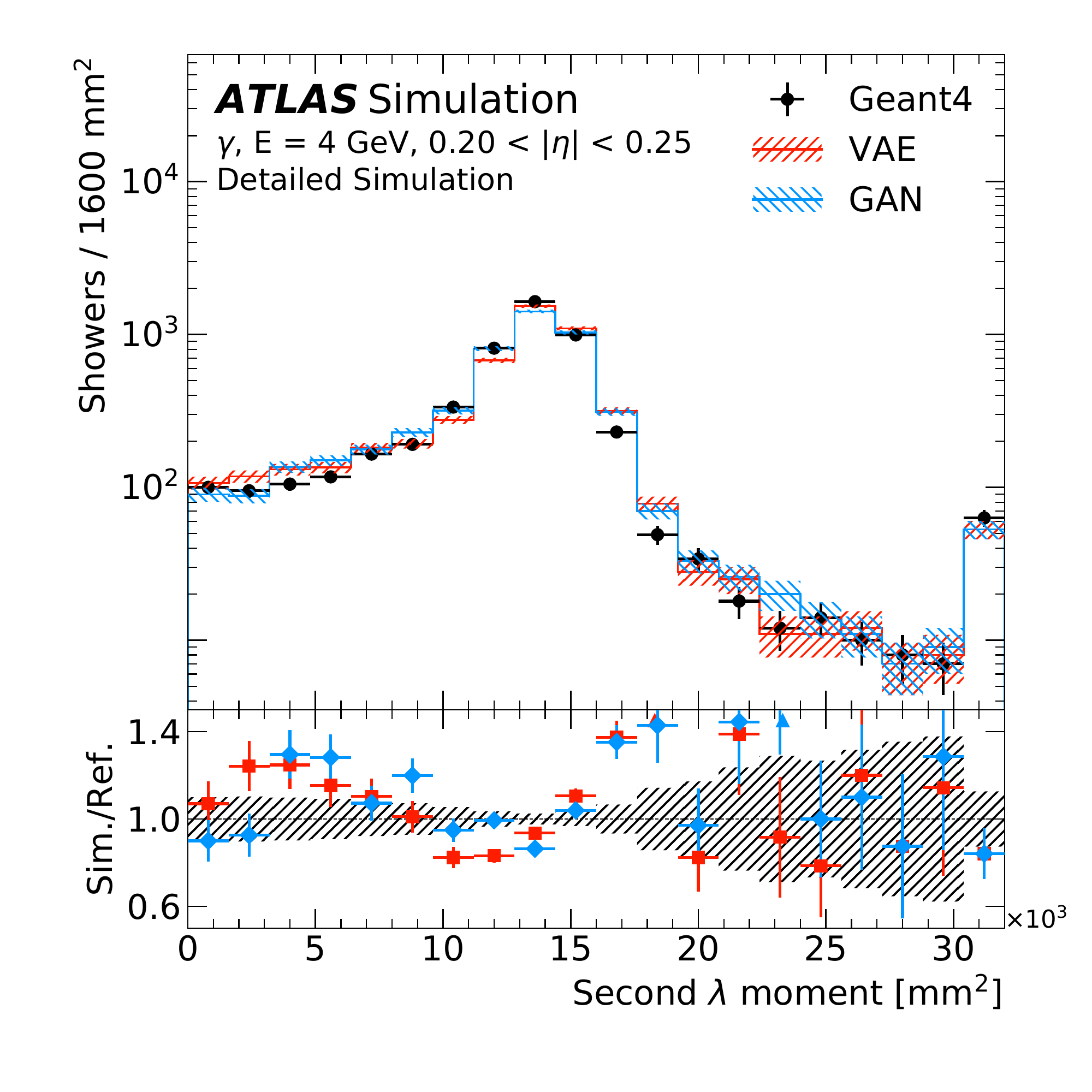}
\caption{}
\label{fig:results:secondlambda:4}
\end{subfigure}
\begin{subfigure}[b]{0.49\textwidth}
\centering
\includegraphics[width=1.0\textwidth]{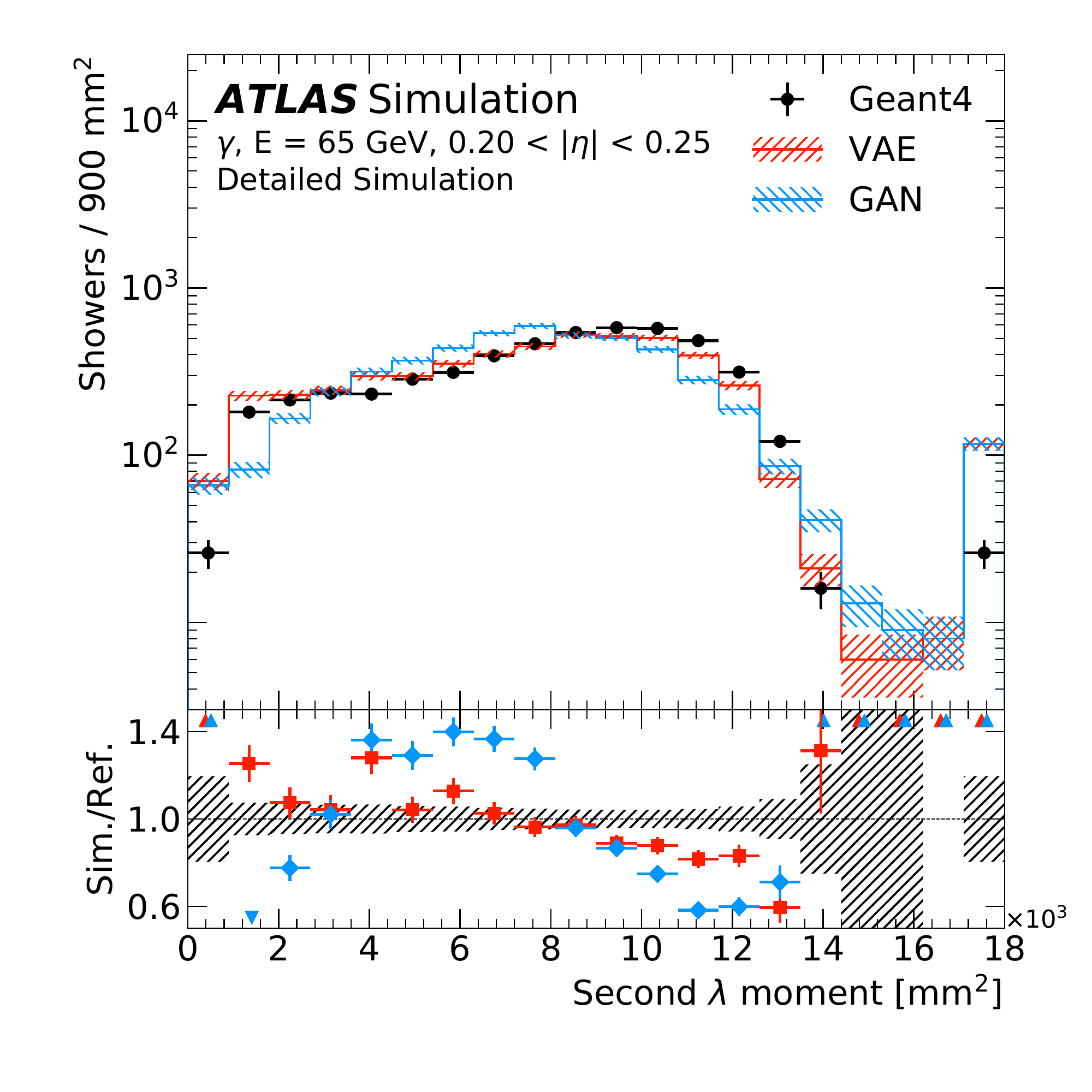}
\caption{}
\label{fig:results:secondlambda:65}
\end{subfigure}
\begin{subfigure}[b]{0.49\textwidth}
\centering
\includegraphics[width=1.0\textwidth]{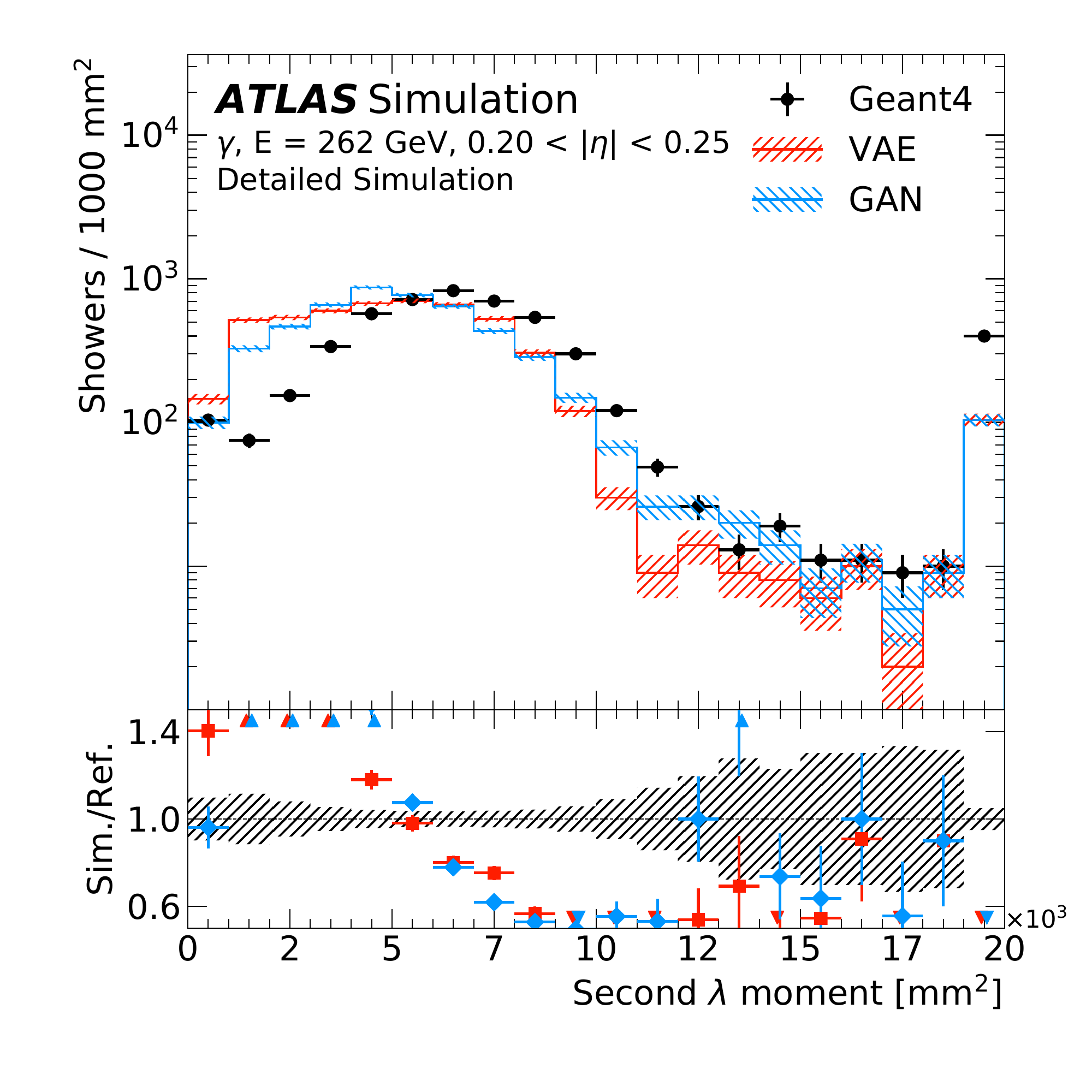}
\caption{}
\label{fig:results:secondlambda:262}
\end{subfigure}
\caption{Reconstructed second $\lambda$ moment of showers for photons with an energy of \subref{fig:results:depth:4} $\SI{4}{\giga\electronvolt}$, \subref{fig:results:depth:65} $\SI{65}{\giga\electronvolt}$ and \subref{fig:results:depth:260} $\SI{262}{\giga\electronvolt}$
. The second $\lambda$ moment for the full detector simulation (black markers) is shown as reference points and compared with that from a VAE (solid red line) and a GAN (solid blue line). The error bars and the hatched bands indicate the statistical uncertainty of the training data and the synthesised samples, respectively. The underflow and overflow are included in the first and last bin of each distribution, respectively.
Arrows at the top and bottom of the ratio indicate when the values for either model in a bin fall outside of the displayed axis range.
All showers are simulated using the full simulation chain, including both the inner detector and the digitisation and crosstalk noise.}
\label{fig:results:secondlambda}
\end{figure}

The average shower shape measured inside the calorimeter layers depends strongly on the longitudinal shower profile in the calorimeter and is used, for example, to distinguish photons from electrons. The modelling of the longitudinal shower development is shown in Figure~\ref{fig:results:depth} for photons with different energies. The reconstructed longitudinal shower centre, referred to as \textit{shower depth}, is the energy-weighted mean of the longitudinal centre positions $d_{i}$ of all calorimeter layers,
\begin{equation*}
d = \frac{1}{E} \sum_{i \in \text{layers}} E_i d_i,
\end{equation*}
with $d_i$ taken from the nominal geometry description.
Both models are able to reproduce the trend between the typical depth of the showers and the incident particle energies, showing that they are able to learn about correlations between energy deposits in different layers. Compared to the GAN, the VAE demonstrates better agreement across all energy points in modelling not just the range of shower depths but also the distributions. This is consistent with the observed performance of the VAE and GAN in modelling the individual layer energies.
Another measure of longitudinal shower shape is the second $\lambda$ moment of the shower, as defined in Ref.~\cite{PERF-2014-07}, which is shown in Figure~\ref{fig:results:secondlambda}. This observable measures the spread of energy in the longitudinal development of the shower, in contrast to the shower depth, which measures the mean position. The second $\lambda$ moment is used in Figure~\ref{fig:results:secondlambda} to compare the models with the detailed \GEANT simulation as it is sensitive to the noise and digitisation of the cells.
It provides a stringent test of the ability of the two models to correctly capture correlations between the energies deposited in the cells of all four layers of the EM calorimeter.
Here the VAE is able to follow the overall distribution for energies up to 65~\GeV, but the GAN is unable to model the observable in all but the lowest-energy showers.
 
\begin{figure}[tb!p]
\centering
\begin{subfigure}[b]{0.49\textwidth}
\centering
\includegraphics[width=1.0\textwidth]{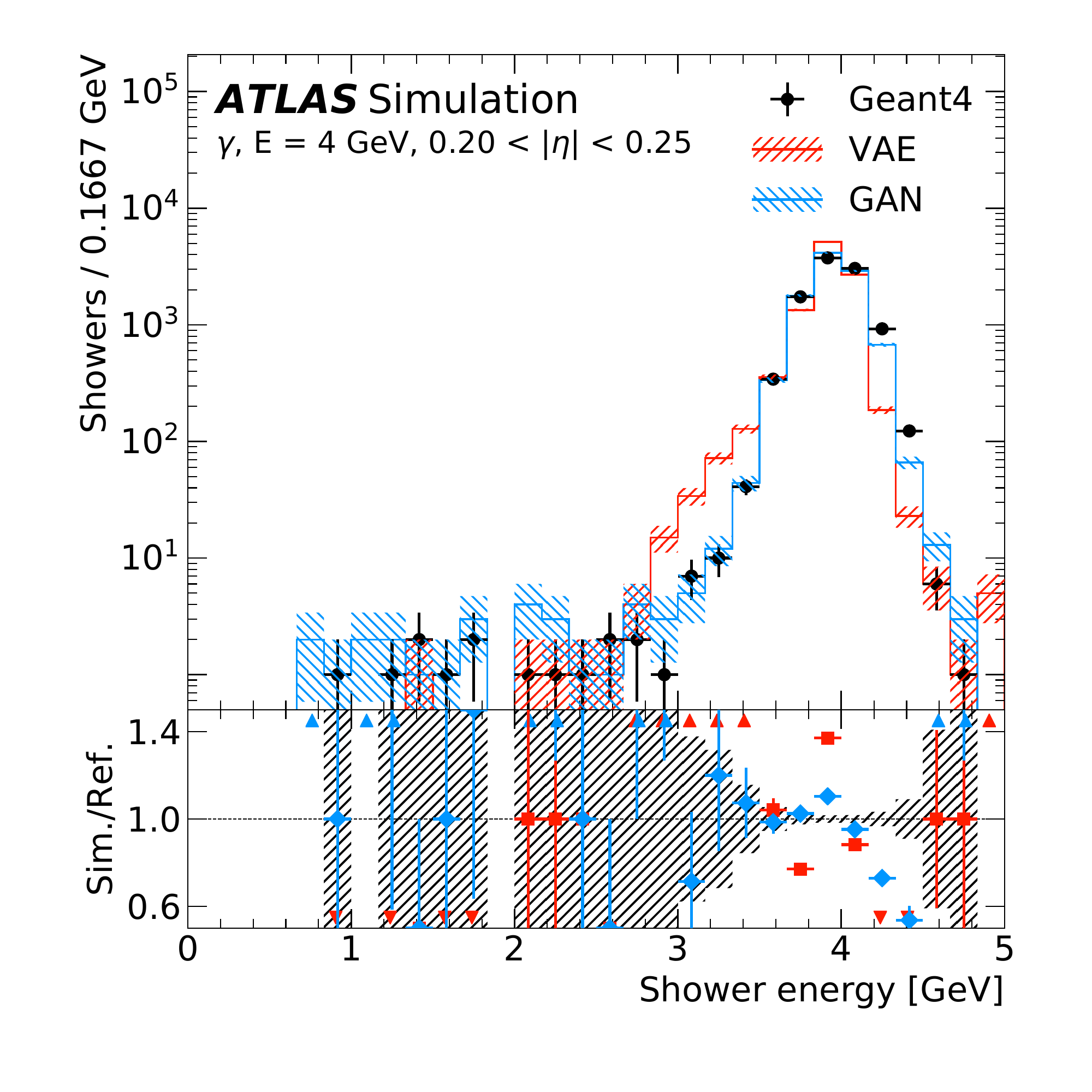}
\caption{}
\label{fig:results:etot:4}
\end{subfigure}
\begin{subfigure}[b]{0.49\textwidth}
\centering
\includegraphics[width=1.0\textwidth]{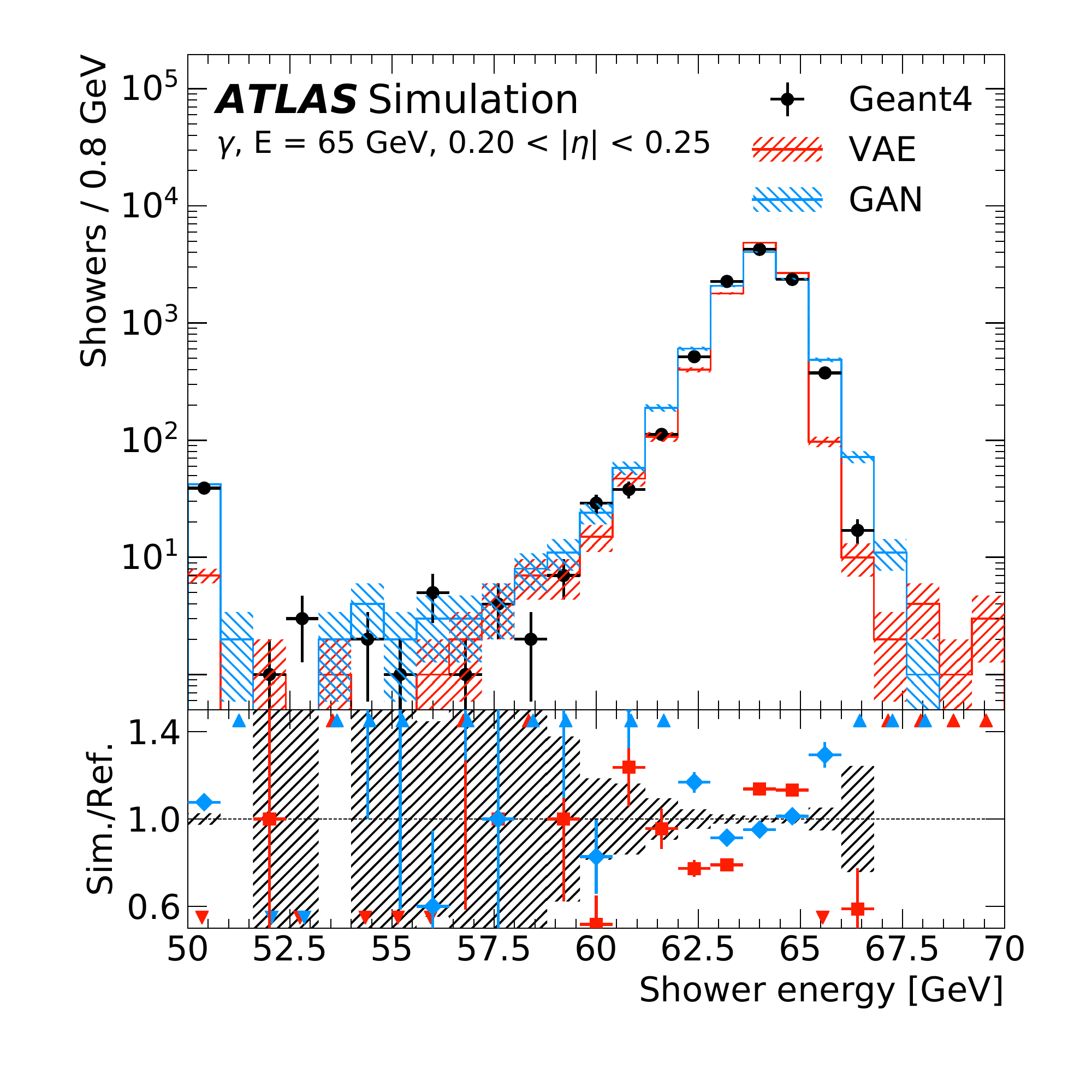}
\caption{}
\label{fig:results:etot:65}
\end{subfigure}\\
\begin{subfigure}[b]{0.49\textwidth}
\centering
\includegraphics[width=1.0\textwidth]{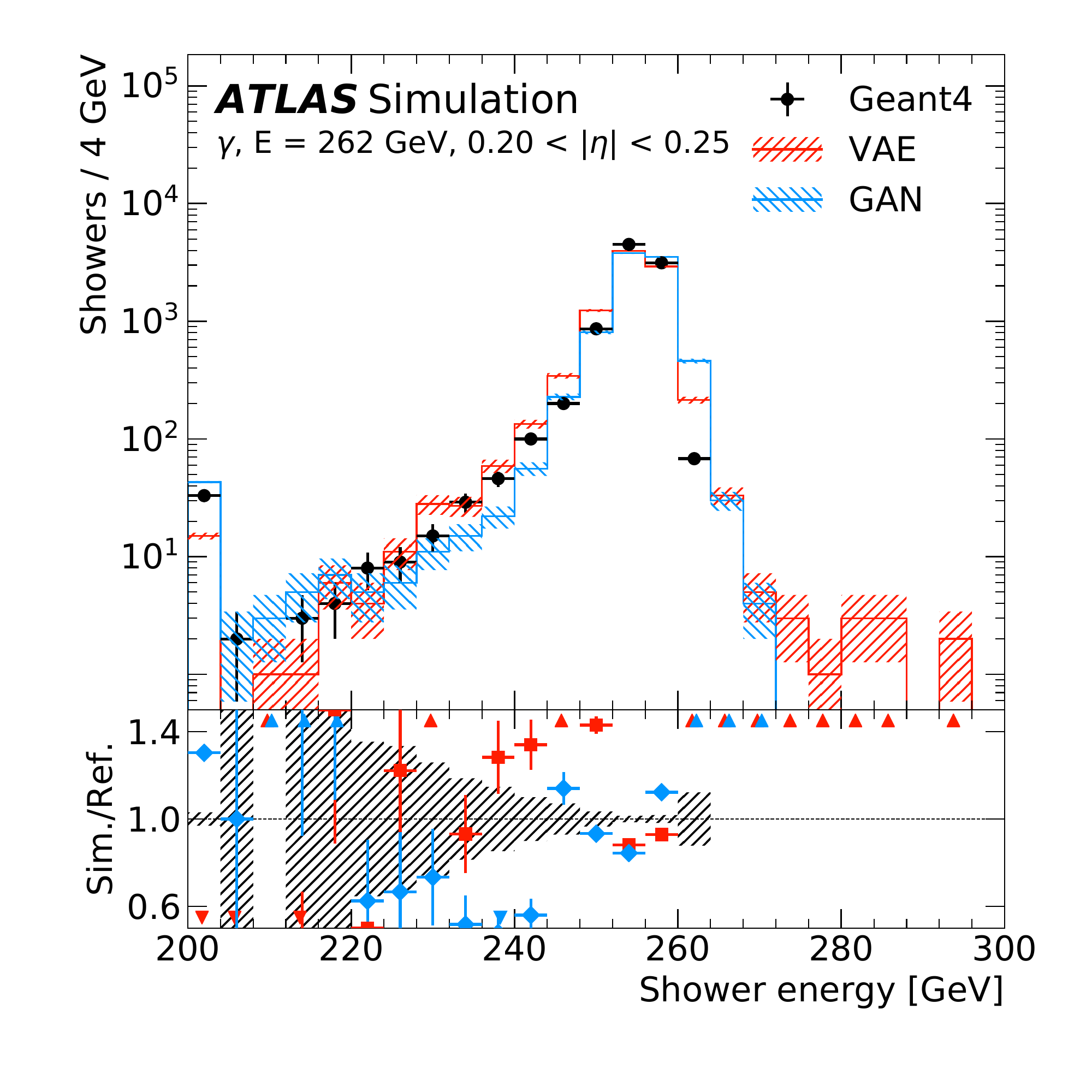}
\caption{}
\label{fig:results:etot:260}
\end{subfigure}
\caption{Total energy response of the calorimeter to photons with an energy of \subref{fig:results:depth:4} $\SI{4}{\giga\electronvolt}$, \subref{fig:results:depth:65} $\SI{65}{\giga\electronvolt}$, and \subref{fig:results:depth:260} $\SI{262}{\giga\electronvolt}$. The calorimeter response for the \GEANT training data (black markers) are shown as reference points and compared with those from a VAE (solid red line) and a GAN (solid blue line). The error bars and the hatched bands indicate the statistical uncertainty of the training data and the synthesised samples, respectively. The underflow and overflow are included in the first and last bin of each distribution, respectively.
Arrows at the top and bottom of the ratio indicate when the values for either model in a bin fall outside of the displayed axis range.
All showers are simulated consistent with the training samples, considering only the energy deposits in the cells of the four electromagnetic calorimeter layers without digitisation and crosstalk noise.}
\label{fig:results:etot}
\end{figure}

\begin{figure}[tb!p]
\centering
\begin{subfigure}[b]{0.49\textwidth}
\centering
\includegraphics[width=0.98\textwidth]{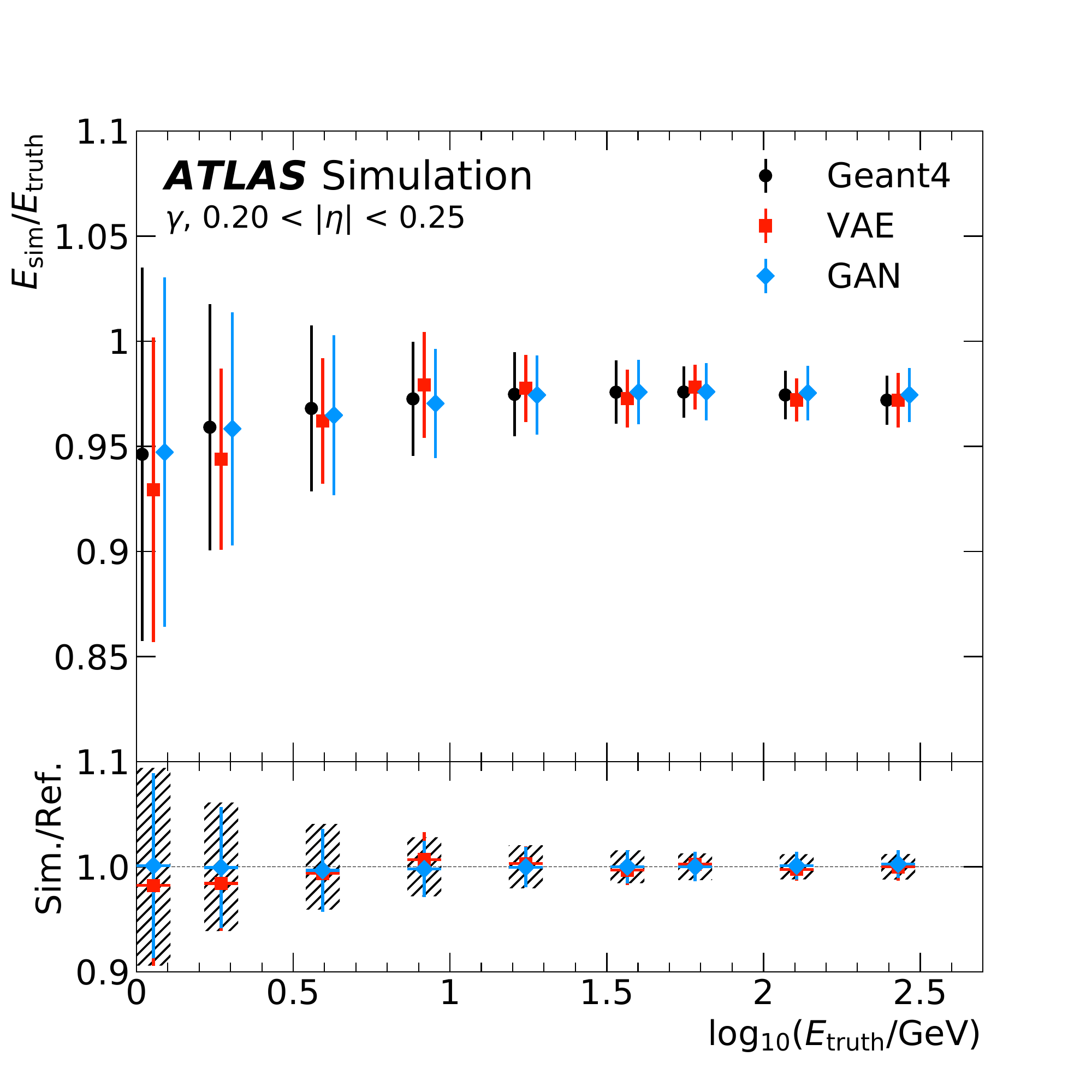}
\caption{}
\label{fig:results:e}
\end{subfigure}
\begin{subfigure}[b]{0.49\textwidth}
\centering
\includegraphics[width=\textwidth]{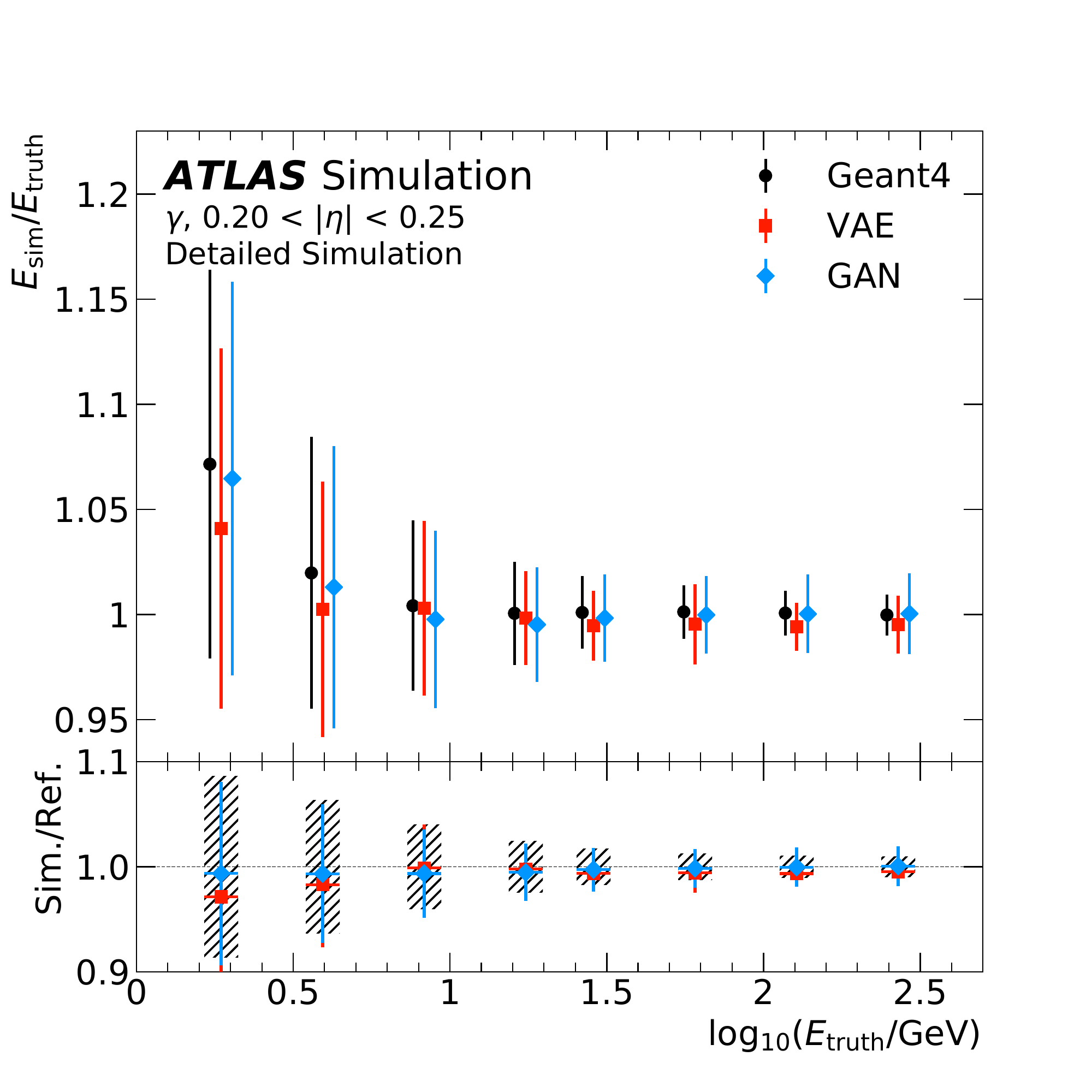}
\caption{}
\label{fig:results:e_athena}
\end{subfigure}
 
\caption{Energy response of the calorimeter as a function of the true photon energy. The calorimeter response for the full detector simulation (black markers) is shown as reference points and compared with that from a VAE (red markers) and a GAN (blue markers). The error bars indicate the resolution for the simulated energy deposits. The statistical standard error of the mean is not shown, as it is too small to be visible at this scale.
Showers are compared with \subref{fig:results:e} training data considering only the energy deposits in the cells of the four electromagnetic calorimeter layers, without digitisation and crosstalk noise, and \subref{fig:results:e_athena} the detailed simulation including digitisation and crosstalk noise.}
\label{fig:results:figure8}
\end{figure}

Figure~\ref{fig:results:etot} shows the total energy response of the calorimeter, $E$, to photons with energies of $\SI{4}{\giga\electronvolt}$, $\SI{65}{\giga\electronvolt}$, and $\SI{262}{\giga\electronvolt}$.
Figure~\ref{fig:results:e} shows the simulated energy as a function of true photon energy, comparing the models' ability to recover the mean shower energy and the energy resolution across all generated showers for a given true energy, testing the reproduction of the stochasticity of the shower energy.
The error bars indicate the resolution of the simulated energy deposits. The statistical standard error of the mean is not shown; it is negligible given the amount of simulated data for each energy point.
The GAN reproduces the mean shower energy simulated by \GEANT for all energy points better than the VAE, with the VAE showing significant deviations from the \GEANT simulation, particularly at low energies, indicating a systematic underestimation.
The GAN also models the energy spread better than the VAE for all energies, although both the GAN and VAE underestimate it at lower energies and overestimate it at higher energies.
The same plot is shown in Figure~\ref{fig:results:e_athena} for reconstructed photons generated using the samples with full detector simulation including digitisation and readout noise. The level of agreement between \GEANT and both the VAE and GAN is similar to that seen in Figure~\ref{fig:results:e}.
In the absence of the energy critic, the energy resolution of the GAN was too large by a factor of two at high energies. Without the reparametrised inputs for the VAE, the spread as measured by the RMS was several times too large.
 
\begin{figure}[tb!p]
\centering
\begin{subfigure}[b]{0.49\textwidth}
\centering
\includegraphics[width=\textwidth]{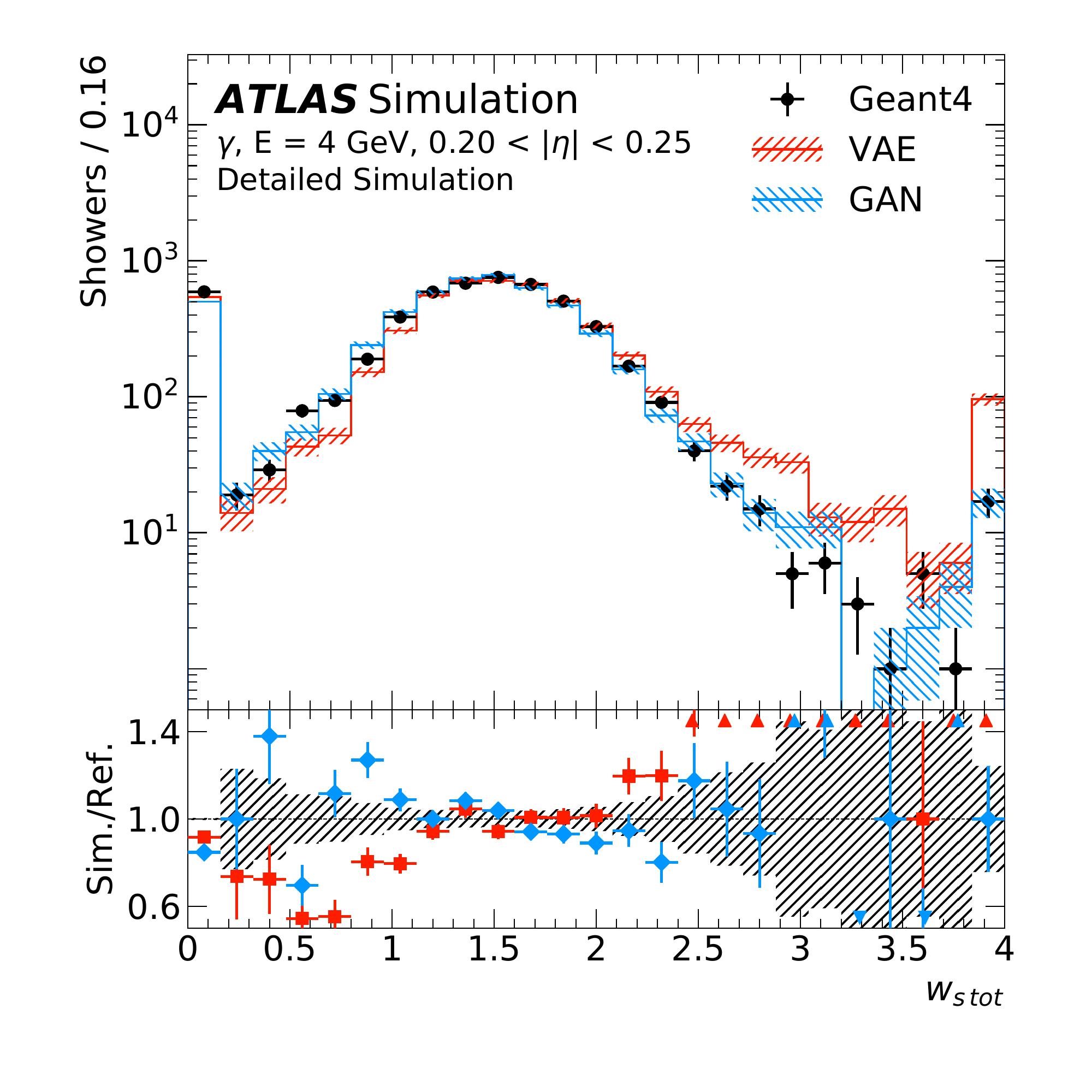}
\caption{}
\label{fig:results:wtots1:4}
\end{subfigure}
\begin{subfigure}[b]{0.49\textwidth}
\centering
\includegraphics[width=\textwidth]{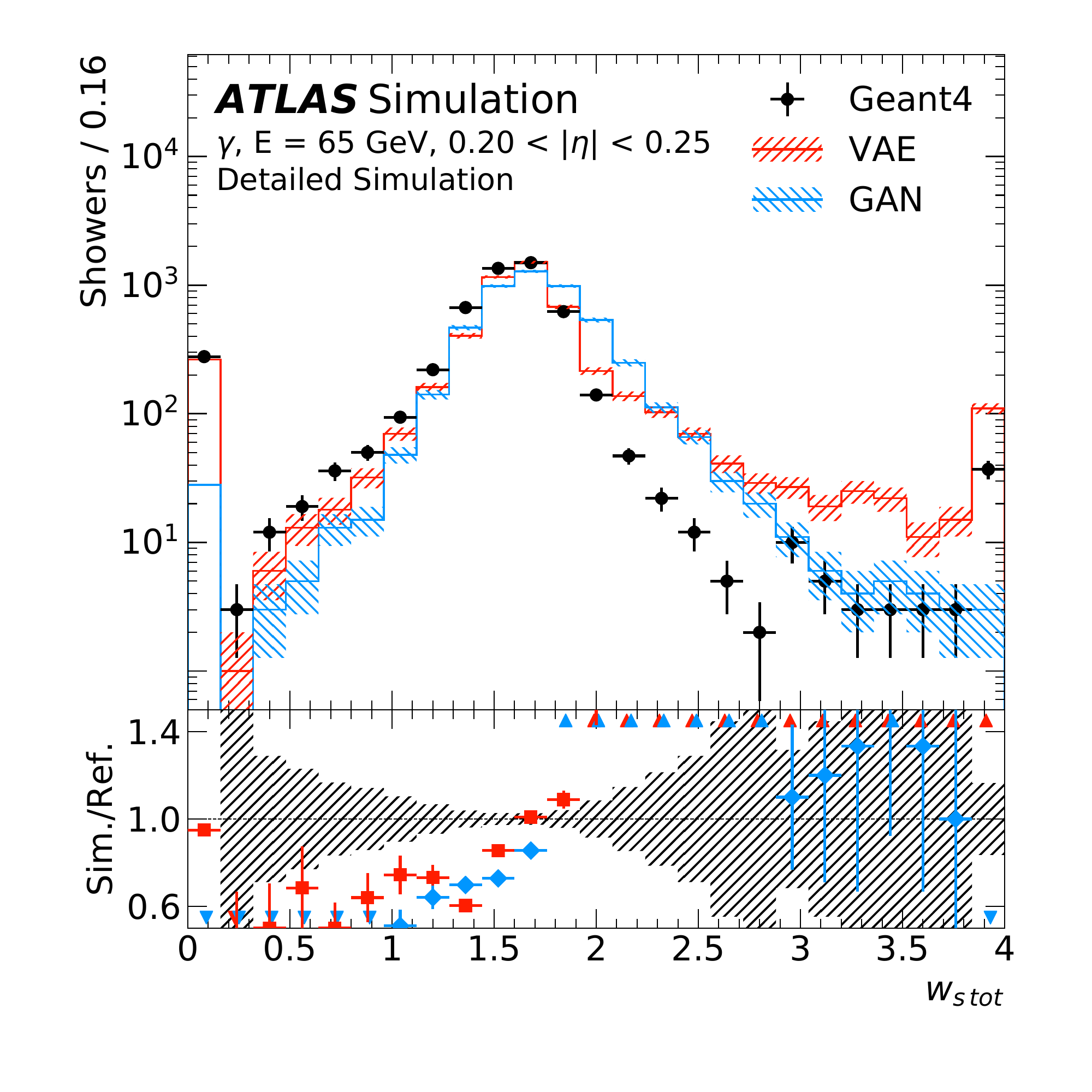}
\caption{}
\label{fig:results:wtots1:65}
\end{subfigure}
\begin{subfigure}[b]{0.49\textwidth}
\centering
\includegraphics[width=\textwidth]{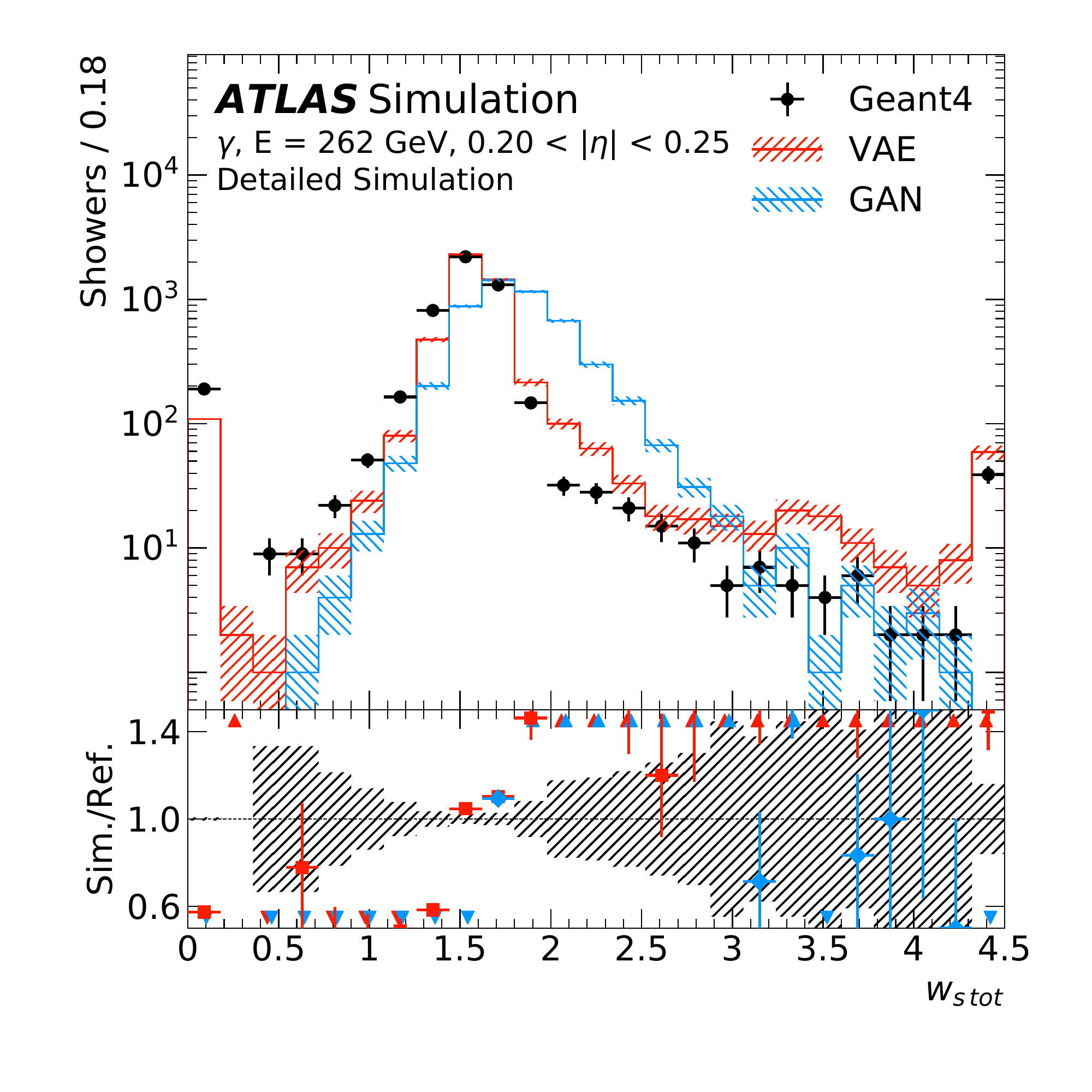}
\caption{}
\label{fig:results:wtots1:260}
\end{subfigure}
\caption{Total lateral shower width $w_{s\,\text{tot}}$ in the front EM layer for photons with an energy of \subref{fig:results:wtots1:4} $\SI{4}{\giga\electronvolt}$, \subref{fig:results:wtots1:65} $\SI{65}{\giga\electronvolt}$, and \subref{fig:results:wtots1:260} $\SI{262}{\giga\electronvolt}$. The calorimeter response for the full detector simulation (black markers) is shown as reference points and compared with that from a VAE (solid red line) and a GAN (solid blue line). The error bars indicate the statistical uncertainty of the detailed simulation and the synthesised samples.
The underflow and overflow are included in the first and last bin of each distribution, respectively.
Arrows at the top and bottom of the ratio indicate when the values for either model in a bin fall outside of the displayed axis range.
All showers are simulated using the full simulation chain, including both the inner detector and digitisation and crosstalk noise.}
\label{fig:results:wtots1}
\end{figure}

\begin{figure}[tb!p]
\centering
\begin{subfigure}[b]{0.49\textwidth}
\centering
\includegraphics[width=\textwidth]{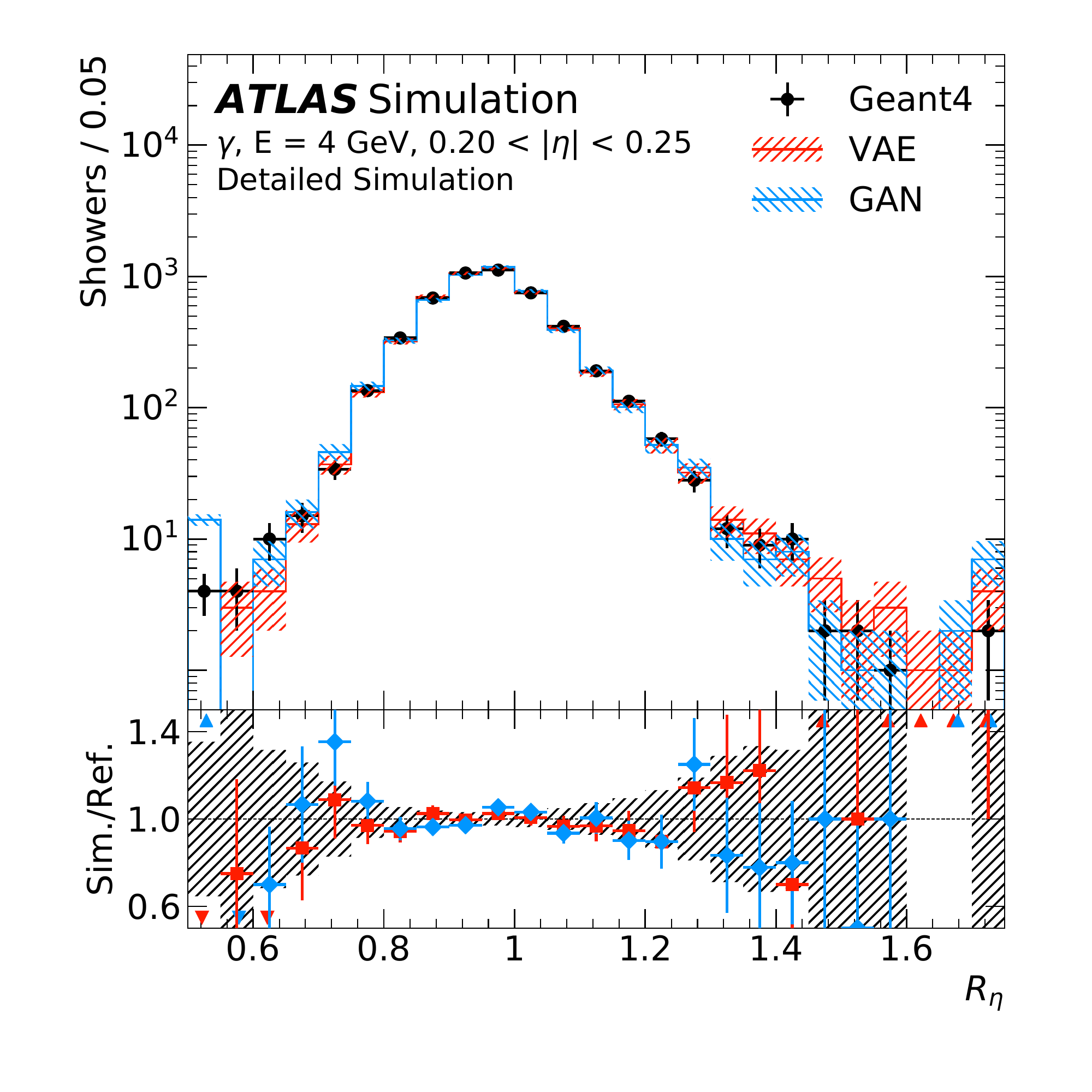}
\caption{}
\label{fig:results:Reta:4}
\end{subfigure}
\begin{subfigure}[b]{0.49\textwidth}
\centering
\includegraphics[width=\textwidth]{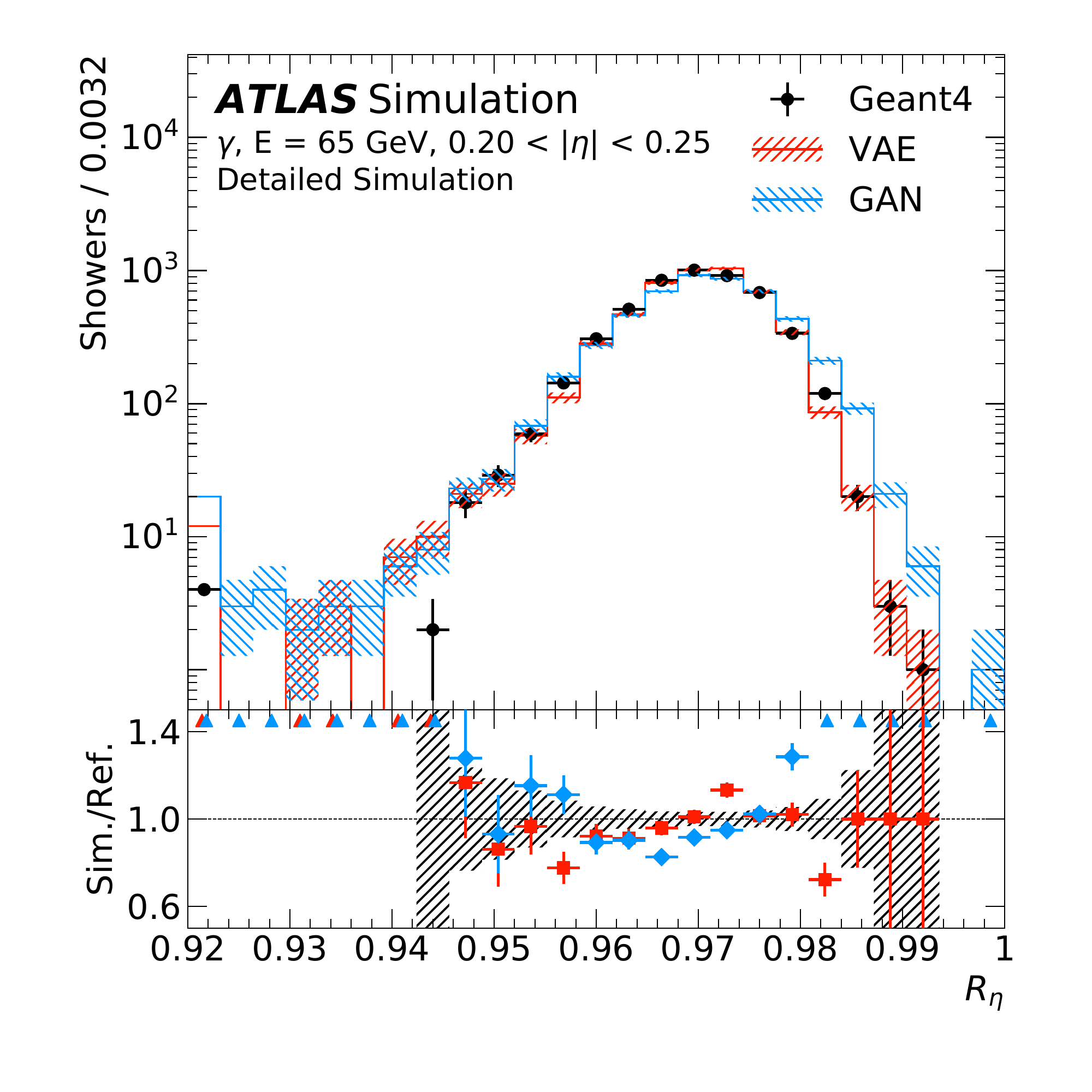}
\caption{}
\label{fig:results:Reta:65}
\end{subfigure}
\begin{subfigure}[b]{0.49\textwidth}
\centering
\includegraphics[width=\textwidth]{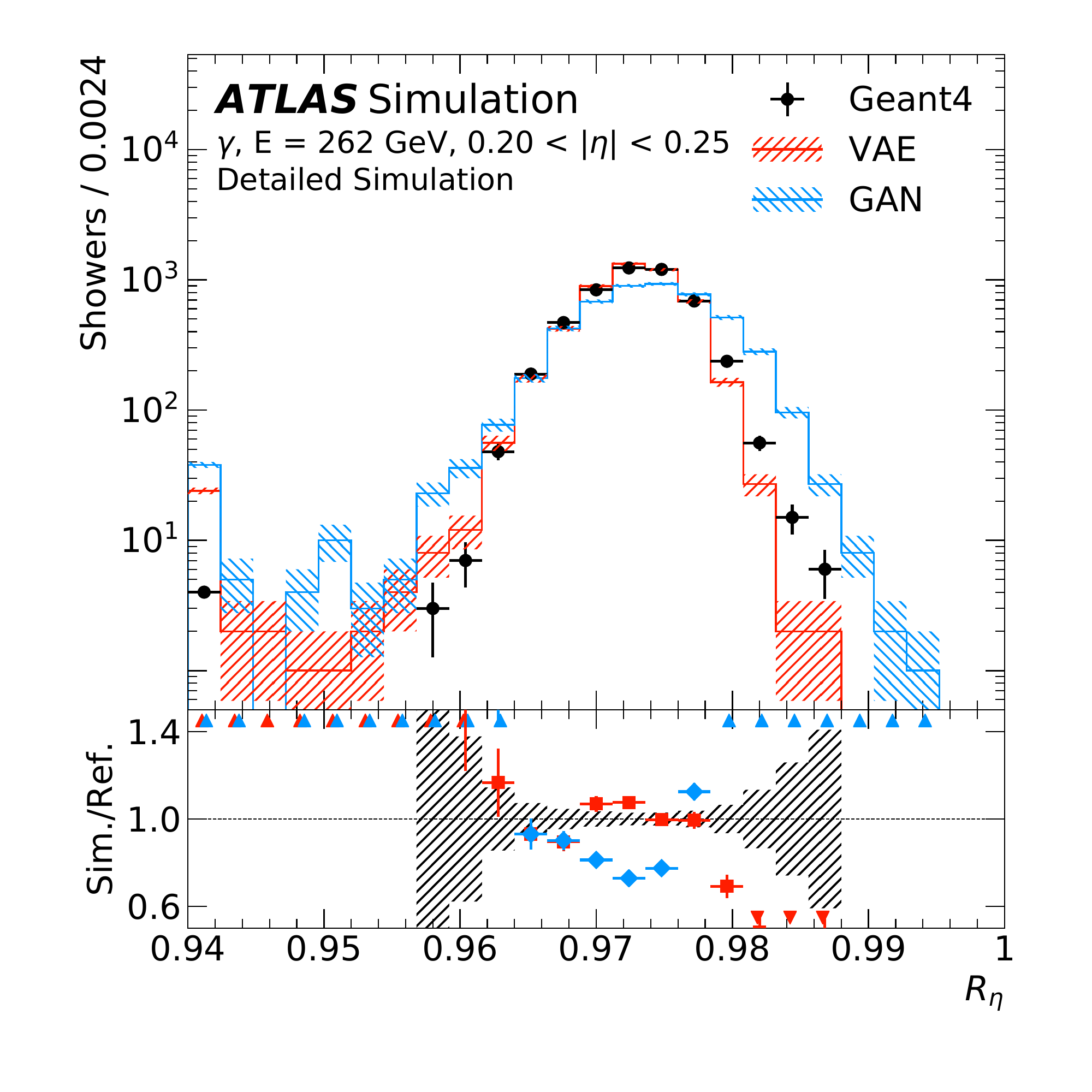}
\caption{}
\label{fig:results:Reta:260}
\end{subfigure}
\caption{The energy ratio $R_\eta$ in the middle EM layer for photons with an energy of \subref{fig:results:Reta:4} $\SI{4}{\giga\electronvolt}$, \subref{fig:results:Reta:65} $\SI{65}{\giga\electronvolt}$, and \subref{fig:results:Reta:260} $\SI{262}{\giga\electronvolt}$. The calorimeter response for the full detector simulation (black markers) is shown as reference points and compared with that from a VAE (solid red line) and a GAN (solid blue line). The error bars indicate the statistical uncertainty of the detailed simulation and the synthesised samples.
The underflow and overflow are included in the first and last bin of each distribution, respectively.
Arrows at the top and bottom of the ratio indicate when the values for either model in a bin fall outside of the displayed axis range.
All showers are simulated using the full simulation chain, including both the inner detector and digitisation and crosstalk noise.}
\label{fig:results:Reta}
\end{figure}

\begin{figure}[tb!p]
\centering
\begin{subfigure}[b]{0.49\textwidth}
\centering
\includegraphics[width=\textwidth]{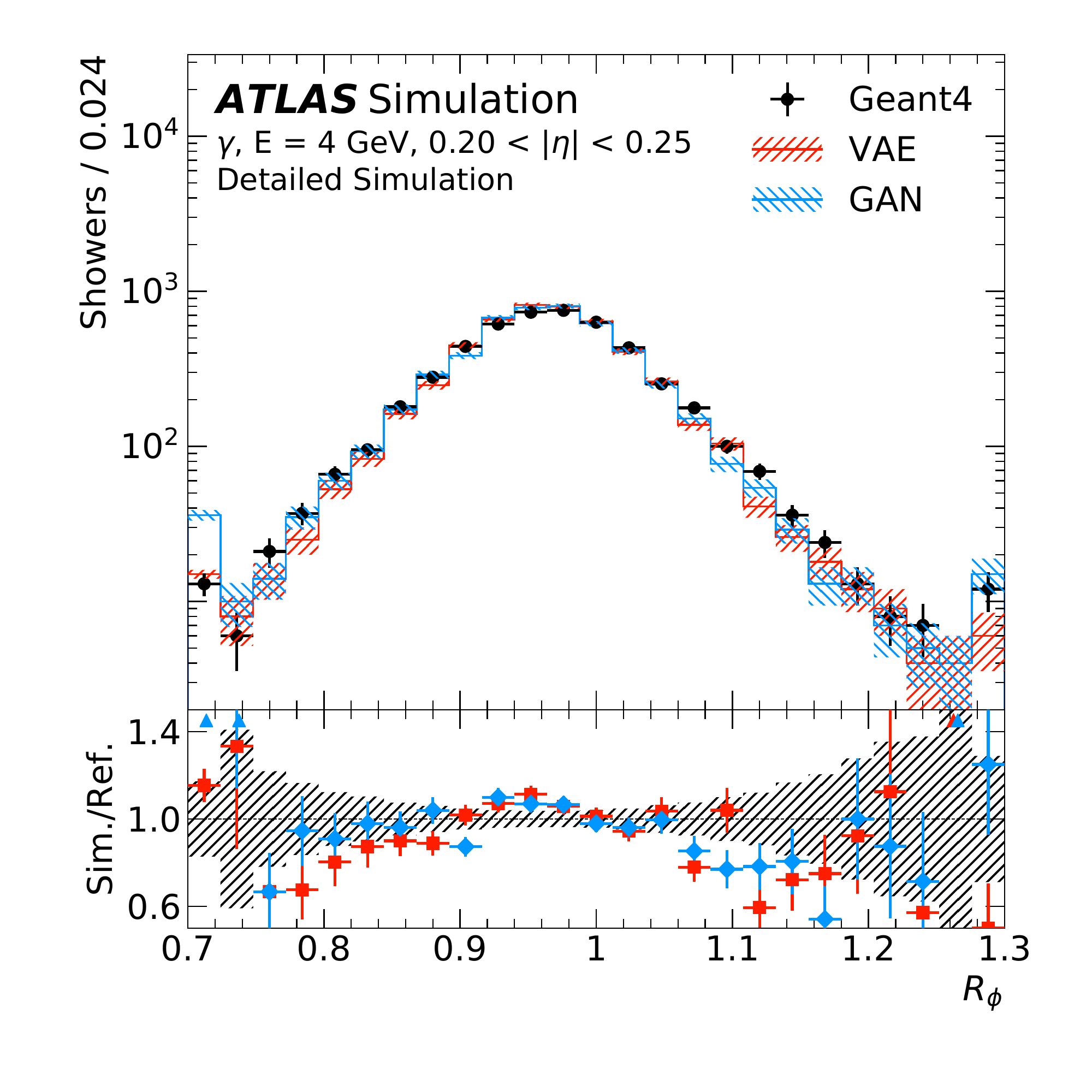}
\caption{}
\label{fig:results:Rphi:4}
\end{subfigure}
\begin{subfigure}[b]{0.49\textwidth}
\centering
\includegraphics[width=\textwidth]{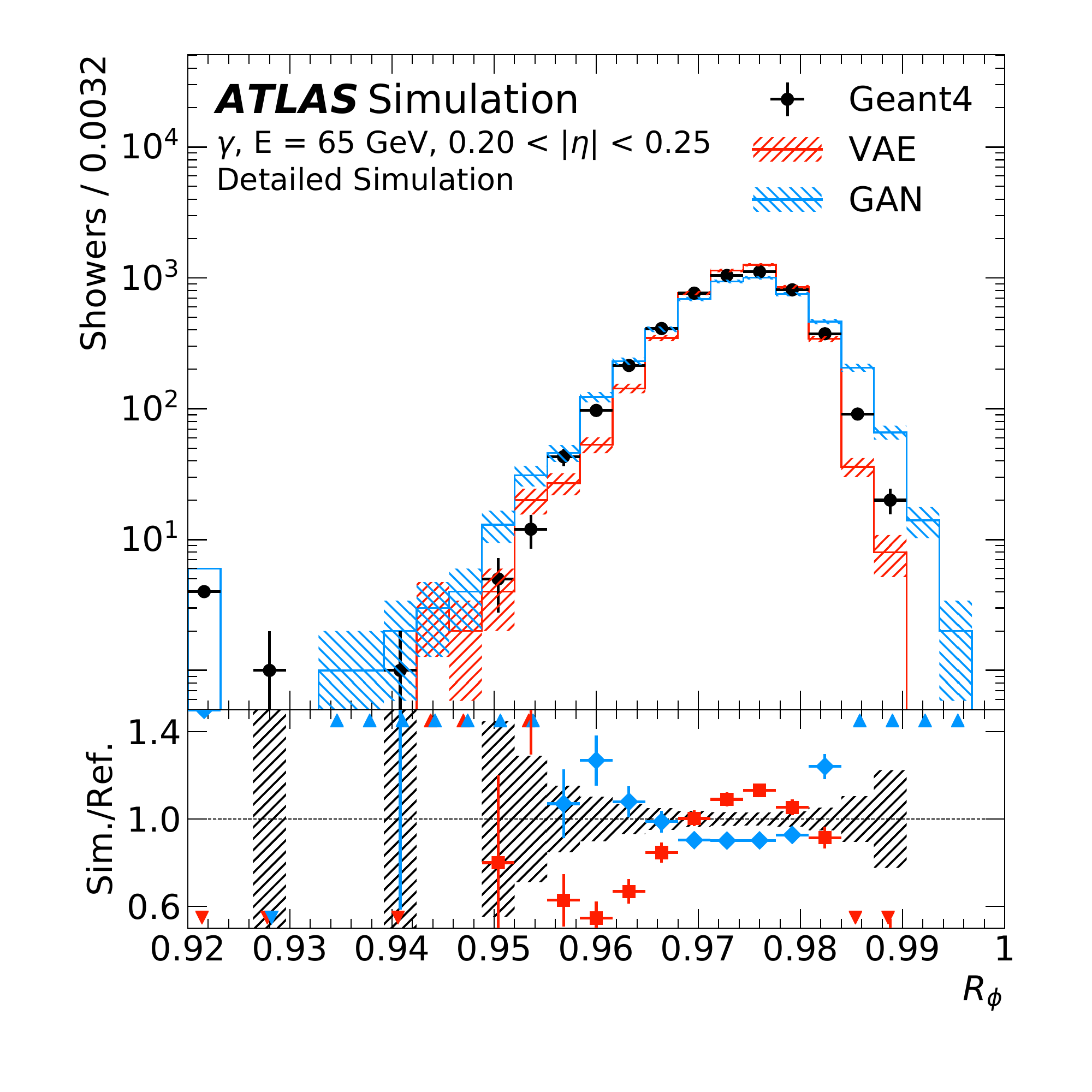}
\caption{}
\label{fig:results:Rphi:65}
\end{subfigure}
\begin{subfigure}[b]{0.49\textwidth}
\centering
\includegraphics[width=\textwidth]{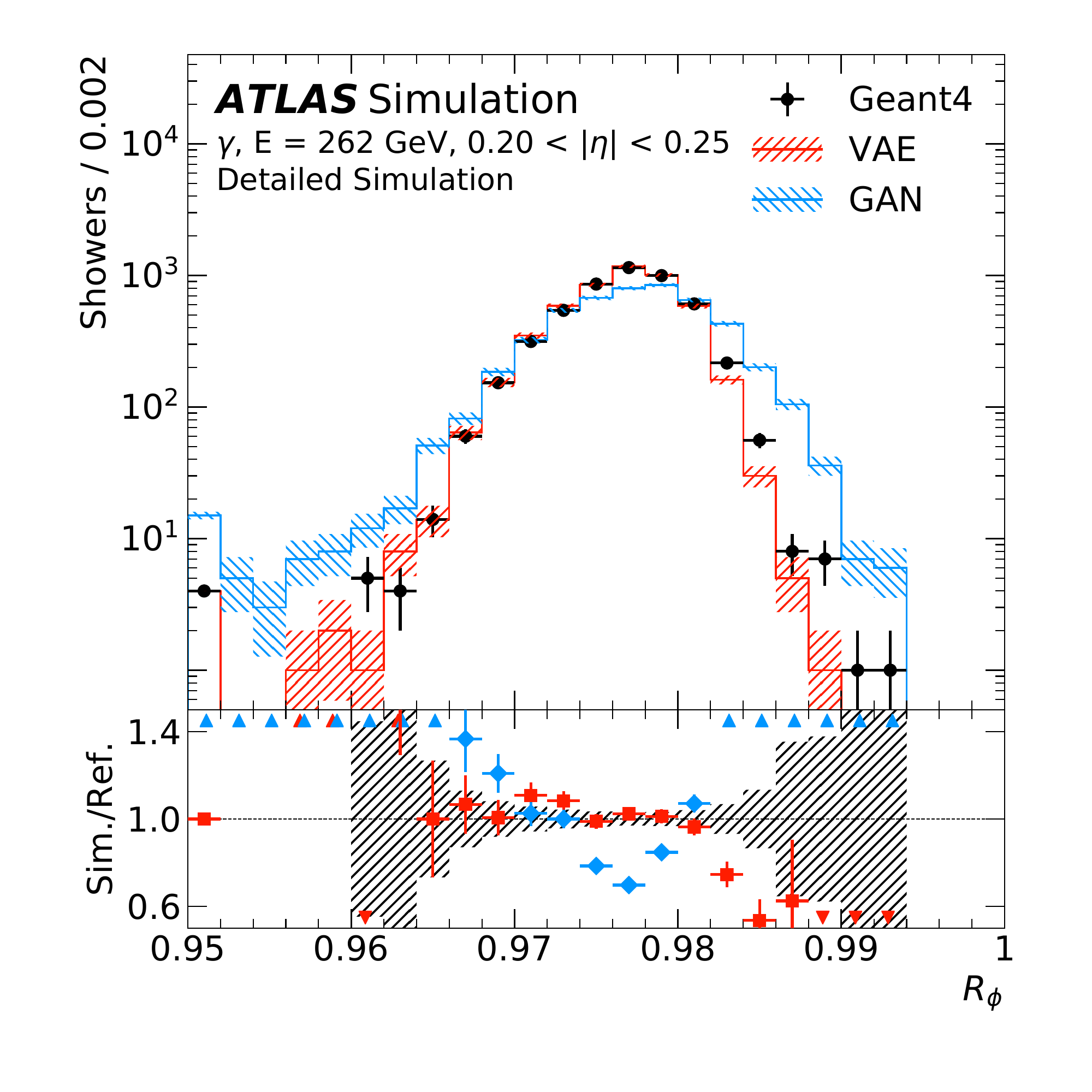}
\caption{}
\label{fig:results:Rphi:260}
\end{subfigure}
\caption{The energy ratio $R_\phi$ in the middle EM layer for photons with an energy of \subref{fig:results:Rphi:4} $\SI{4}{\giga\electronvolt}$, \subref{fig:results:Rphi:65} $\SI{65}{\giga\electronvolt}$, and \subref{fig:results:Rphi:260} $\SI{262}{\giga\electronvolt}$. The calorimeter response for the full detector simulation (black markers) is shown as reference points and compared with that from a VAE (solid red line) and a GAN (solid blue line). The error bars indicate the statistical uncertainty of the detailed simulation and the synthesised samples.
The underflow and overflow are included in the first and last bin of each distribution, respectively.
Arrows at the top and bottom of the ratio indicate when the values for either model in a bin fall outside of the displayed axis range.
All showers are simulated using the full simulation chain, including both the inner detector and digitisation and crosstalk noise.}
\label{fig:results:Rphi}
\end{figure}

\begin{figure}[tb!p]
\centering
\begin{subfigure}[b]{0.49\textwidth}
\centering
\includegraphics[width=\textwidth]{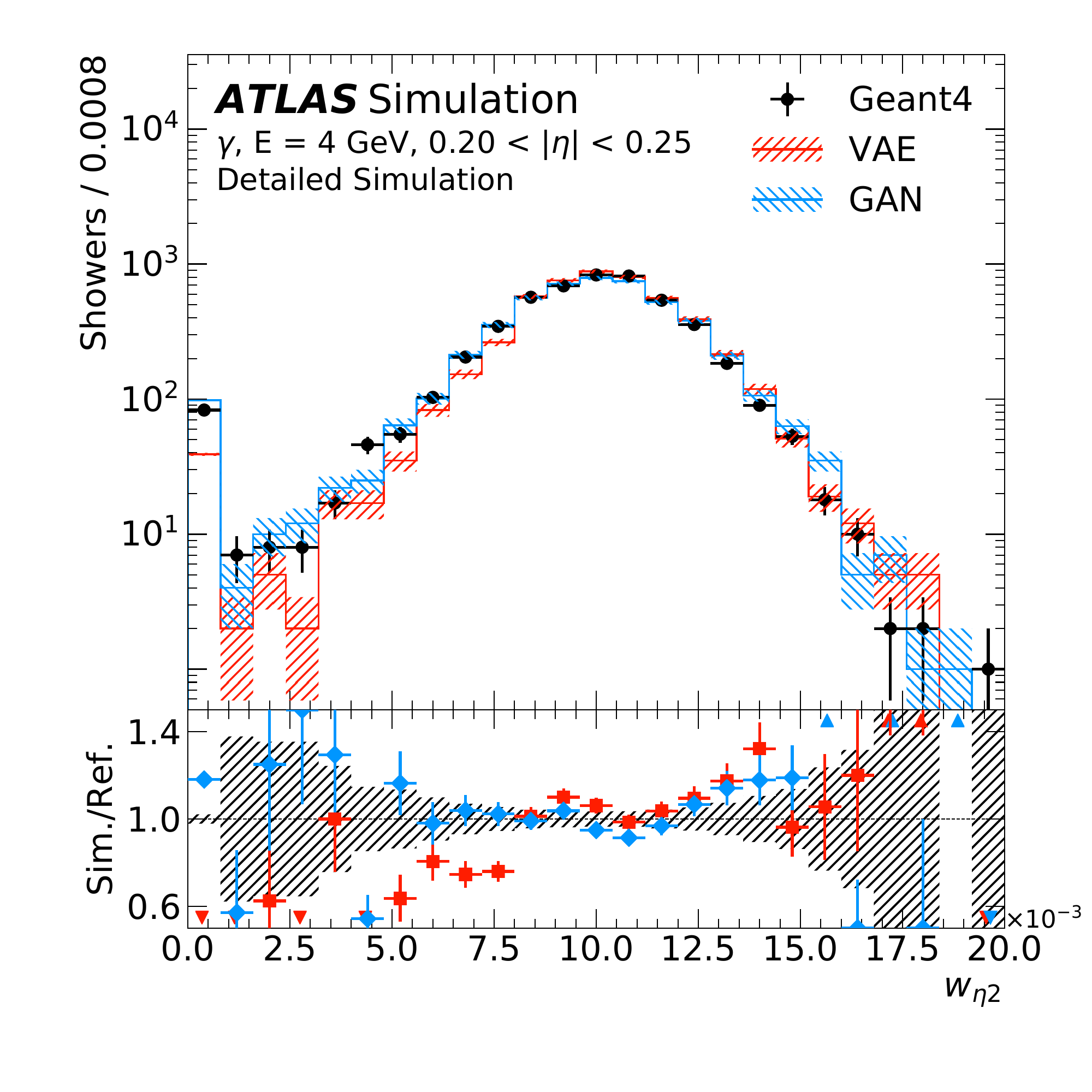}
\caption{}
\label{fig:results:weta2:4}
\end{subfigure}
\begin{subfigure}[b]{0.49\textwidth}
\centering
\includegraphics[width=\textwidth]{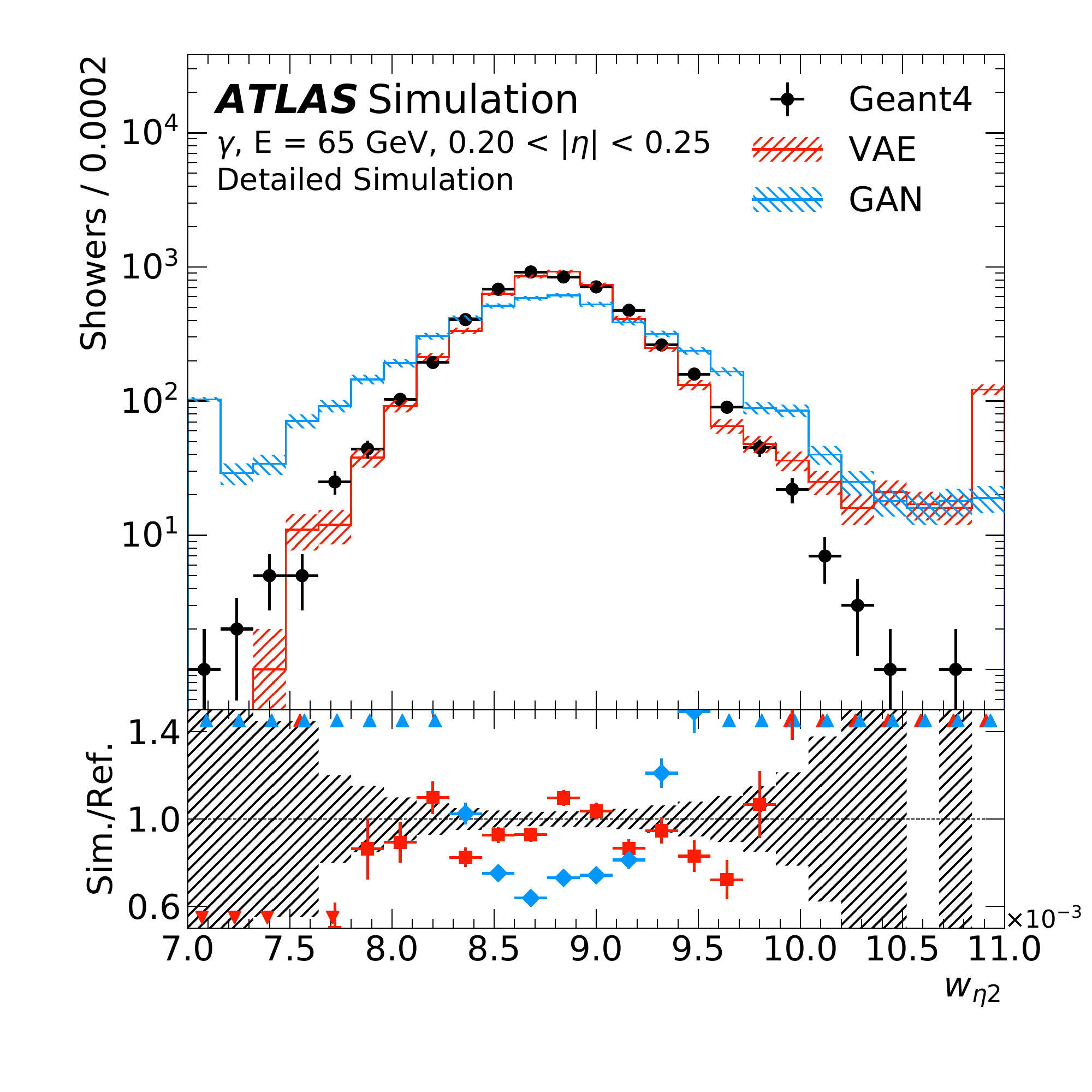}
\caption{}
\label{fig:results:weta2:65}
\end{subfigure}
\begin{subfigure}[b]{0.49\textwidth}
\centering
\includegraphics[width=\textwidth]{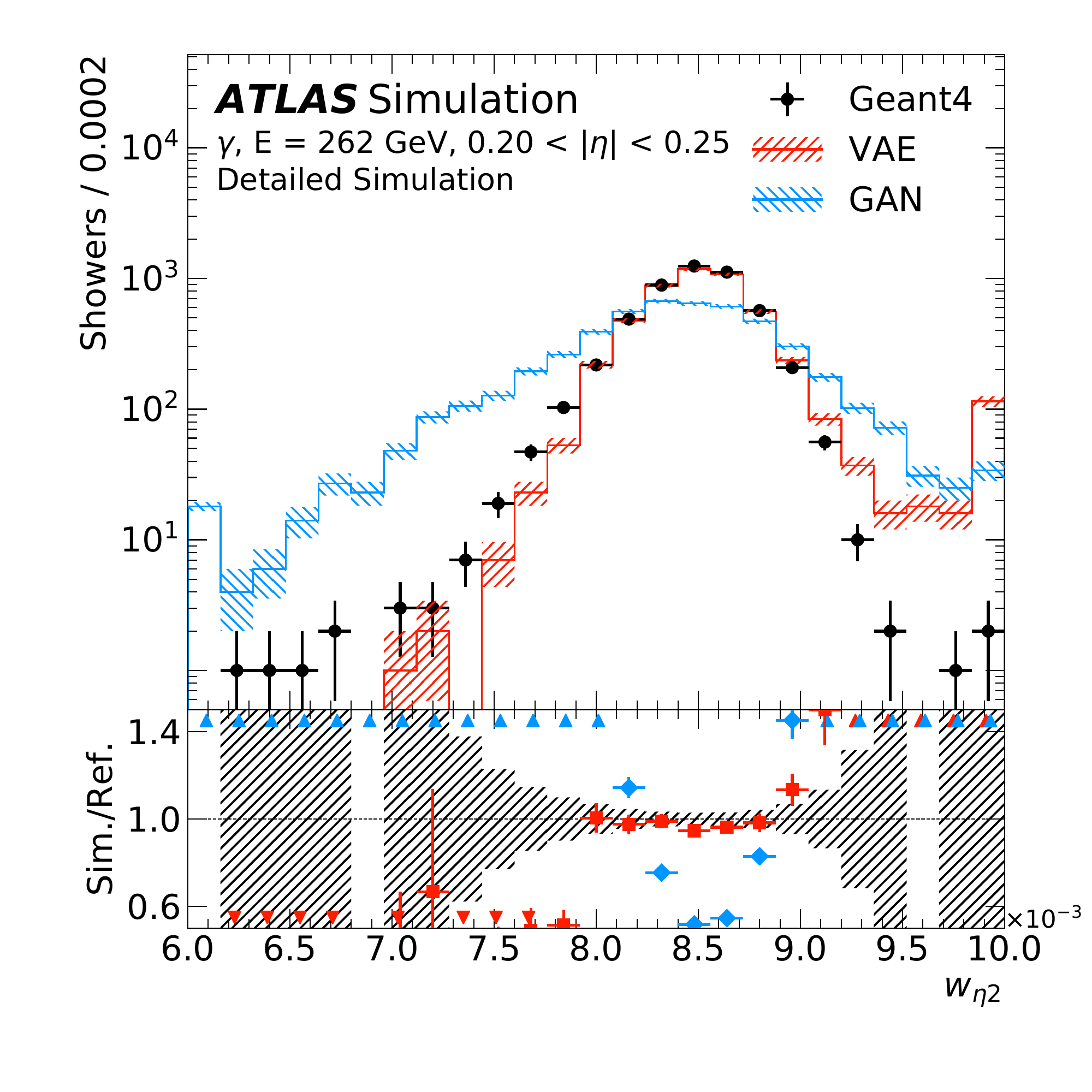}
\caption{}
\label{fig:results:weta2:260}
\end{subfigure}
\caption{Lateral shower width $w_{\eta2}$ in the middle EM layer for photons with an energy of \subref{fig:results:weta2:4} $\SI{4}{\giga\electronvolt}$, \subref{fig:results:weta2:65} $\SI{65}{\giga\electronvolt}$, and \subref{fig:results:weta2:260} $\SI{262}{\giga\electronvolt}$. The calorimeter response for the full detector simulation (black markers) is shown as reference points and compared with that from a VAE (solid red line) and a GAN (solid blue line). The error bars indicate the statistical uncertainty of the detailed simulation and the synthesised samples.
The underflow and overflow are included in the first and last bin of each distribution, respectively.
Arrows at the top and bottom of the ratio indicate when the values for either model in a bin fall outside of the displayed axis range.
All showers are simulated using the full simulation chain, including both the inner detector and the digitisation and crosstalk noise.}
\label{fig:results:weta2}
\end{figure}

In addition to the detector energy response and general shower shape, there are further distributions which must be well modelled in order for the generative models to be considered for use in the simulation chain. These distributions are used to identify the nature of the incident particle which produced the shower in the ATLAS detector, and describe the substructure of the shower \cite{Aad:2019tso}.
The performance of the two generative models in attempting to reproduce four of these distributions, namely $w_{s\,\text{tot}}$, $R_\eta$, $R_\phi$ and $w_{\eta2}$, is shown in Figures~\ref{fig:results:wtots1}, \ref{fig:results:Reta}, \ref{fig:results:Rphi}, and \ref{fig:results:weta2} respectively. The four variables describe the lateral shape of the shower in the front and middle layers of the EM calorimeter, and definitions of the variables can be found in Ref.~\cite{Aad:2019tso}. For these comparisons, the detailed simulation is used.
The GAN and VAE show a similar level of agreement with the \GEANT simulation at low energies, but at higher energies the VAE reproduces the distributions more accurately than the GAN. 
 
Another measure of performance is the time taken by a model to generate a new shower relative to the time taken by the full simulation in \GEANT. For the full simulation, the amount of time required to simulate a single particle scales linearly with the particle energy.
For both the VAE and the GAN, there is no energy dependence because the showers are generated in the same manner for all energy points.
Furthermore, the time required by the models to generate a photon shower is an order of magnitude less for photons with an energy of $\SI{8}{\giga\electronvolt}$ and over two orders of magnitude less at $\SI{262}{\giga\electronvolt}$. The memory footprint of the generative models is also very small, with only about $\SI{5}{\mega\byte}$ needed to generate showers for photons. This is small enough to encourage development of such methods for a wider region of the detector.
As is the case for the current fast simulation approaches, the generative models simulate showers quickly enough for other aspects of the simulation code to dominate the required computation time.
 
\subsection{Modelling of particle impact position within a cell}
 
In the case of the GAN, two conditioning parameters are provided to the model in order to account for the impact position of the incoming particle with respect to the centre of the impact cell, extrapolated from its true trajectory. This offset can have an effect on the shower shape, with the showers on average not being centred on the centre of the impact cell which was used to select the window, but instead on the true impact point within that cell.
 
\begin{figure}[tb!p]
\centering
\begin{subfigure}[b]{0.49\textwidth}
\centering
\includegraphics[width=1.0\textwidth]{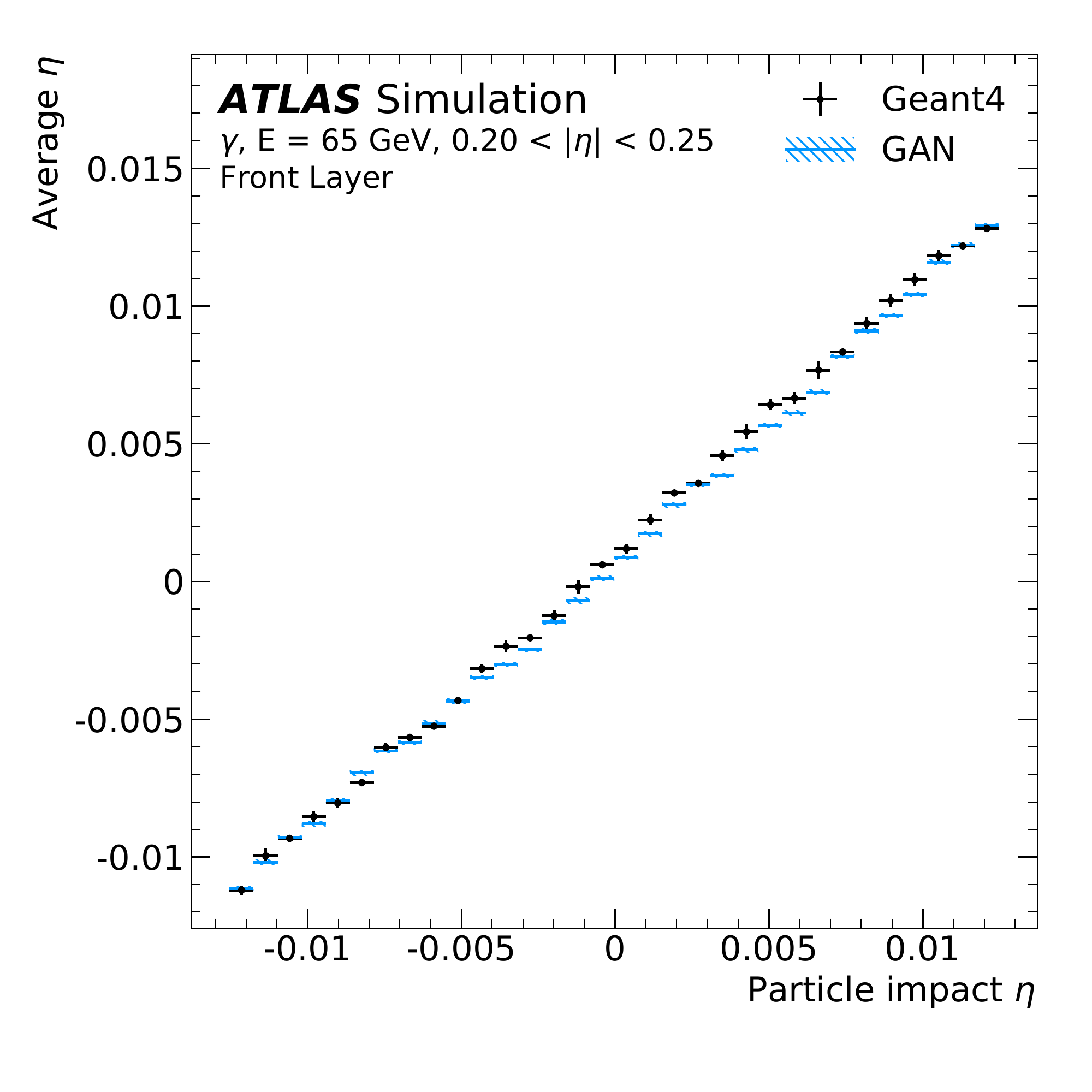}
\caption{}
\label{fig:results:Splot:front}
\end{subfigure}
\begin{subfigure}[b]{0.49\textwidth}
\centering
\includegraphics[width=1.0\textwidth]{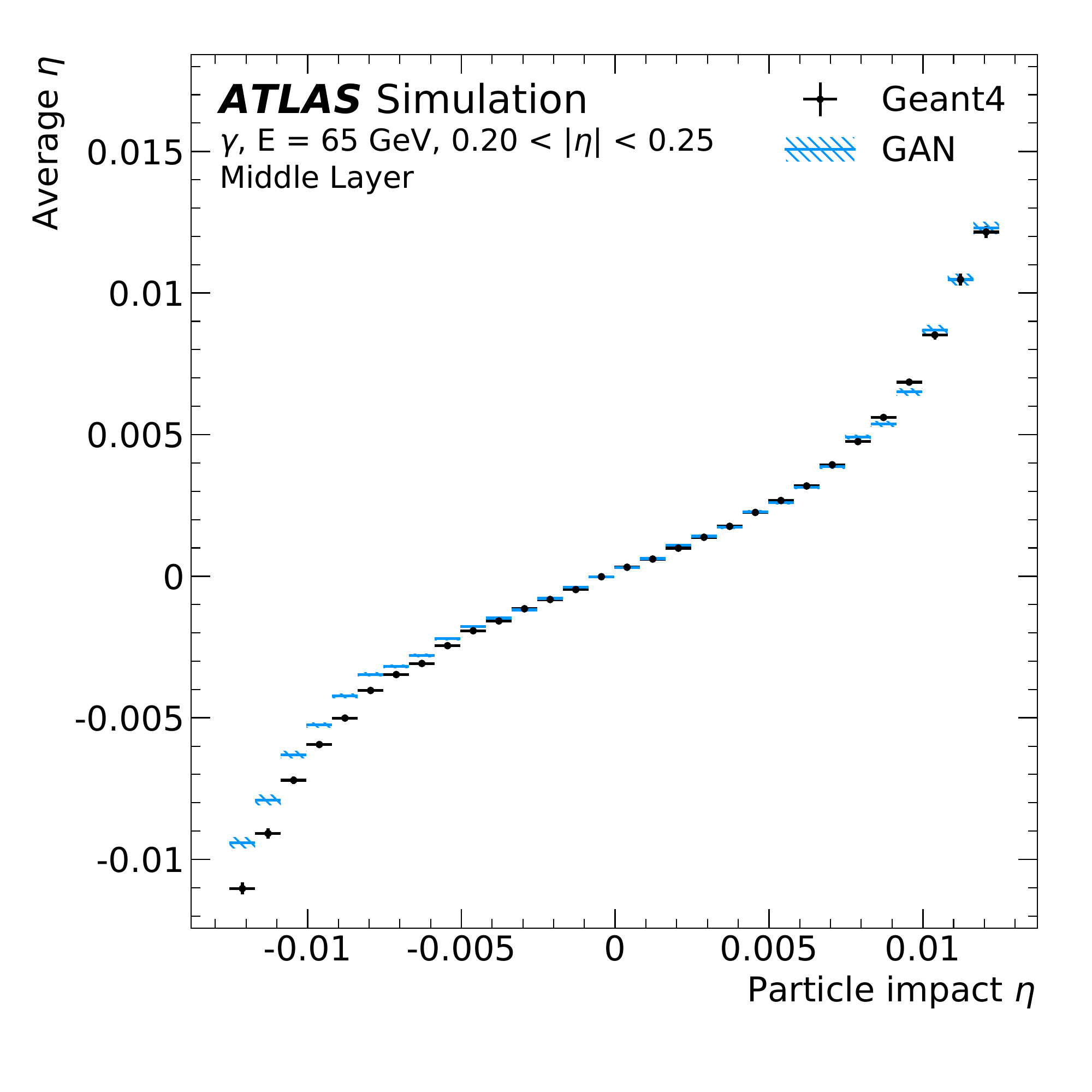}
\caption{}
\label{fig:results:Splot:middle}
\end{subfigure}
\caption{Average reconstructed energy-weighted shower centre in $\eta$ as a function of the impact point displacement in $\eta$ relative to the centre of the impact cell for the \subref{fig:results:Splot:front} front and \subref{fig:results:Splot:middle} middle layers for photons with an energy of $\SI{65}{\giga\electronvolt}$. The distribution for the \GEANT training sample (black markers) is shown as reference points and compared with the GAN (blue markers).
The VAE is not shown on the plots, because for all relative impact values the weighted shower centre is zero with an equal spread.
All showers are simulated consistent with the training samples, considering only the energy deposits in the cells of the four electromagnetic calorimeter layers without digitisation and crosstalk noise.}
\label{fig:results:Splot}
\end{figure}

To visualise the performance of these two conditioning parameters, Figure~\ref{fig:results:Splot} shows the average energy-weighted shower centre in $\eta$, with respect to the centre of the impact cell, as a function of the true displacement of the incoming particle from the centre of the impact cell in the middle layer.
Although there is not perfect agreement, the trend is well modelled with `S' shapes for each impact cell. This level of agreement results in an improvement in the lateral development of generated showers in comparison with models trained without the impact position information.
For the VAE, which was not conditioned on these parameters, the average position is always the centre of the impact cell in each layer for all values of the particle impact $\eta$. In the front layer, this appears as a step function with seven steps, one for each cell, and a flat distribution in the middle layer as the impact position is always in the centre cell.
 
\subsection{Interpolation and extrapolation}
 
A key consideration when training a generative model is its ability to infer and model distributions for points which were not seen during training. In the case of shower simulation, one of the key properties of an incoming particle is its energy, which is used as a conditioning input in both of the architectures presented in this paper. However, in total only nine discrete energies were used in the simulation of the samples used to train the networks, whereas there is a continuous distribution of possible energies.
 
A test of the performance of these generative models is how well showers for particles with novel MC-truth energies are modelled. To test this, two cases are considered, interpolation and extrapolation. Two new energy points are considered, 25~\GeV, which falls roughly halfway between two energy points (16~\GeV\ and 32~\GeV), and 524~\GeV, which is twice as large as the highest energy point seen during training.
New showers are generated at these energies using the VAE and GAN and compared with showers simulated using the full simulation chain.
Generative models are expected to perform well at interpolation. However, they are not expected to deliver the same level of performance when extrapolating outside of the known distributions.
 
\begin{figure}[tb!p]
\centering
\begin{subfigure}[b]{0.49\textwidth}
\centering
\includegraphics[width=1.0\textwidth]{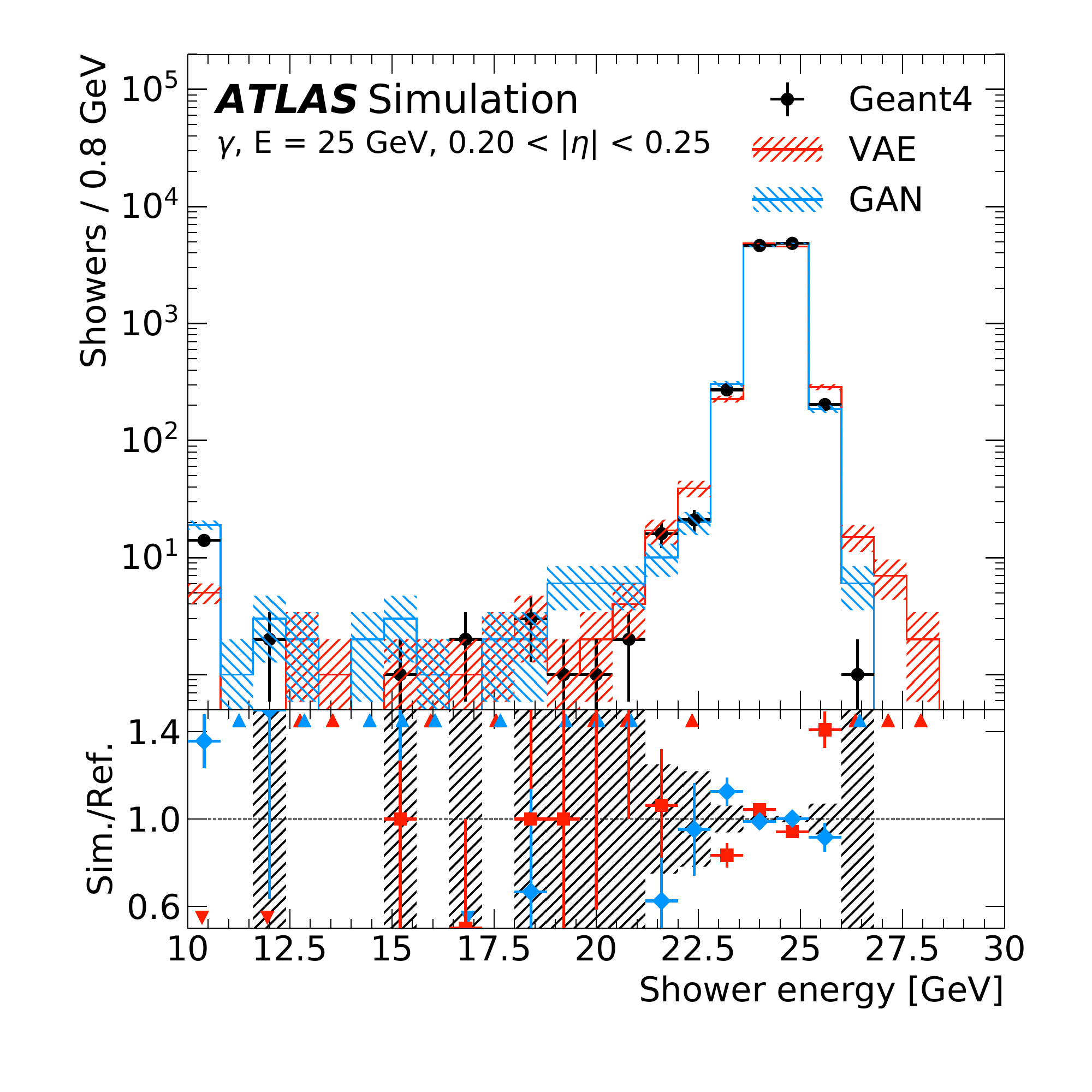}
\caption{}
\label{fig:results:interp:e}
\end{subfigure}
\begin{subfigure}[b]{0.49\textwidth}
\centering
\includegraphics[width=1.0\textwidth]{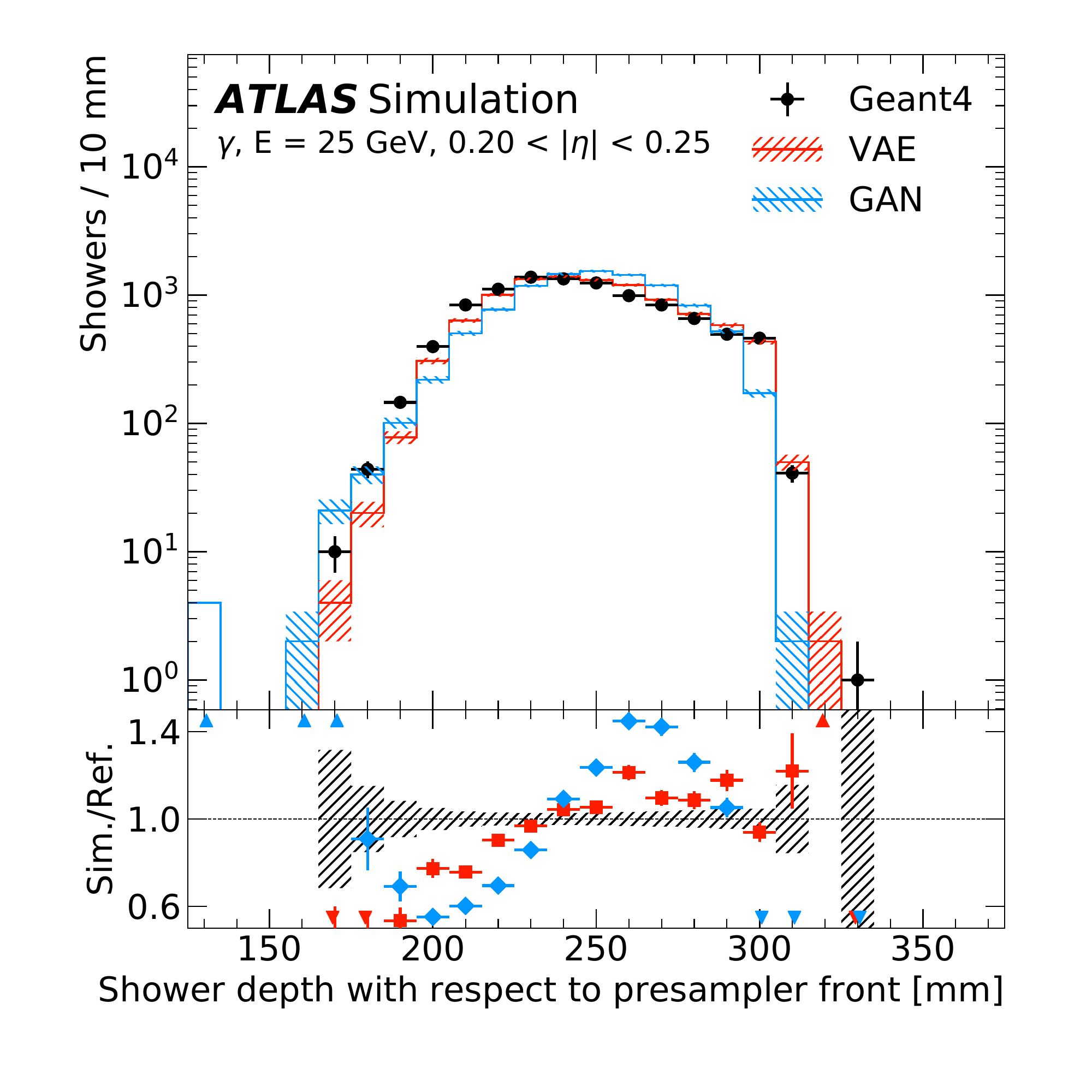}
\caption{}
\label{fig:results:interp:depth}
\end{subfigure}
\caption{The \subref{fig:results:interp:e} total energy response of the calorimeter  and \subref{fig:results:interp:depth} reconstructed longitudinal shower centre (shower depth) for photons with an energy of $\SI{25}{\giga\electronvolt}$. This energy is not seen during training and demonstrates the ability of the models to interpolate to other energies.
The distributions for the \GEANT training data (black markers) are shown as reference points and compared with those from a VAE (solid red line) and a GAN (solid blue line). The error bars and the hatched bands indicate the statistical uncertainty of the training data and the synthesised samples, respectively. The underflow and overflow are included in the first and last bin of each distribution, respectively.
Arrows at the top and bottom of the ratio indicate when the values for either model in a bin fall outside of the displayed axis range.
All showers are simulated consistent with the training samples, considering only the energy deposits in the cells of the four electromagnetic calorimeter layers without digitisation and crosstalk noise.}
\label{fig:results:interp}
\end{figure}
 
\begin{figure}[tb!p]
\centering
\includegraphics[width=0.49\textwidth]{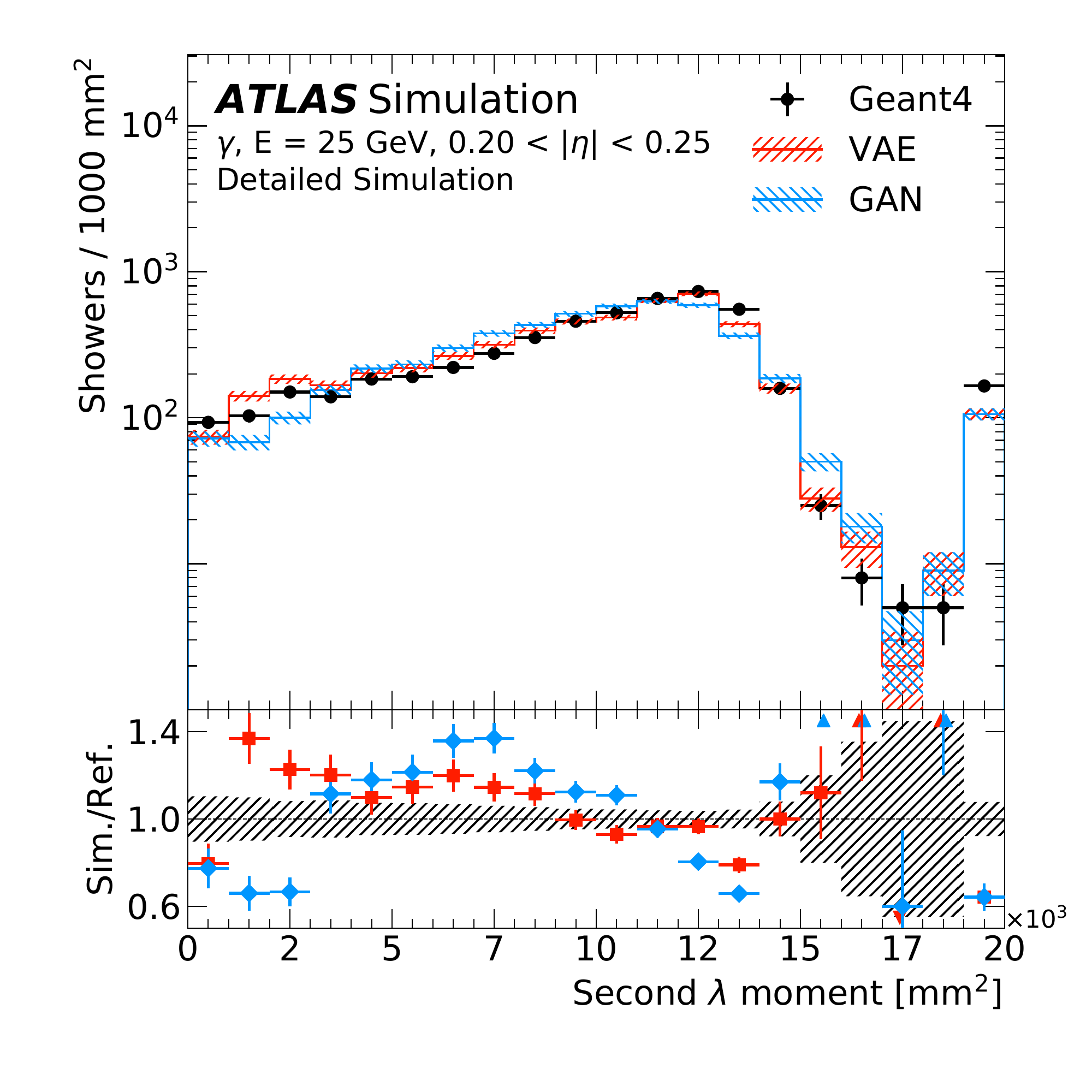}
\caption{Reconstructed second $\lambda$ moment of showers for photons with an energy of $\SI{25}{\giga\electronvolt}$. This energy is not seen during training and demonstrates the ability of the models to interpolate to other energies.
The distributions for the full detector simulation (black markers) are shown as reference points and compared with those from a VAE (solid red line) and a GAN (solid blue line). The error bars and the hatched bands indicate the statistical uncertainty of the detailed simulation and the synthesised samples, respectively.
The underflow and overflow are included in the first and last bin of each distribution, respectively.
Arrows at the top and bottom of the ratio indicate when the values for either model in a bin fall outside of the displayed axis range.
All showers are simulated using the full simulation chain, including both the inner detector and the digitisation and crosstalk noise.}
\label{fig:results:interp:lambda2}
\end{figure}
 
\begin{figure}[tb!p]
\centering
\begin{subfigure}[b]{0.49\textwidth}
\centering
\includegraphics[width=\textwidth]{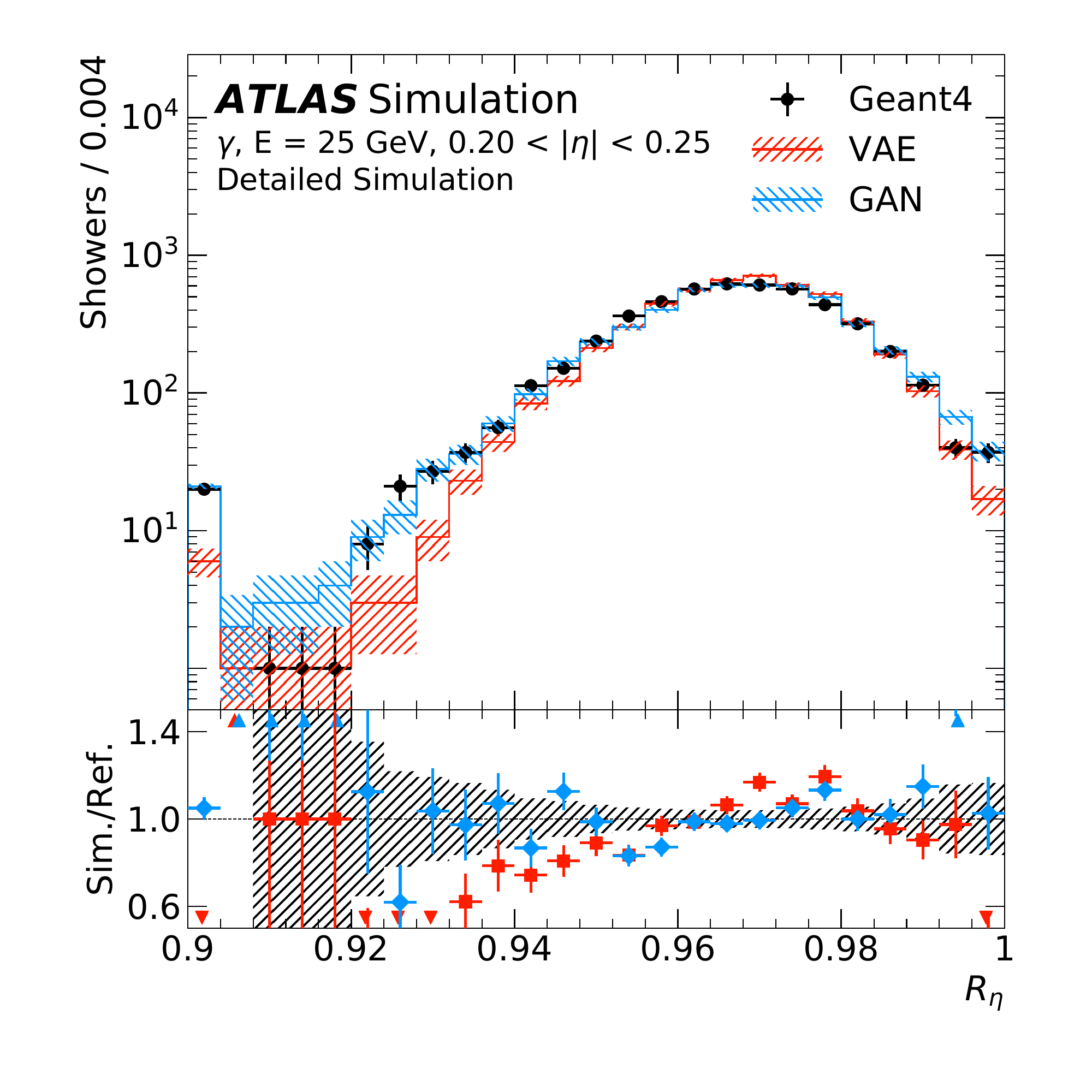}
\caption{}
\label{fig:results:Reta:25}
\end{subfigure}
\begin{subfigure}[b]{0.49\textwidth}
\centering
\includegraphics[width=\textwidth]{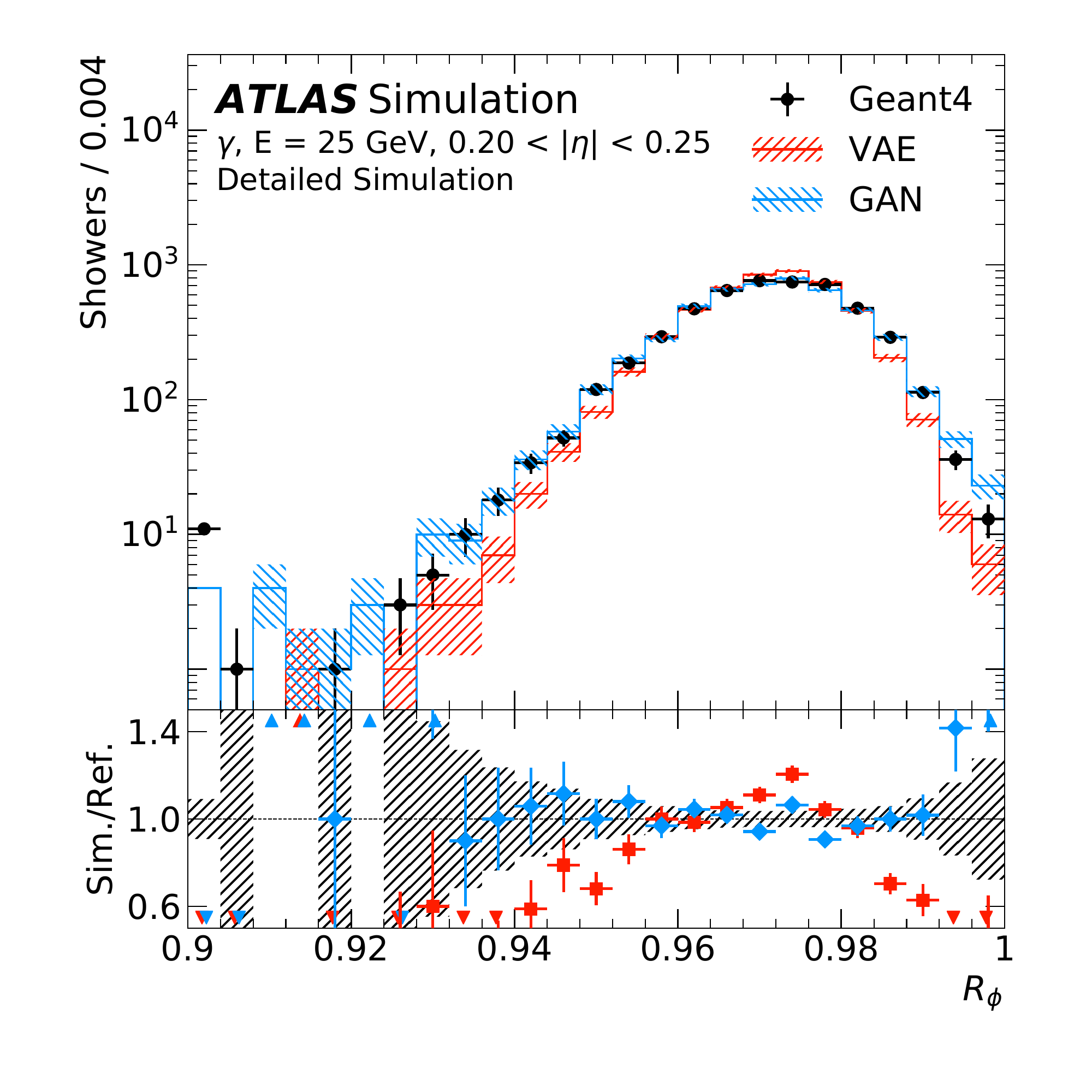}
\caption{}
\label{fig:results:Rphi:25}
\end{subfigure}
\caption{Energy ratios \subref{fig:results:Reta:25} $R_\eta$ and \subref{fig:results:Rphi:25} $R_\phi$ in the middle EM layer for photons with an energy of $\SI{25}{\giga\electronvolt}$.
This energy is not seen during training and demonstrates the ability of the models to interpolate to other energies.
The distributions for the full detector simulation (black markers) are shown as reference points and compared with those from a VAE (solid red line) and a GAN (solid blue line). The error bars indicate the statistical uncertainty of the detailed simulation and the synthesised samples. The underflow and overflow are included in the first and last bin of each distribution, respectively.
Arrows at the top and bottom of the ratio indicate when the values for either model in a bin fall outside of the displayed axis range.
All showers are simulated using the full simulation chain, including both the inner detector and the digitisation and crosstalk noise.}
\label{fig:results:interp:Retaphi}
\end{figure}

To assess the interpolation performance, showers for photons with an MC-truth energy of $\SI{25}{\giga\electronvolt}$ are generated using the VAE and GAN, and compared with showers produced using the \GEANT simulation. The total energy response of the detector and the reconstructed longitudinal shower centre for these showers are shown in Figure~\ref{fig:results:interp} compared with the training data. The second $\lambda$ moments are shown in Figure~\ref{fig:results:interp:lambda2} compared with the detailed simulation. Figure~\ref{fig:results:interp:Retaphi} shows the $R_\eta$ and $R_\phi$ shower shape observables used in electron and photon identification compared with the detailed simulation.
The level of performance is the same as seen for neighbouring energy points used to train the models, demonstrating that the models are able to interpolate to energy points not seen during training.
 
\begin{figure}[tb!p]
\centering
\includegraphics[width=0.49\textwidth]{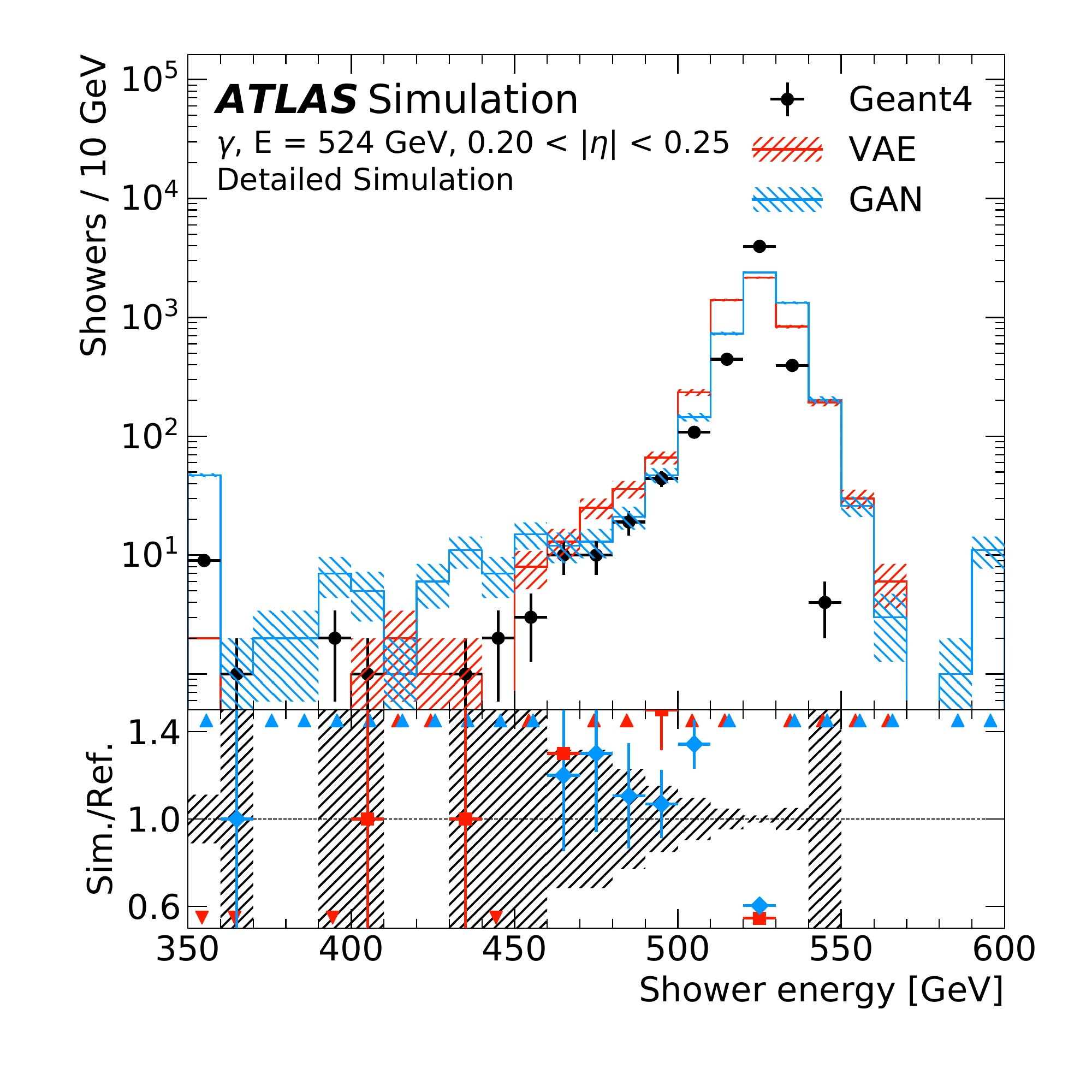}
\caption{Total energy response of the calorimeter 
for photons with an energy of $\SI{524}{\giga\electronvolt}$.
This energy is not seen during training and illustrates the extent to which the models are able to extrapolate to higher energies.
The calorimeter response for the full detector simulation (black markers) is shown as reference points and compared with that from a VAE (solid red line) and a GAN (solid blue line). The error bars and the hatched bands indicate the statistical uncertainty of the detailed simulation and the synthesised samples, respectively.
The underflow and overflow are included in the first and last bin of each distribution, respectively.
Arrows at the top and bottom of the ratio indicate when the values for either model in a bin fall outside of the displayed axis range.
All showers are simulated using the full simulation chain, including both the inner detector and the digitisation and crosstalk noise.}
\label{fig:results:extrap}
\end{figure}

\begin{figure}[tb!p]
\centering
\begin{subfigure}[b]{0.49\textwidth}
\centering
\includegraphics[width=\textwidth]{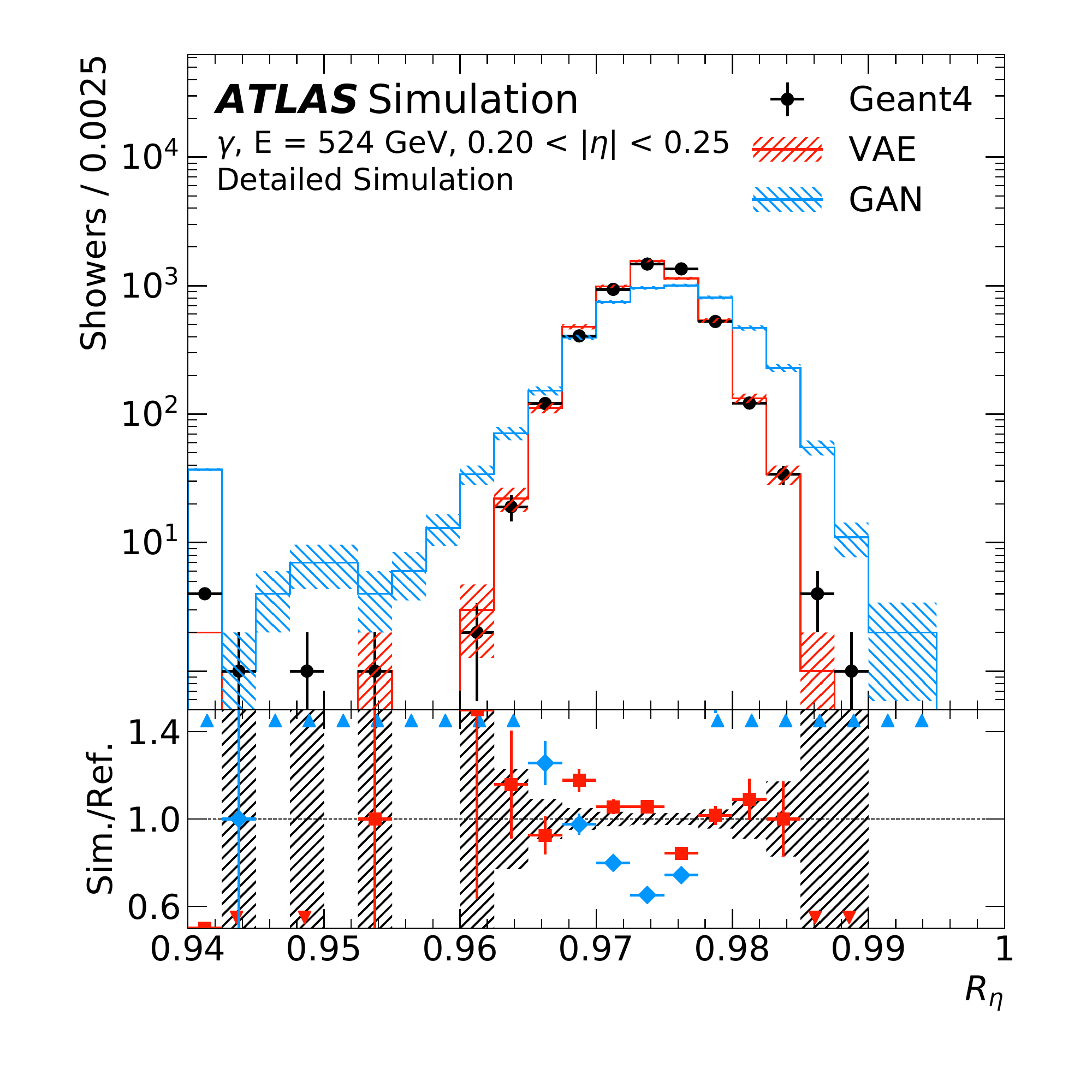}
\caption{}
\label{fig:results:Reta:524}
\end{subfigure}
\begin{subfigure}[b]{0.49\textwidth}
\centering
\includegraphics[width=\textwidth]{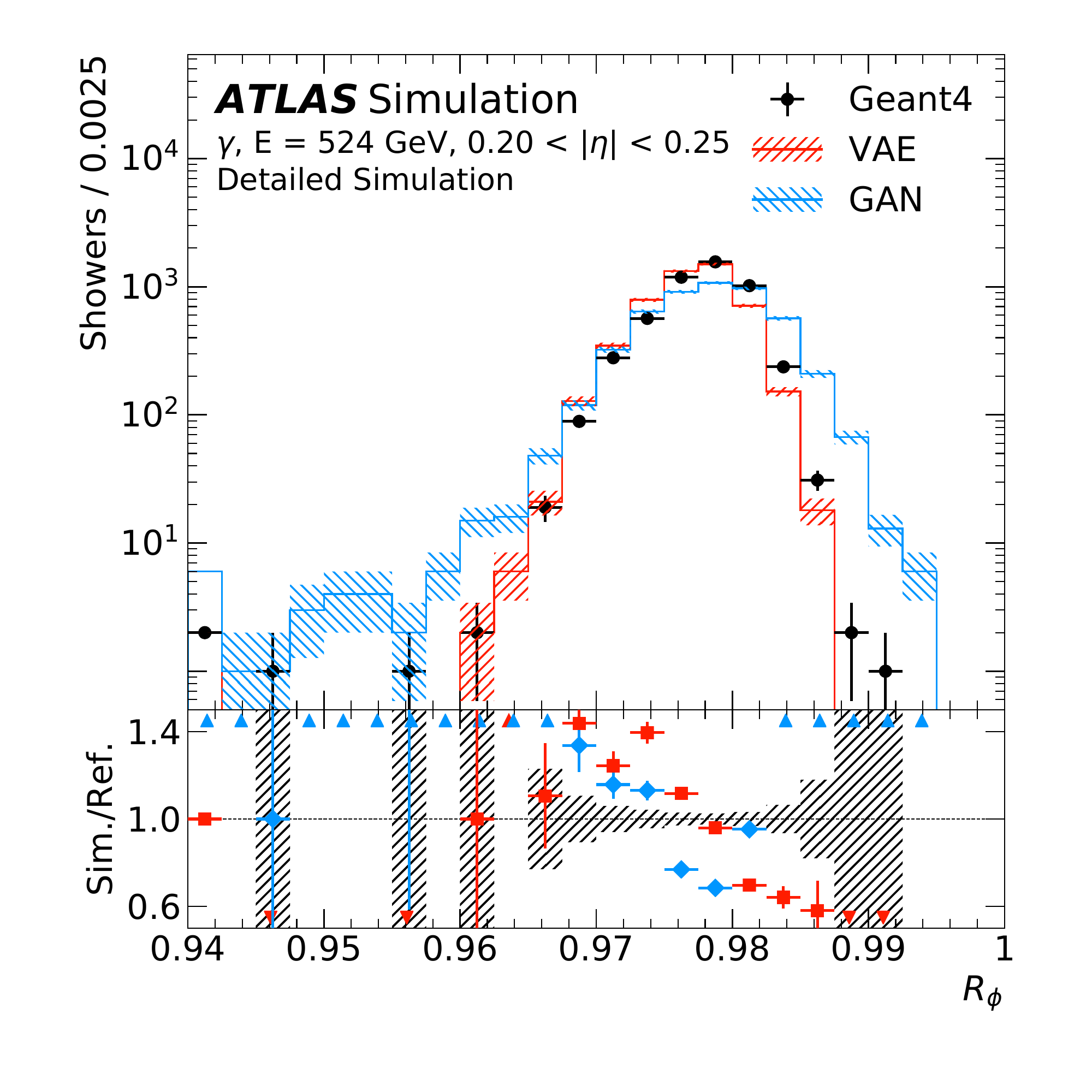}
\caption{}
\label{fig:results:Rphi:524}
\end{subfigure}
\caption{Energy ratios \subref{fig:results:Reta:524} $R_\eta$ and \subref{fig:results:Rphi:524} $R_\phi$ in the middle EM layer for photons with an energy of $\SI{524}{\giga\electronvolt}$.
This energy is not seen during training and demonstrates the ability of the models to interpolate to other energies.
The distributions for the full detector simulation (black markers) are shown as reference points and compared with those from a VAE (solid red line) and a GAN (solid blue line). The error bars indicate the statistical uncertainty of the detailed simulation and the synthesised samples. The underflow and overflow are included in the first and last bin of each distribution, respectively.
Arrows at the top and bottom of the ratio indicate when the values for either model in a bin fall outside of the displayed axis range.
All showers are simulated using the full simulation chain, including both the inner detector and the digitisation and crosstalk noise.}
\label{fig:results:extrap:Retaphi}
\end{figure}

To test the extrapolation performance, showers are generated for photons with an MC-truth energy of $\SI{524}{\giga\electronvolt}$. The chosen energy point is a factor of two larger than any of the photon energies seen during training. Figure~\ref{fig:results:extrap} shows the total calorimeter energy response in comparison with the detailed simulation.
The total energy response is not as well modelled as the highest energy point used in training, seen in Figure~\ref{fig:results:etot:260}, illustrating the difficulty in extrapolating to higher energies, although both models generate showers with energies close to the target values.
One way to address this challenge is to include showers at additional energy points in the training which exceed the expected kinematic end point for showers in the ATLAS calorimeter. Fewer examples are needed in the extrapolation region than in the area of interest, but this does illustrate one of the drawbacks of using generative models for simulation.
Figure~\ref{fig:results:extrap:Retaphi} shows the $R_\eta$ and $R_\phi$ shower shape observables for this energy point, with similar agreement to the detailed simulation observed as for photons generated at $\SI{262}{\giga\electronvolt}$.
 
\section{Challenges}
\label{sec:challenges}
 
Developing these models has highlighted challenges from both the physics context and machine learning domain. While many of these were successfully overcome in this paper, there remain outstanding challenges for future approaches to address.
 
One of the main difficulties in generative modelling is defining what is the best model. Because the models are trained on the cell energies, and particle showers are stochastic in nature, it is not possible to look at examples of individual simulated showers and determine if the model performs well. Instead, distributions need to be studied to determine not how real a single shower looks, but rather whether the probability density functions for all shower types are accurately modelled, given the underlying correlations.
Furthermore, as there are a large number of physics distributions that must be considered, it is not trivial to combine these into a single quantitative score of performance.
It is in particular difficult to assign a value of relative importance for individual distributions, especially as often what is important are specific features within individual distributions, such as the sharp drop off in the total shower energy distribution. Furthermore, different observables can be more or less important at different energy scales.
 
Whilst the models in this work were being developed, performance was most commonly evaluated through qualitative inspection and comparison using a wide range of physics motivated observables.
Aggregate scores calculated with methods such as the Frechet inception distance~\cite{fidscore} show promise in quantitatively assessing the modelling performance of generative models in high energy physics~\cite{Kansal1,Kansal2}, however it does not form a definitive measure and although inputs to the metric can be chosen to target certain observables, this does not dictate which aspects and features of distributions it will consider as important.
Another direction of promise is using classifiers to compare the relative performance of generative models for detector simulation in a more systematic approach has become more common~\cite{Krause:2021ilc,Krause:2021wez,Mikuni:2022xry,Krause:2022jna,Diefenbacher:2023vsw,Buckley:2023rez,Das:2023ktd,Pang:2023wfx,Amram:2023onf}.
However, although this approach can consider many observables simultaneously, there are still several considerations.
For example, when training a binary classifier to separate the generated showers from full simulation, if a model is very far from the target distributions little insight can be gained from their application, and it is difficult to compare between models.
Furthermore, the choice of inputs to the classification networks play a large role in the inferred performance, and the relative importance of input features within a classifier cannot be determined in advance.
Finally, although such an approach can rank models based on similarity to the target distributions, it does not say how suitable for physics applications a model is.
Therefore, although it is possible to rule out poorly performing models with some quantitative metrics, human judgement has to be used to determine which model performs best overall.
Often this results in manual investigation of a large number of distributions.
 
As the performance of generative models improves, studying how they compare in downstream tasks for particle reconstruction will become more important.
In the example of photon shower generation this would, for example, include measuring the performance in photon and electron identification algorithms compared to \GEANT, as well as the diphoton invariant mass in Higgs boson decays.
As future models will also be integrated into the ATLAS simulation framework to evaluate their performance outside of an idealised setting,
this provides an opportunity to study a wide range of quantitative metrics, including classifier-based approaches, 
and determining a set of metrics which are most relevant to various aspects of the full simulation and analysis chain.
For example, it would be particularly interesting to study how well various metrics correlate to the impact on the energy resolution of reconstructed physics object, and the identification efficiency with algorithms optimsed using \GEANT for the detector simulation.
This insight would very valuable whilst monitoring and optimising the performance of models during development.
However, as the tension between the current VAE and GAN are visible on a log scale, we leave the detailed study of these metrics to future work.

The difficulty of selecting the best-performing model also contributes to one of the challenges in optimising a GAN.
Since GANs do not converge in training and are constantly evolving, finding the best-performing model for the task requires finding not only the best hyperparameters but also the optimal epoch of each training.
This is often called `epoch picking' and is a common technique in the application of GANs.
In combination with their long training times, having to qualitatively evaluate a large number of models and epochs made it difficult to explore the full hyperparameter space and identify the best-performing GAN.
 
In the case of VAEs, which are trained to minimise a target objective, the process of optimising the model is simpler; only the hyperparameters need to be optimised and the epoch where the training objective is minimised can be safely assumed to be the best epoch. Furthermore, as they are also much simpler in structure, VAEs are typically faster than GANs to train and therefore optimise. This presents one potential advantage of VAEs relative to GANs.
However, the difficulty of qualitatively choosing the best-performing model over a large number of distributions remains a challenge.

Another challenge is encountered when extending the techniques to include other regions of the detector.
Individual layers do not cover the whole $\eta$ range, and different combinations of layers cover different regions of the detector.
As a result, the granularity of the cells in each respective layer does not remain constant across the $\eta$ range, and the number of relevant layers for the simulation varies depending on particle type and displacement in $\eta$.
To overcome this challenge, the changing composition of layers and non-constant cell granularities will need to be addressed.
One approach to the changing cell geometry is to move to a polar coordinate voxelisation, centred on the extrapolated particle trajectory, which can remain constant across $\eta$. This approach has been adopted in the latest version of the ATLAS fast calorimeter simulation for pions~\cite{atlfast3}. The granularity of these voxels will need to be optimised to preserve shower observables after assigning the energy deposits to cells. In total, the ATLAS calorimeter has 24 different layers of cells, which cover different $\eta$ regions. Here, modelling the showers in the boundary regions might prove particularly challenging, especially where the depths of layers can change significantly.

\FloatBarrier


\FloatBarrier
\section{Conclusion}
\label{sec:conclusion}

This paper presents a novel application of generative models for simulating particle showers in the ATLAS electromagnetic calorimeter. Two models, a VAE and a GAN, are trained to learn the response of the EM calorimeter to unconverted photons with energies between approximately $\SI{1}{\giga\electronvolt}$ and $\SI{262}{\giga\electronvolt}$ in the range $0.20 < |\eta| < 0.25$. 
The typical computational time required to generate new showers with either the VAE or the GAN is up to two orders of magnitude less than in the full \GEANT simulation, with a very small memory footprint of order 5~MB needed to model the slice of the detector covered by these models.
The VAE is able to model the shape of the showers' distributions better on average than the GAN, whilst the GAN is able to model the total energy of the showers more accurately. Furthermore, due to additional conditional parameters, the GAN is able to capture underlying correlations with the impact position of the incoming particle not learned by the VAE.
 
The complexity of the geometry of the ATLAS EM calorimeter, in particular the correlations between neighbouring cells, within one layer or in depth, is learned directly from the training samples without being reflected in the architecture, which bodes well for the generality of the technique. 
For the two models presented here, a key area of future improvement for the GAN is in the longitudinal shower development, whereas for the VAE it lies in accurately modelling the total energy of the shower.
Additionally, for the reconstruction and identification of electrons and photons, there are several key distributions which would need to match the detailed simulation from \GEANT more accurately in order to avoid introducing further sources of disagreement between simulated showers and those recorded by the ATLAS detector in particle collisions.
Nonetheless, the techniques presented in this paper demonstrate the feasibility of using such models for fast calorimeter simulation for the ATLAS experiment in the future and show a possible way to complement current techniques in order to achieve the required accuracy for physics analysis.


\section*{Acknowledgements}


We thank CERN for the very successful operation of the LHC, as well as the
support staff from our institutions without whom ATLAS could not be
operated efficiently.
 
We acknowledge the support of
ANPCyT, Argentina;
YerPhI, Armenia;
ARC, Australia;
BMWFW and FWF, Austria;
ANAS, Azerbaijan;
CNPq and FAPESP, Brazil;
NSERC, NRC and CFI, Canada;
CERN;
ANID, Chile;
CAS, MOST and NSFC, China;
Minciencias, Colombia;
MEYS CR, Czech Republic;
DNRF and DNSRC, Denmark;
IN2P3-CNRS and CEA-DRF/IRFU, France;
SRNSFG, Georgia;
BMBF, HGF and MPG, Germany;
GSRI, Greece;
RGC and Hong Kong SAR, China;
ISF and Benoziyo Center, Israel;
INFN, Italy;
MEXT and JSPS, Japan;
CNRST, Morocco;
NWO, Netherlands;
RCN, Norway;
MEiN, Poland;
FCT, Portugal;
MNE/IFA, Romania;
MESTD, Serbia;
MSSR, Slovakia;
ARRS and MIZ\v{S}, Slovenia;
DSI/NRF, South Africa;
MICINN, Spain;
SRC and Wallenberg Foundation, Sweden;
SERI, SNSF and Cantons of Bern and Geneva, Switzerland;
MOST, Taiwan;
TENMAK, T\"urkiye;
STFC, United Kingdom;
DOE and NSF, United States of America.
In addition, individual groups and members have received support from
BCKDF, CANARIE, Compute Canada and CRC, Canada;
PRIMUS 21/SCI/017 and UNCE SCI/013, Czech Republic;
COST, ERC, ERDF, Horizon 2020 and Marie Sk{\l}odowska-Curie Actions, European Union;
Investissements d'Avenir Labex, Investissements d'Avenir Idex and ANR, France;
DFG and AvH Foundation, Germany;
Herakleitos, Thales and Aristeia programmes co-financed by EU-ESF and the Greek NSRF, Greece;
BSF-NSF and MINERVA, Israel;
Norwegian Financial Mechanism 2014-2021, Norway;
NCN and NAWA, Poland;
La Caixa Banking Foundation, CERCA Programme Generalitat de Catalunya and PROMETEO and GenT Programmes Generalitat Valenciana, Spain;
G\"{o}ran Gustafssons Stiftelse, Sweden;
The Royal Society and Leverhulme Trust, United Kingdom.
 
The crucial computing support from all WLCG partners is acknowledged gratefully, in particular from CERN, the ATLAS Tier-1 facilities at TRIUMF (Canada), NDGF (Denmark, Norway, Sweden), CC-IN2P3 (France), KIT/GridKA (Germany), INFN-CNAF (Italy), NL-T1 (Netherlands), PIC (Spain), ASGC (Taiwan), RAL (UK) and BNL (USA), the Tier-2 facilities worldwide and large non-WLCG resource providers. Major contributors of computing resources are listed in Ref.~\cite{ATL-SOFT-PUB-2023-001}.


\clearpage

\printbibliography
 
\clearpage
 
\begin{flushleft}
\hypersetup{urlcolor=black}
{\Large The ATLAS Collaboration}

\bigskip

\AtlasOrcid[0000-0002-6665-4934]{G.~Aad}$^\textrm{\scriptsize 101}$,
\AtlasOrcid[0000-0002-5888-2734]{B.~Abbott}$^\textrm{\scriptsize 119}$,
\AtlasOrcid[0000-0002-7248-3203]{D.C.~Abbott}$^\textrm{\scriptsize 102}$,
\AtlasOrcid[0000-0002-2788-3822]{A.~Abed~Abud}$^\textrm{\scriptsize 36}$,
\AtlasOrcid[0000-0002-1002-1652]{K.~Abeling}$^\textrm{\scriptsize 55}$,
\AtlasOrcid[0000-0002-2987-4006]{D.K.~Abhayasinghe}$^\textrm{\scriptsize 94}$,
\AtlasOrcid[0000-0002-8496-9294]{S.H.~Abidi}$^\textrm{\scriptsize 29}$,
\AtlasOrcid[0000-0002-9987-2292]{A.~Aboulhorma}$^\textrm{\scriptsize 35e}$,
\AtlasOrcid[0000-0001-5329-6640]{H.~Abramowicz}$^\textrm{\scriptsize 150}$,
\AtlasOrcid[0000-0002-1599-2896]{H.~Abreu}$^\textrm{\scriptsize 149}$
% ...
\bigskip
\\

% ...
$^{29}$Physics Department, Brookhaven National Laboratory, Upton NY; United States of America.\\
% ...
$^{35}$$^{(a)}$Facult\'e des Sciences Ain Chock, R\'eseau Universitaire de Physique des Hautes Energies - Universit\'e Hassan II, Casablanca;$^{(b)}$Facult\'{e} des Sciences, Universit\'{e} Ibn-Tofail, K\'{e}nitra;$^{(c)}$Facult\'e des Sciences Semlalia, Universit\'e Cadi Ayyad, LPHEA-Marrakech;$^{(d)}$LPMR, Facult\'e des Sciences, Universit\'e Mohamed Premier, Oujda;$^{(e)}$Facult\'e des sciences, Universit\'e Mohammed V, Rabat;$^{(f)}$Institute of Applied Physics, Mohammed VI Polytechnic University, Ben Guerir; Morocco.\\
$^{36}$CERN, Geneva; Switzerland.\\
% ...
$^{55}$II. Physikalisches Institut, Georg-August-Universit\"{a}t G\"ottingen, G\"ottingen; Germany.\\
% ...
$^{94}$Department of Physics, Royal Holloway University of London, Egham; United Kingdom.\\
% ...
$^{101}$CPPM, Aix-Marseille Universit\'e, CNRS/IN2P3, Marseille; France.\\
$^{102}$Department of Physics, University of Massachusetts, Amherst MA; United States of America.\\
% ...
$^{119}$Homer L. Dodge Department of Physics and Astronomy, University of Oklahoma, Norman OK; United States of America.\\
% ...
$^{149}$Department of Physics, Technion, Israel Institute of Technology, Haifa; Israel.\\
$^{150}$Raymond and Beverly Sackler School of Physics and Astronomy, Tel Aviv University, Tel Aviv; Israel.\\
% ...
$^{a}$ Also Affiliated with an institute covered by a cooperation agreement with CERN.\\
$^{b}$ Also at Borough of Manhattan Community College, City University of New York, New York NY; United States of America.\\
$^{c}$ Also at Bruno Kessler Foundation, Trento; Italy.\\
$^{d}$ Also at Center for High Energy Physics, Peking University; China.\\
$^{e}$ Also at Centro Studi e Ricerche Enrico Fermi; Italy.\\
$^{f}$ Also at CERN, Geneva; Switzerland.\\
% ...

\end{flushleft}


\end{document}